%% file: main.tex
\pdfoutput=1

\RequirePackage{rotating}

\documentclass[acmsmall,nonacm=true,screen]{acmart}

\usepackage{verbatim}
\usepackage{xspace}
\usepackage{tcolorbox}
\usepackage{booktabs}

\usepackage{tikz}
\usepackage{pgfplots}
\pgfplotsset{width=10cm,compat=1.9}

\usepackage{pdflscape}

\usepackage{fontawesome}
\usepackage{xcolor}

\usepackage{lookup}
\input{aliases.tex}

\usetikzlibrary{arrows.meta,
                chains,
                positioning}

\usepackage{mybchart}
\usepackage{tabularx}

\usepackage{stackengine}

\begin{document}

\input{participants.tex}

\title{A Tale of Two Cities: Software Developers Working from Home During the COVID-19 Pandemic}

\author{Denae Ford}
\email{denae@microsoft.com}
\affiliation{%
  \institution{Microsoft Research}
  \streetaddress{1 Microsoft Way}
  \city{Redmond}
  \state{WA}
  \postcode{98052}
}
\author{Margaret-Anne Storey}
\email{mstorey@uvic.ca}
\affiliation{%
  \institution{University of Victoria}
  \streetaddress{PO Box 3055, STN CSC}
  \city{Victoria}
  \state{BC}
  \country{Canada}
  \postcode{V8W 3P6}
}
\author{Thomas Zimmermann}
\email{tzimmer@microsoft.com}
\affiliation{%
  \institution{Microsoft Research}
  \streetaddress{1 Microsoft Way}
  \city{Redmond}
  \state{WA}
  \postcode{98052}
}
\author{Christian Bird}
\email{cbird@microsoft.com}
\affiliation{%
  \institution{Microsoft Research}
  \streetaddress{1 Microsoft Way}
  \city{Redmond}
  \state{WA}
  \postcode{98052}
}
\author{Sonia Jaffe}
\email{sonia.jaffe@microsoft.com}
\affiliation{%
  \institution{Microsoft Corp.}
  \streetaddress{1 Microsoft Way}
  \city{Redmond}
  \state{WA}
  \postcode{98052}
}
\author{Chandra Maddila}
\email{chmaddil@microsoft.com}
\affiliation{%
  \institution{Microsoft Research}
  \streetaddress{1 Microsoft Way}
  \city{Redmond}
  \state{WA}
  \postcode{98052}
}
\author{Jenna L.\ Butler}
\email{jennbu@microsoft.com}
\affiliation{%
  \institution{Microsoft Research}
  \streetaddress{1 Microsoft Way}
  \city{Redmond}
  \state{WA}
  \postcode{98052}
}
\author{Brian Houck}
\email{Brian.Houck@microsoft.com}
\affiliation{%
  \institution{Microsoft Corp.}
  \streetaddress{1 Microsoft Way}
  \city{Redmond}
  \state{WA}
  \postcode{98052}
}
\author{Nachiappan Nagappan}
\email{nnagappan@acm.org}
\affiliation{%
  \institution{Facebook}
  \streetaddress{1 Hacker Way}
  \city{Menlo Park}
  \state{CA}
  \postcode{94025}
}

\begin{abstract}
The \covid pandemic has shaken the world to its core and has provoked an overnight exodus of developers that normally worked in an office setting to working from home. 
The magnitude of this shift and the factors that have accompanied this new unplanned work setting go beyond what the software engineering community has previously understood to be remote work. %
To find out how developers and their productivity were affected, \BEGINADDED we distributed two surveys (with a combined total of \surveytotalresponses responses that answered all required questions)
\ENDADDED ---weeks apart to understand the presence and prevalence of the benefits, challenges, and opportunities to improve this special circumstance of remote work.
From our thematic qualitative analysis and statistical quantitative analysis, we find that there is a \emph{dichotomy} of developer experiences influenced by many different factors (that for some are a benefit, while for others a challenge). 
For example, a benefit for some was being close to family members  but for others having family members share their working space and interrupting their focus, was a challenge.
Our surveys led to powerful narratives from respondents and
revealed the scale at which these experiences exist to provide insights as to how the future of (pandemic) remote work can evolve.
\end{abstract}

\authorsaddresses{%
Authors' addresses: Denae Ford, denae@microsoft.com; Thomas Zimmermann, tzimmer@microsoft.com; Christian Bird, cbird@microsoft.com; Sonia Jaffe, sonia.jaffe@microsoft.com; Chandra Maddila, chmaddil@microsoft.com; Jenna L.\ Butler, jennbu@microsoft.com; Brian Houck, Brian.Houck@microsoft.com, Microsoft, 1 Microsoft Way, Redmond, WA, USA, 98052; Margaret-Anne Storey, mstorey@uvic.ca, University of Victoria, PO Box 3055, STN CSC, Victoria, BC, Canada, V8W 3P6; Nachiappan Nagappan, nnagappan@acm.org, Facebook, 1 Hacker Way, Menlo Park, CA, USA, 94025. Margaret-Anne Storey worked on this project while being a vendor at Microsoft. Nachiappan Nagappan worked on this project while being a researcher at Microsoft.
}

\maketitle

\renewcommand{\shortauthors}{D.~Ford, M.-A.~Storey, T.~Zimmermann, C.~Bird, S.~Jaffe, C.~Maddila, J.~Butler, B.~Houck, and N.~Nagappan}

\section{Introduction}

\begin{quote}
\emph{Charles Dickens' ``A Tale of Two Cities'' begins, ``It was the best of times, it was the worst of times.'' Adapting Dickens' line to leading an engineering team during the global pandemic, I'd say ``We're doing very well, we're barely hanging in there.''} \hfill --- Shane O'Flynn~\cite{oflynn2020lead}
\end{quote}

Software engineering is a complex technical, knowledge-based task that requires focused, uninterrupted work~\cite{meyer2014software} while still coordinating and collaborating with other developers and stakeholders~\cite{lanubile2007collaboration}. Despite a need for intense periods of coordination, collaboration, and communication for managing intricate dependencies within and across systems, there are compelling success stories of how developers can effectively develop high-quality, complex software in a distributed fashion. Successful open source communities, globally distributed software projects and fully remote software companies are all testaments to distributed and remote work.  Decades of engineering tools (such as version control and continuous integration tools) and knowledge sharing tools (such as email, Stack Overflow, and Wikipedia) were conceived and designed by developers for developers to manage the collaborative and distributed nature of software engineering~\cite{storey2016social}.

Despite having rich tools to support distributed development, many software companies believe there are significant advantages to working in a co-located fashion, with many advocating for close proximity among developers (such as in shared team rooms)~\cite{teasley2002rapid}. 
Some of the claimed benefits for co-location are seamless coordination, increased creativity, faster learning, and projects that are easier to manage~\cite{conboy2009creativity,espinosa2002shared}.

There are numerous studies that give insights about the benefits and challenges of distributed corporate development work compared with co-located development~\cite{al2008comparative,bird2009does,gupta2009use}, but these studies tend to focus on specific teams or investigate the velocity or quality of the code developed using quantitative analysis of system trace data. They do not capture the experiences of developers in large companies that have had to switch (over night) from a mostly co-located mode to remote work from home. And yet this is what happened for many developers worldwide with the pandemic forcing technology companies to close their offices. 

The COVID-19 pandemic has undoubtedly been a global human and economic disaster, but it has also led to interesting and unexpected revelations about how we work, play and live.
Although working from home during a pandemic is not the same as working from home during ``normal times''~\footnote{\url{https://twitter.com/shanselman/status/1252040170783641600}}, it is nevertheless an opportunity to study the results from a ``natural experiment'' and compare the benefits and challenges developers may experience in terms of their development productivity in these two modes.
\BEGINADDED
Our work reveals a ``tale of two cities'' effect, where some developers flourished working at home, while others did not fare as well.
\ENDADDED

\BEGINADDED 
The aim of this research is to understand how and why developers' perceived productivity may have changed over the weeks following the initial work from home directive, and to identify the benefits and challenges they experienced. 
We report on the very early experiences developers found working from home and their experiences after several weeks of working in this mode. 
The paper is based on data collected through two surveys with a combined total of 3,634 responses. 
The first survey was among employees working at Microsoft in Washington State (the location of the company's headquarters), consisted of 30 questions, and received 1,369 responses.
The second survey was among Microsoft employees in the United States, consisted of 42 questions, and received 2,265 responses.
Our initial survey uncovered the main benefits and challenges they faced, and the second survey revealed the frequency/impact of those benefits/challenges. From the second survey, we also uncovered developers' self-reported changes in productivity since working from home (WFH) and how the benefits and challenges they experienced associate with those changes in productivity. 
From our survey responses, we also identified improvements that can be made to support the WFH experience.
\ENDADDED 

\BEGINADDED 
Notably, we found dichotomous experiences of engineers (developers and program managers) working from home (WFH) during the pandemic from the same factors.  For example, for some no commute was a benefit, but for others the lack of a commute removed a period of relaxation or preparation from their day. 
These differential experiences were in some cases driven by engineers' personal contexts (e.g., if they have school-age children or space for a home office) and the characteristics of their work (such as reliance on team members). 
However, some factors were experienced as both a benefit and a challenge by some of the same participants (such as time flexibility being both a blessing and a curse).
\ENDADDED

The main takeaways from our paper are as follows: 
\begin{itemize}
\item Productivity, when measured using engineering system data, appears to be stable or slightly improved, but some developers, at least initially, appeared to thrive and report being more productive, while others face significant challenges with remote work and feel they are not as productive.
\item Some factors (such as schedule flexibility, proximity to family members and more time for work) lead to dichotomous experiences across developers and even by individual developers. 
\item Organizations can support remote work by understanding the varied experiences of developers and the challenges their employees may face, and that there are actionable recommendations they can follow to support developers working from home now or in the future as part of a hybrid model.
\end{itemize}

Our paper is organized as follows. In Section~\ref{sec:background}, we provide a background of literature that has studied developer performance and productivity.
In Section~\ref{sec:methodology}, we describe the methodology we followed for the two surveys we conducted over the first few months of the pandemic. 
In Section~\ref{sec:change}, we discuss the changes in productivity reported across the surveys. In Section~\ref{sec:benefits}, we report the main benefits encountered, and in Section~\ref{sec:challenges}, we report the main challenges encountered as developers \wfh.  
Section~\ref{sec:improvements} summarizes the key recommendations that developers suggested organizations should follow to improve the situation for developers \wfh. 
In Section~\ref{sec:discussion}, we delve into the main factors and discuss how these may play a role in the dichotomous experiences of developers. We also discuss how engineering system output data can help triangulate the impact of working from home and how the pandemic may shape the future of remote work.
In Section~\ref{sec:relatedwork}, we discuss research on \BEGINADDED remote work (pre-pandemic) as well as \ENDADDED more recent related works about programmers during the pandemic.
Finally, we conclude our paper in Section~\ref{sec:conclusions}.

\BEGINADDED
\section{Background: Developer Performance and Productivity}
\ENDADDED
\label{sec:background}

Understanding developer productivity in software engineering has seen great interest from research and industry, as improving developer productivity may lead to faster development speed, higher quality code, and also higher developer satisfaction. Some of the factors we uncovered in our study overlap earlier research, but the abrupt change for individual developers and the entire organization to \wfh reveal new benefits and challenges of \wfh, as we discuss later in the paper. A concern during the shift to working from home during the pandemic is that both developer productivity and well-being may have been negatively affected. Existing research has led to insights about developer performance, productivity, satisfaction and developer well-being. 

In terms of performance, system engineering activity metrics can provide important signals about developer activity and productivity.
Wagner and Ruhe's review of the literature summarize studies that use performance measures such as \textit{lines of code} or \textit{function points} as proxies to productivity~\cite{WR18}.
However, within the area of organizational behavior, performance and productivity are acknowledged to be related~\cite{SHMM79}, with higher levels of performance leading to higher levels of productivity, but they are also recognized as distinct concepts. 
Indeed, many researchers and practitioners emphasize that developer productivity cannot and should not be measured by engineering metrics alone as development work is not mechanized work that can be assessed using a single metric. In fact, doing so may be detrimental to overall and long-term development objectives~\cite{ko2019}.
For example, developers spend time mentoring newcomers, reviewing each other's work informally, and learning new skills. 

Through their systematic literature review, Wagner and Ruhe~\cite{WR18} also identified 51 factors that influence productivity. 
In addition to the identified technical factors that seem to dominate productivity studies in software engineering, they also distilled a number of soft factors that focus on aspects such as organizational culture and working environment.

Using a different lens, Meyer et al.~\cite{MFMZ14} investigated how developers perceive and think about their own productivity.
Through a survey, observations, and interviews, their study brought to the surface that a developer's sense of how productive they are, may be distorted by how many interruptions and context switches they experience.
Other research about \BEGINADDED perceived \ENDADDED developer productivity reported that the quality of one's work environment plays a major role~\cite{johnson2019workenviro}, while the effectiveness of a manager~\cite{KBZBDG19} is also an important factor on \BEGINADDED perceived \ENDADDED productivity.

More recent research has expanded these factors to include additional elements that may influence \BEGINADDED perceived \ENDADDED productivity and satisfaction, or that can be used in certain contexts to predict productivity.
In an earlier study~\cite{storey2019theory} we conducted with developers at Microsoft, the factors that more closely associate with one's satisfaction with their self-reported productivity include \textit{job satisfaction}, \textit{doing impactful work}, \textit{having autonomy over one's work}, the \textit{ability to complete tasks}, the \textit{quality of the engineering system}, the \textit{ability to complete tasks}, \textit{personal technical skills}, and their \textit{work environment}.
Predictive factors~\cite{murphyhill2019predicts} include
\textit{job enthusiasm}, \textit{peer support for new ideas}, and \textit{getting job feedback}, while ``use of remote work to concentrate'' showed the lowest variance across three large software companies in terms of self-reported productivity. 

\BEGINADDED
In our research, we build on this previous work by inquiring about factors already shown to be important for productivity while also allowing new factors that are particular to the pandemic (such as having children at home) to emerge from our study. We discuss our methodology and the surveys in the next section. 
\ENDADDED

\section{Methodology}
\label{sec:methodology}

We investigated the experiences of software engineers at Microsoft during the height of the \covid pandemic in the United States using a set of online surveys. Figure~\ref{fig:study-timeline} outlines the timeline of our study from March--May 2020.

\input{figures/timeline}

\paragraph{Research Context} The first presumptive positive COVID-19 case in King County, WA (which includes the Microsoft headquarters) was reported on February 29, 2020. In the late afternoon of March 4, Microsoft informed its employees that \emph{``Consistent with King County guidance, we are recommending all employees who are in a job that can be done from home should do so''}~\cite{kurt:email:2020}. On March 11, the schools were closed in Washington State (which was made permanent for the school year on April 8). In April, many other restrictions were introduced, and by April 27, the outbreak has reached its peak in Washington State. Some restrictions on outdoor activities were later lifted but social distancing was still recommended. The pandemic also affected the rest of the US: at the end of March, 42 states and a total of 308 million people (94\% of the US population) were under stay at home orders~\cite{stay-at-home-orders}. At the time of writing this paper, Microsoft and other large tech companies had extended their recommendations to work from home until at least the Fall of 2020 and in some cases even Summer 2021~\cite{amazon-wfh,google-wfh}.

\paragraph{Research Questions} To understand the effect of WFH on software engineers through our surveys, we answer the following research questions: 
\begin{description}
\setlength{\itemsep}{0pt}
\setlength{\parskip}{0pt}
    \item[\textbf{RQ1}] {How has engineers' self-reported \textbf{productivity changed} since WFH?}
   \item[\textbf{RQ2}] {What are the \textbf{benefits} engineers experience when working from home? How have these benefits affected productivity since WFH?}
    \item[\textbf{RQ3}] {What are the \textbf{challenges} engineers face when working from home? How have these challenges impacted productivity since WFH?}
    \item[\textbf{RQ4}] {What \textbf{recommendations} should be made to companies whose engineers may wish to work from home?} 
\end{description}

To answer these research questions, we distributed two anonymous surveys to understand the experiences of software developers during the pandemic, their prevalence, and the effect on their work. 

\subsection{\surveyone: Washington State}

Our first survey was designed to understand the types of experiences software developers were having during the pandemic.
In this survey, we included a closed-answer question on how productivity has changed with a five-point scale for the responses. Through a following open-ended question, participants were asked to explain their response.

\begin{itemize}
    \item Compared to working in office, \textbf{how has your productivity changed}?     (Q13)~\footnote{Q13 indicates the question numbers in our survey instrument} \newline {\small\emph{(significantly less productive / less productive / about the same / more productive / significantly more productive)}}
    \item Please share details about your answer to the previous question on how your productivity has changed.
    (Q14)
\end{itemize}

Although there has been previous research to understand the factors that affect developer productivity (as discussed above), we anticipated that different benefits and challenges may be more relevant in this period of unexpected and mass transition to working from home for an entire organization. Therefore, the first survey was mainly exploratory to investigate if new factors  would emerge through the following open-ended questions:  
\begin{itemize}
    \item What is \textbf{good} about working from home? 
    (Q15)
    \item What is \textbf{bad} about working from home? 
    (Q16)
    \item What \textbf{challenges} have you encountered working from home? 
    (Q17)
    \item What could be \textbf{improved} about how we do work from home at Microsoft? 
    (Q22)
\end{itemize}

In addition, the survey included questions about Internet connectivity, interruptions and distractions, work times, meetings, and commuting. The full survey is available as supplemental material~\cite{supplemental-materials}.

\paragraph{Survey Distribution.}

This survey was distributed to Microsoft employees in King County during the week of March 16, 2020 (approximately two weeks after the advice to work from home). 
Each day the survey was sent to 1,000 randomly selected
employees in King County, for a total of 5,000 employees (3,500 developers, 1,500 program managers). We received \surveyoneresponses survey responses, \BEGINADDED with all required questions answered, \ENDADDED for a response rate of 27\% (comparable to the response rates of many other software engineering surveys~\cite{punter-isese-2003,smith-chase-2013}). To encourage participation, survey respondents could enter a raffle of multiple \$100 Amazon.com gift certificates. No reminder emails were sent.

\paragraph{Data Analysis}
For the open-ended responses to this first survey, we used an open-coding approach, iterating and refining through multiple rounds of independent coding of an initial sample of responses. 
We coded all of the open-ended questions listed above, Codes for positive aspects of working from home (benefits, RQ2), negative aspects of working from home (challenges, RQ3) and improvements (RQ4) emerged across the questions. 

After several iterations coding and discussing codes, we finalized a unified coding scheme with codes, code definitions and code categories (see Appendix~\ref{appendix:codebook}). We applied these finalized codes to a random selection of 400 responses (see Table~\ref{tab:codecount}). No new codes emerged during this process. %
To improve the reliability of our codes, an external researcher used our coding scheme to code a subset of our sample (100) showing an agreement of 81.9\%.

The final coding scheme contained 32 codes organized into the following six themes. (The complete list of codes and descriptions can be found in Appendix~\ref{appendix:codebook}.)

\begin{itemize}
    \item \emph{Beyond work} -- Effects of work from home on non-work aspects of respondents' lives, such as proximity to family, distribution of finances, and access to food.
    \item \emph{Collaboration} -- Aspects of respondents' collaborative tasks of respondents, including challenges being creative with others, being blocked waiting on others, and a range of interactions with co-workers.
    \item \emph{Communication} -- Work-related communication, including channels used, frequency, duration, planned versus ad-hoc, and the result of missing communication.
    \item \emph{Well-being} -- Responses related to the welfare of respondents, including changes to flexibility of schedule and location, and the effects of working from home on health (physical, mental, and emotional) and personal comfort.
    \item \emph{Work} -- Responses related to a direct effect on respondents' technical work output, including codes related to productivity, motivation, and factors affecting focus and distraction.
    \item \emph{Work environment} -- Aspects of the setting in which the respondents accomplish their work when working from home, including the existence or lack of reliable internet connectivity, ergonomically sound furniture, satisfactory hardware, and dedicated space.
\end{itemize}

Note that some responses were coded with multiple codes when participants raised multiple points. For example, the following response was assigned multiple \framebox{codes}:
\myquotewithcodez{1.~Avoiding commute, hence more productive, save on fuel (environment friendly). 2.~Comfort of home  (Take a nap of about 20 mins in the noon which powers up my rest of the day work) 3.~Avoid time spent in getting ready to office (10-20 mins per day)}{11}{R_1f6NoOYKjXY9Nm1}{\mycodez{Commute}\mycodez{EcologicalImpact}\mycodez{PersonalComfort}\mycodez{Break}}

We show the frequency of the main codes from the 400 responses in Table~\ref{tab:codecount} in the Appendix. However, these counts do not represent an accurate description of which benefits or challenges may be more important, and which ones may affect productivity more or less as these are open-ended questions.\footnote{Quantifying inherently qualitative data such as responses to open-ended questions carries some limitations. For example, when the Pew Research Center asked about the single issue that mattered most in deciding how participants voted for president, 35\% responded the economy in an open-ended question; however, when the economy was explicitly offered in a multiple-choice question, 58\%, more than half, chose the economy. \url{https://www.pewresearch.org/methods/u-s-survey-research/questionnaire-design/}}

\subsection{\surveytwo: United States}
\label{ss:surveytwo}

To investigate the importance and frequency of the reported benefits and challenges from the first survey and their association with self- reported productivity, we designed and deployed a second survey. 
Rather than open-ended questions, we included closed-answer questions for the factors that emerged from our coding of the first survey. 
These closed questions asked about benefits and how important they were to the respondent, as well as challenges and the impact of those challenges. 
In addition, this second survey inquired about additional benefits and challenges.\footnote{If a question was identical between \surveyone and \surveytwo, we use the same question number (e.g., Q13). For a closed-answer question in \surveytwo that was based on an open-ended question in \surveyone, we append an asterisk (*) to the question number (e.g., Q15* is based on Q15).}

\begin{itemize}
    \item Compared to working in office, \textbf{how has your productivity changed}?  (Q13) \newline {\small\emph{(significantly less productive / less productive / about the same / more productive / significantly more productive)}}
    \item What \textbf{benefits} have you experienced working from home and how \textbf{important} are these benefits?
    (Q15*) \newline {\small\emph{(I don't experience this benefit / I experience this benefit but it's *not* important to me / I experience this benefit and it's *important* to me / I experience this benefit and it's *very important* to me)}}
    \item What work-related \textbf{challenges} have you experienced working from home and how \textbf{impactful} are these challenges? 
    (Q17*) \newline {\small\emph{(I don't experience this challenge / I experience this challenge but it's a *minor issue* for me / I experience this challenge and it's a *major issue* for me)}}
    \item What could be improved about working from home (WFH)? Choose up to three (3) items. 
    (Q22*)
\end{itemize}

As items for the questions, we identified a list of 15 benefits (B1..B15), 20 challenges (C1..C20), and 12 improvements/suggestions (S1..S12) based on the thematic analysis of the responses to \surveyone. The items were displayed in random order within a question. The full survey is available as supplemental material~\cite{supplemental-materials}. 

\paragraph{Survey Distribution} This survey was distributed to 9,000 engineers (consisting of developers, program managers and data scientists) across the entire US over a period of three weeks (a different sample of 3,000 employees was selected for each week). There was no overlap between the samples in \surveyone and \surveytwo.
We received \surveytworesponses responses for a response rate of 25\%, \BEGINADDED with all required questions answered \ENDADDED (comparable to the response rates of many other software engineering surveys~\cite{punter-isese-2003,smith-chase-2013}). To encourage participation, survey respondents could enter a raffle of multiple \$100 Amazon.com gift certificates. No reminder emails were sent.

Collecting data across three weeks, and using the same question as in the first survey, allowed us to compare the answers to the closed question about change in productivity so that we could detect if there were any significant changes in productivity (RQ1) as people adapted to or found it harder working from home over time.  

\paragraph{Data Analysis\label{ss:lassodetails}} For the quantitative data in the second survey, we present descriptive statistics about the selected benefits and their importance, and the challenges and their impact. \RADDED{We performed a simple subgroup analysis for management responsibility (people manager vs.\ individual contributor) and job role (software engineers vs.\ program manager). We did not perform a subgroup analysis for gender because gender information was only collected in the final week of the survey.}

We also considered the association of the benefits (Q15*), challenges (Q17*), and suggested improvements (Q22*) with the reported changes in productivity (Q13). We used Wilcoxon Mann Whitney and Fisher Exact Value tests~\cite{lehmann2006testing} to check for statistically significant differences. To reduce false discoveries due to multiple hypothesis testing, p-values were adjusted with the Benjamini Hochberg correction~\cite{benjamini1995controlling}; \BEGINADDED we recorded the original p-values for each hypothesis test and then computed the adjusted p-values using the \emph{p.adjust()} function in R.\ENDADDED

To identify benefits and challenges that are most strongly related to productivity and analyze whether there are important interactions between them, we used a \emph{least absolute shrinkage and selection operator (Lasso)} analysis~\cite{tibshirani1996regression}. \BEGINADDED Lasso is a regression analysis method designed to select the most important explanatory variables from a set of variables. It works well in the presence of many explanatory variables.  Lasso analysis has been applied in economics and finance~\cite{hunt:welfare,finkelstein2016sources} and was found to improve prediction and interpretability of models. It was also found to select sometimes neglected variables~\cite{TIAN201589,coad2019catching}. \ENDADDED

We ran two analyses: for benefits (Q15*) and for challenges (Q17*). In both analyses, the dependent variable was whether a participant reported that productivity stayed the same/increased/significantly increased. \BEGINADDED Since the outcome is binary, we ran a logit analysis. For potential explanatory variables, we included direct effects and interactions \ENDADDED for whether a respondent reported each benefit as important or very important (for the analysis of Q15*) or each challenge as a major issue (for Q17*). \BEGINADDED With the interactions, we had 240 potential explanatory variables for the benefits regression (420 for challenges). Even with a relatively large survey, that is too many to meaningfully include in  a single analysis. The Lasso approach selects the most important variables by first maximizing the likelihood with a penalty for the absolute value of the coefficients. The penalty makes it so a variable is only given a non-zero coefficient if it explains enough of the variation in the dependent variable to outweigh the penalty.  As a standard practice, we ran the analysis with a range of penalty parameters and used the maximum penalty parameter that gets a mean cross-validation error within one standard error of the minimum. Then, using the variables with non-zero coefficients given that penalty parameter,\footnote{\BEGINADDED Since there is randomness in the sample-splitting for cross-validation, this procedure can result in a different set of variables if run with different sample splits. We ran it 50 times and used the variables that were consistently selected in at least 95\% of the runs to avoid the variable selection being affected by randomness.} we ran a standard logit analysis (logistic regression) \ENDADDED to get coefficients that are not shrunk by the penalty procedure. \BEGINADDED We report the logit coefficients as marginal effects for interpretability. Since all variables in the logit analysis are dichotomous, the marginal effect reports the difference between changing the variable from 0 (absent) to 1 (present).\ENDADDED

\subsection{Limitations}

We describe the threats to validity and limitations of our study.

\paragraph{External Validity} 
Single-case empirical studies have historically been shown to contribute to scientific discovery~\cite{Flyvbjerg06} and have delivered insights in the social sciences~\cite[pp. 95]{Kuper04}. 
The company we studied employs tens of thousands of software engineers that work on diverse products across many domains (operating systems, databases, cloud software, software tools, to productivity software), and  use many tools and diverse processes.
By studying a single company, we were able to control for many factors that otherwise may influence the employee experience during the pandemic, such as the region and the company's response to COVID-19.
We do not claim that our results are representative of the views of all software engineers and companies in general. 

\BEGINADDED
Likewise, we do not include the demographics of race, gender, nor age in this work. This is an opportunity for future work to highlight the experiences of marginalized individuals who have been disproportionately impacted by the pandemic~\cite{cdc2021healthequity-race, cdc2021covidwomen, wang2021struggle}.
\ENDADDED

It is important to keep in mind that remote work during a pandemic is not the same as regular remote work. While some findings will be specific to the pandemic (e.g., lack of childcare as a challenge because schools and day cares were closed), not all findings are specific to the pandemic. We discuss implications of our work for the future of remote work in Section~\ref{subsec:future-work}.

As with any survey, there may have been non-response bias, i.e., the results might not be representative of the population because the participants disproportionately possess certain traits which affect the outcome. In addition, our survey was advertised as a ``Work From Home Survey'' and therefore could have been subject to self-selection bias, e.g., participants might have been more likely to participate in the survey if they were more strongly affected by work from home (negatively or positively). To reduce non-response and self-selection bias, we kept the surveys as short as possible, were transparent about the survey length (single-page survey), provided an incentive to participate (raffle), and kept the surveys anonymous.

\paragraph{Construct Validity}

Although we could have used existing surveys that inquire about developer productivity and their experience working from home (see Section~\ref{sec:background}), we recognized from early reports that new factors specific to the pandemic were playing a bigger role in developer productivity and experience (such as not having child care, and the stress of the pandemic). Nothing like this has happened before and so we used an initial open-ended survey to study the factors emerging during this phenomenon, and then conducted a second survey to quantify the change in productivity and frequency/impact of challenges/benefits encountered. 

Measuring perceived productivity with a single question has limitations.
We consider self-reported changes in their perceived
productivity, \BEGINADDED as objective productivity metrics in software
development remain elusive due to the creative and collaborative nature of development work.
We chose single response items to keep the survey length reasonable because shorter questionnaires have been found to receive higher response rates~\cite{deutskens2004response}. 
Self-rated, single-item response items for productivity have also been found to correlate with objective productivity measures for software engineers~\cite{murphyhill2019predicts}. Later in the paper (in Section~\ref{sec:discussion}) we show and discuss an analysis of productivity at the company data by by mining software data collected through the engineering systems to triangulate our findings from the surveys.
\ENDADDED

\paragraph{Internal Validity}
There are additional biases from our survey. Respondents may have shared what they wanted management to hear (in terms of suggested improvements in particular, but also for benefits and challenges), and the wording of our questions may have led to certain responses. 
Furthermore, our analysis may have been biased by our own experiences (as we were also working from home) \BEGINADDED and most of us shared the same employer as the respondents.\ENDADDED{}  We tried to offset these by using additional coders and by having experts review our survey from outside our team.

Finally, our study involved a single research method (a survey). However, we tried to offset this limitation by considering the findings from other ongoing studies at the company using different methods (including objective quantitative analysis of system data). 

\BEGINADDED
There was no overlap between the samples in both surveys to reduce survey fatigue among participants. As a result, the second survey was limited in capturing how things had changed since the first survey (although some extrapolation is possible). To study the long-term impact of work from home, other research designs have used diary studies~\cite{butler2021}. The benefit of survey designs like ours is that they require less commitment and time from participants, and as a result, have more participants, which makes it easier to observe statistical effects.
\ENDADDED

\BEGINADDED
\paragraph{Conclusion Validity}
The Lasso approach is designed to select the variables with the strongest relationship to the outcome and uses cross-validation to avoid over-fitting, but it is still possible to find a statistically significant relationship when there is none (since p-values below .01 are still greater than zero).  There may also be other relationships in the data that are important but not captured in the results because the sample size is not large enough or there is too much noise relative to the effect size. The relationships we uncover are all correlational; we cannot distinguish causal effects using the survey.
\ENDADDED
\section{Change in Productivity \BEGINADDED(RQ1)\ENDADDED}
\label{sec:change}

In this section, we address the research question ``How has engineers' self-reported productivity changed since WFH?'' (RQ1).
In both surveys, we asked participants how their productivity has changed compared to working in office. The results are shown in Table~\ref{tab:productivity}.

\begin{itemize} 
\item In both surveys, the majority of participants reported that their productivity had not changed or had even improved (62\%-68\%). However, a substantial portion of participants (32\%-38\%) reported that they were less productive
\item The percentage of people reporting to be less productive consistently dropped over the study period:\ from 38\% in Survey 1 to 30\% in the last week (W3) of Survey 2. This suggests that some (but not all) people found ways to restore their productivity to their original levels.
\end{itemize}

Ralph et al.~\cite{ralph2020pandemic} found evidence that developers have lower perceived productivity while working from home during the \covid pandemic. Our findings also support this result but offer a more nuanced view: initially in Survey 1, more people reported lower productivity (38\%) than higher productivity (30\%); however, this later changed in Survey 2, when more people reported higher productivity (\BEGINADDED 37\%\ENDADDED) than lower productivity (\BEGINADDED 32\%\ENDADDED). Similar observations have been made by Forsgren~\cite{forsgren2020covid} and Bao et al.~\cite{bao2020does}. We will discuss these papers in more detail in the related work (Section~\ref{sec:relatedwork}).

\RADDED{We performed a simple \emph{subgroup analysis} of change in productivity across both surveys combined. We found no statistically significant difference between people managers (32\% more productive) and individual contributors (35\%). However, the difference between software engineers (31\%) and program managers (40\%) was statistically significant. At Microsoft, program managers define and design products and services, collaborating with many different stakeholders, including engineering teams. Their responsibilities span the entire product/service life cycle~\cite{amandasong,pmatmicrosoft}.} 

It is important to recognize that at an \emph{individual} level, people are affected differently by work from home: productivity can decrease, stay the same, or improve depending on a variety of challenges and benefits. In the next two sections, we discuss the higher level themes in terms of the challenges and benefits experienced that emerged from our qualitative analysis of the additional survey questions.

\input{tables/productivity}

\section{Benefits \BEGINADDED(RQ2)\ENDADDED}
\label{sec:benefits}

In this section, we address the research question \emph{``What are the benefits engineers experience when working from home? How have these benefits affected productivity since WFH?''} (RQ2). To identify the benefits, we analyzed the responses to the open-ended questions in \surveyone, and to quantify the association with productivity, we used the responses to \surveytwo.

\subsection{\surveyone: Benefits Experienced Working from Home}

Respondents identified a wide range of benefits working from home. In this section, we discuss themes that were frequently mentioned in \surveyone or later emerged as significant in the productivity analysis based on \surveytwo. 

\wfhcategory{Commute.} 
Most respondents pointed out benefits related to their well-being. In particular, over half of the participants mentioned the lack of commute as a positive aspect of working from home.

\myquote{No commute worries. I can focus on my job instead of checking the traffic reports and worrying.}{287}{R_25Eez8sloLGliJq}

On average, respondents reported a daily, round trip commute of 67 minutes (median 60 minutes), which is comparable to other people living in the Puget Sound area~\cite{seattle-commute}. In extreme cases, respondents mentioned a daily commute of more than four hours. \myquotex{I love saving the 14-18hrs/week of commuting and being home when my daughter gets home from school.}{NNN}{R_2EsqYDXocagFMUo}

The time saved on commuting led to a wide range of other benefits. It allowed respondents to work more (\textbf{Work hours}) but also spend time with their families (\textbf{Family, children, pets}) and focus on healthy activities such as physical exercise and more sleep (\textbf{Healthy habits}).

\myquote{More time with Kids due to reduced commute times. Can workout more when the sun is out.}{245}{R_2dsTV5Z8nG9DmW7}

\myquote{Removes the stress of a 30 minute commute to work and saves that wasted time for either more sleep or more work. Let's me sleep in a bit and work in a more rested state than if I got up earlier due to the commute.}{336}{R_dg0qwkWIWkdqEEx}

\myquote{My health has improved. I am getting more sleep and rest, since there is no commute involved. I can work at my own pace, in a more comfortable environment (my home). }{78}{R_3kfQjXS8yJog7uN}

\wfhcategory{Schedule flexibility.}
Another frequently mentioned benefit was schedule flexibility. Working from home allowed respondents to plan their day more freely and work at different times than they did before. Activities that required respondents to be physically present at home at certain times were easier to do, for example, accepting deliveries, laundry, or physical exercise. 

\myquote{I can dictate my own schedule, take breaks when I want to, prioritize self-care throughout the day (running errands, taking a quick nap, exercising, calling family + friends)}{211}{R_1rqdpRwbgJ6nsYT}

\myquote{Working from home gives me more flexibility to do things when I take 10 minutes break twice a day. I can finish up loading the laundry or dishwasher, I can lie down to straighten up my back if I wish without compromising the productivity.}{244}{R_03AB2ycRjm2WuFX}

\myquote{I can sometimes take a break and do some chores that are often more difficult to do later in the day (e.g. laundry, where all the machines are usually being used in my building).}{356}{R_2dvkxLJ9ceA2JrC}

The flexibility in schedule often led to seamless integration of work and life, where participants completed chores while waiting for builds or during short work breaks.

\myquote{Sometimes an idea clicks in the middle of the night, and with wfh, implementing that idea is literally 2 seconds away}{198}{R_2RUQZMv2QoUGbdA}

\myquote{Being able to quickly task switch for other non-work related tasks and quickly return to work.}{328}{R_31sBkX3q0PZV6o6}

\myquote{I feel like I can solve problems more easily since I don't feel constrained by a clock. I can start a job and cook dinner, then come back to check the job results while I leave something in the oven or when I'm done cooking.}{258}{R_2dKZf5IG6crztdq}

\smallskip\textbf{Focus} and \textbf{Interruptions and Distractions} were the second and third most frequent codes.
Fewer distractions and interruptions at home made it easier for respondents to concentrate and focus for longer periods of time. Having fewer meetings, the ability to continue work during remote meetings, a quiet work environment, and more control over interruptions further contributed to \myquotex{Undistracted focus time.}{387}{R_1o23gvJUzZhuRnW} for the respondents.

\myquote{I get interrupted less and am able to focus on tasks more without distraction.}{223}{R_3iQgCcULw8KZOS8}

\myquote{More focus time, ability to go heads down without distractions of being in an open office constantly or needing to move focus rooms every hour if I am working on something that requires prolonged quiet time. }{230}{R_vDI2a6huvHjrgHf}

\myquote{Less distraction from others, especially coming from an open office scenario. Teams meetings for some meetings where very little participation is required lets you continue to work while ``attending'' and listening in, which is better than being idle in a conference room.}{101}{R_sROzubcZSHDdNgB}

\myquote{There were plenty of distractions at work too with my office set up.  So obviously colleagues interrupting me is easier to manage now as you have to answer the teams chat or email or phone call vs. someone just coming into your office or dragging you to a meeting.  It feels more in my control now.  }{384}{R_1rD0XIqTLYEcnpq}

\wfhcategory{Work Environment.} Several respondents pointed out that they preferred their  environment at home compared to the office, for example, because the environment is more quiet, more spacious, has a window, more sunlight, or closer bathrooms. Respondents also liked having more privacy and more control at home, for example, over room temperature or decorations, which is more difficult in a shared environment.

\myquote{I enjoy the relaxing nature of being in the comfort of my own home. I like being able to use my nice chair at home, and other nice comforts my home can offer. [\dots] My wife is the only other person here, and she respects my zone while I'm working and gives me a nice quiet environment.}{178}{R_2YRhr7ZndVNEfRZ}

\myquote{The environment is much nicer. It is quiet with few distraction than my normal open office. I also have a window which gives me natural light and a nature view neither of which are present in my open office. This helps improve my mood and makes me more productive.}{232}{R_1iaOAgsLf8pTvX2}

\myquote{My apartment is much less dreary than the office. Visuals and decorations are not just about ``looking cool'', but have a deep effect on how well/fast/creatively/how long I can think. In contrast the office has mostly blank white walls, no windows, and every hallway in the building is identical. Just being in the monotony of that environment is mentally draining, which drops my productivity.}{355}{R_3DERDufB5YtfC5b}

\myquote{I work in an open floor area and had to always be careful what was on my screens that others shouldn't see, no concerns now with that.}{348}{R_1immlUzAEXPS5Yc}

Participants also enjoyed the ability to wear comfortable clothes, listen to loud music, and work at their own pace (\textbf{Personal comfort}) in their home work environments.

\myquote{I'm in the comfort of my own home (I can wear lounge clothing, play music, etc).}{195}{R_2aFDcTv5D89AvCd}

\myquote{No need makeup, suitable dress, and few unnecessary social except meeting. Saving time. Not worry about if anyone will look at me when the moment don't want to be looked. More concentrate on work.}{393}{R_6hxPRq3I1dVhE8V}

\myquote{I feel more comfortable and have more privacy. I feel less pressured to do work and get to work on my own pace.}{88}{R_WvCqhVGQNqXOFnr}

\wfhcategory{Family, Children, and Pets}. Respondents liked being close to their families, children, and pets. They appreciated being able to see them during breaks or lunch and that they could take care of family needs when required. 

\myquote{Being at home with family, especially with a toddler and baby.  I get to spend a bit of time each day every few hours to just say hi and be around them, even if just for a few brief minutes.}{329}{R_3oLL90vAb5X36TW}

\myquote{Work breaks are fulfilling if you have family members around. }{344}{R_3RscSpCHH7ci6Cl}

\wfhcategory{Money.} Several participants pointed out that working from home saved them money because of the lack of commute and eating home-made food.

\myquote{I save money on food because I'm eating more out of the refrigerator than spending money on lunch every day.}{102}{R_3PnPgxkqeEEggfI}

\myquote{No wasted time \& money on commute.}{3}{R_1K2yfUfcFB1wjLn}

\input{tables/mybarcharts}
\input{tables/benefits}

\subsection{\surveytwo: Benefits and Productivity}

From the themes that emerged in \surveyone, we inferred a list of 15 characteristic benefits (B1..B15) that we included in \surveytwo. The results are displayed in Table~\ref{tab:benefits}. The benefits are sorted and numbered in descending order of frequency.

\paragraph{Frequency and Importance of Benefits.}

We make the following observations from Table~\ref{tab:benefits} about the prevalence and importance of benefits:
\begin{itemize}
\item The \emph{Prevalence} column shows the frequency of the benefits. The most frequently reported benefits (B1-B5) were \emph{less time on commute} (96\%), \emph{spending less money} (84\%), \emph{flexible work hours} (81\%), \emph{closer to family} (81\%), and \emph{more comfortable clothing} (80\%).

\item The \emph{Importance} column shows the percentage of participants who indicated a benefit to be important or very important if they experienced it. Almost all benefits were rated as important by the majority of participants who experienced them. The benefits most frequently rated as important were \emph{better work life balance} (95\%, B11), \emph{better focus time} (93\%, B7), and \emph{more physical activity} (91\%, B15). The benefits less frequently rated as important were \emph{more comfortable clothing} (48\%, B5), \emph{more breaks} (64\%, B10), and \emph{spending less money} (66\%, B2). 

\end{itemize}

\paragraph{Relation between Benefits and Productivity}

\BEGINADDED The \emph{Delta} column of Table~\ref{tab:benefits} shows the difference between the percentage of respondents who reported that their productivity stayed ``about the same'' or increased (``more productive'', ``significantly more productive'') for those who did \emph{not} experience the benefit vs.\ those who did experience the benefit. \ENDADDED We make the following observations:

\begin{itemize}

\item \BEGINADDED Almost all \ENDADDED benefits had a positive delta on productivity change. This means that, on average, respondents who experienced a benefit also reported being more productive when working from home. The delta was \BEGINADDED only negative \ENDADDED for the benefit \emph{more breaks} (\BEGINADDED --1.3 percentage points\ENDADDED, B10), \BEGINADDED but the difference was not statistically significant.\ENDADDED 

\item Not all benefits had a statistically significant productivity delta: there was no statistical difference in productivity change for the benefits \emph{less time on commute} (B1) and \emph{more breaks} (B10). 

\item The benefits with the highest productivity delta were \emph{better focus time} (\BEGINADDED +42.6\ENDADDED, B7), \emph{less distractions or interruptions} (\BEGINADDED +35.4\ENDADDED, B8), and \emph{better work environment} (\BEGINADDED +34.1\ENDADDED, B12). All of these have been found to be significant predictors of productivity in the past~\cite{murphyhill2019predicts,storey2019theory,meyer2014software,johnson2019workenviro}.

\end{itemize}

\BEGINADDED
\paragraph{\RADDED{Subgroup analysis}} \RADDED{We compared the following subgroups to identify differential characteristics in the survey data.}
\begin{itemize} 
\item People Managers vs.\ individual contributors. Compared to individual contributors, people managers were \emph{less} likely to indicate \emph{better work life balance} (--14.3 percentage points), \emph{flexible work hours} (--13.2), \emph{more breaks} (--12.2), and \emph{better focus time} (--7.5) as benefits. They were \emph{more} likely to indicate \emph{being closer to family} (+9.1 percentage points) as a benefit.
\item \RBEGINADDED Software engineers vs.\ program managers. Compared to program managers, software engineers were \emph{less} likely to indicate \emph{being closer to family} (--13.8 percentage points), \emph{spending less money on commute, food, etc.} (--12.2), \emph{more physical activity} (--10.3), \emph{better work environment at home} (--7.4), and \emph{more control over work} (--6.8) as benefits. 
\end{itemize}
\ENDADDED

\input{tables/benefitslasso}

\paragraph{Lasso Analysis} %
We ran a  Lasso analysis to see which benefits were most strongly associated with productivity change in a combined model and to check whether interactions between the benefits matter \BEGINADDED (details in Section \ref{ss:lassodetails}). The marginal effects of the logit analysis \ENDADDED with the Lasso-selected variables, shown in Table \ref{tab:benefitsLASSO}, are similar to the pairwise relationships. The likelihood that people's productivity stayed the same or improved was increased by the benefits of \emph{better focus time} (\BEGINADDED+18.9\%\ENDADDED) and \emph{less distractions or interruptions} (\BEGINADDED+10.6\%\ENDADDED). The coefficient for \emph{better work environment} at home was not statistically significant, however, the interaction between \emph{better work environment} and \emph{less time spent on commute} (\BEGINADDED+15.7\%) was\ENDADDED. People who reported benefiting from both \emph{less time spent on commute} and \emph{more time to complete their work} were also more likely to report that their productivity remained the same or increased (\BEGINADDED +12.2\%\ENDADDED). \BEGINADDED As were people who benefited from both \emph{less distractions or interruptions} and \emph{closer to family} (+8.9\%).\ENDADDED

\section{Challenges \BEGINADDED(RQ3)\ENDADDED}
\label{sec:challenges}

In this section, we address the research question \emph{``What are the challenges engineers face when working from home? How have these challenges affected productivity since WFH?''} (RQ3). To identify the challenges, we analyzed the responses to the open-ended questions in \surveyone, and to quantify the association with productivity, we used the responses to \surveytwo.

\subsection{Survey 1: Challenges Experienced Working from Home}

The survey respondents shared a wide range of challenges and many respondents indicated they experienced multiple challenges working from home. In this section, we discuss challenges that were frequently mentioned in \surveyone or later emerged as significant in the productivity analysis based on \surveytwo.

\wfhcategory{Connectivity.} 
Of all the challenges, problems with connectivity was the most frequent challenge shared. This included access to remote desktops, special access workstations, and internet bandwidth. Respondents mentioned they experienced slow speeds due to a high number of users on their internet connections:
\myquote{Periodic internet disruption due to wi-fi router and modem resetting due to 2 VPN connections for my wife \& I, and our kids being online for school work.}{25}{R_3ER3V3aSCaZkbS7}
Internet connectivity speed went beyond one's home internet and became a challenge when a colleague's internet connection was not as resilient.
\myquote{Remote desktop connectivity issues. Coworkers with spotty internet quality are hard to meet with.}{101}{R_sROzubcZSHDdNgB}%

Respondents also described workarounds they used to rectify their connectivity issues:
\myquote{The VPN / Remote tools are not great, crashes often. I especially dont want to Intune my personal device, so working remotely have been challenging with the redmondts gateway down more than 50\% of the time, and the WVD features crashing / disconnecting / not allowing correct alt-tab etc.}{157}{R_2y2RYhzd2Ogysvl}
Although many attempted to find a resolution, at times the connectivity issues they experienced felt out of their control.

\wfhcategory{Family, Children, Pets.}
One of the most frequent challenges respondents shared concerned proximity to family life. Being physically co-located with family members, housemates, children, or pets encouraged some to change their work habits:
\myquote{Staying focused, especially with young kids around.  Normally I would only work from home for a few hours occasionally after the kids went to bed.  That is the only time I currently feel like I can be productive.}{346}{R_2tgmuEtLpdUmeDt}
The additional interactions with family have even to be more mindful of supporting family ``child care schedules'' (267).
Respondents also mentioned how there was an implicit expectation of being engaged that often felt at odds with work:
\myquote{Family in the house means there is also expectations from them to spend time or help around. }{390}{R_C9QURDTfT96z7AR}
The challenge of being physically present with family but mentally focused on other tasks is an experience that was hard to resolve.

\wfhcategory{Communication Channels.}
Another frequent challenge that respondents reported having was with the channels they used to communicate with their team members. One issue with communication channels was the increased friction to get ahold of a colleague versus simply walking over to their office:
\myquote{The hardest thing has been that standard communications/questions and general collaboration take about 2-3 times as long. Something I could just pop over to someone's office to ask now requires an online chat or an email and the response is much slower.}{211}{R_1rqdpRwbgJ6nsYT}
Likewise, for some there was also a higher frequency of using instant messages which made some participants feel like they should be highly responsive at all times:
\myquote{ I also feel like there is no "down time" away from work. I constantly get emails/messages/asks and sometime I have to respond right away.}{288}{R_32PHtzXJggT42hZ}
Managing multiple communication channels and the expectation to be very responsive on many of these channels presents an additional layer of interactions that does not adapt to every respondent's working style.

\wfhcategory{Work Environment.}
During the pandemic, most respondents' work environments were their homes, however, the experience drew comparable challenges with in-office work settings.
\myquote{Tuning out distractions (which is a similar problem I've faced working in Open Spaces), finding the space to set up my home workspace.}{32}{R_1i2U3Oa5z0qd3D3}

Many respondents were also not prepared to work from home and improvised their work settings:
\myquote{I did not own a desk and chair so currently improvising with dining table. Not sure if I want to invest in or have space for expensive home office equipment. I miss having multiple screens but do not have space at home to set up.}{137}{R_Y4bK9fRAQNF03kt}

As respondents missed their work office settings, they found themselves under new constraints ranging from financial to square footage when trying to create a comparable home office setting.

\wfhcategory{Interruptions and Distractions.}
When software developers are working in the office, interruptions and distractions often come from colleagues stopping by their desks. In a remote work setting, respondents described the advantage and disadvantage of only being available online:
\myquote{Interruptions and concentration as I can [only] be reached on Teams and by email vs someone walking over for a question. Harder to keep tabs on direct reports.}{348}{R_1immlUzAEXPS5Yc}

However, in this special remote work environment, a new set of distractions emerged from people they live with (e.g., spouse, children, etc): \myquotex{Constant distractions especially from kids who are bored at home}{351}{R_3EbfzmvK96BSYJI}. For some respondents, this created a similarly distracting environment they had to manage in open office settings: \myquote{Tuning out distractions (which is a similar problem I've faced working in Open Spaces), finding the space to set up my home workspace.}{32}{R_1i2U3Oa5z0qd3D3}

\wfhcategory{Healthy Habits.} When respondents described challenges with reduced physical activity, they often mentioned their movements between physical meeting locations which no longer happened:\myquotex{Sitting for a long time is hard on the body.  At work, I'm up and around, moving more.  At home all meetings are online so I never (hardly) move....}{3}{R_1K2yfUfcFB1wjLn}. When respondents did find an opportunity to move, it was either only to the restroom or for more coffee so that they can sit down for longer periods of time: \myquote{Since I don't do my daily bike ride I sometimes feel I just sit the whole day, and only do very few steps to the toilet [and] coffee machine}{301}{R_Uo0MGDk1CyXVmqB}
The reduction in what participants referred to as healthy habits also affect their work-life routines.

\wfhcategory{Work-life Balance} and \textbf{Routine}.
Respondents described their work-life boundaries blurring outside of the typical eight-hour work day, running late into the evenings:
\myquotex{Unless I impose a strict regimen, I feel like I am working for a lot more hours sometimes way into the night - the line between home and work gets far more blurry.}{212}{R_1FEcpXZJ8McnSJm}

Respondents also reflected on routines they previously had to distinguish boundaries for that were now lost: 
\myquote{To find my time boundaries. Very easily you can end up working much more hours because you don't have the signals of "Time to leave the desk", you don't have the time to decompress your mind in traffic, for example. You just jump from personal to work tasks (and vice-versa) so fast. }{276}{R_2Y9XvduDVpracQo}
In summary, the lost transition time and lack of physical movement between work and home removed a boundary they had before. 

In the next subsection of our paper,  we report on the association of these challenge with the \surveytwo respondents' productivity.

\input{tables/challenges}

\subsection{Survey 2: Challenges and Productivity}

From the themes that emerged in \surveyone, we inferred a list of 20 characteristic challenges (C1..C20) that we included in \surveytwo. The results are displayed in Table~\ref{tab:challenges}. The challenges are sorted and numbered in descending order of frequency.

\paragraph{Frequency and Impact of Challenges}
We make the following observations from Table~\ref{tab:challenges} about the prevalence and impact of challenges:

\begin{itemize}
\item The Prevalence column shows the frequency of the challenges. The most frequently reported challenges (C1-C5) were \emph{missing social interactions} (83\%), \emph{lack of work-life boundaries} (78\%), \emph{poor ergonomics} (70\%), \emph{less awareness of colleagues work} (65\%), and \emph{less physical activity} (65\%).

\item The Impact column shows the percentage of participants that indicated a challenge to be a major issue among the participants who experienced it. The challenges most frequently rated as impactful are \emph{lack of childcare} (58\%, C16), \emph{poor ergonomics} (52\%, C3), and \emph{less physical activity} (51\%, C5). 
The challenges less frequently rated as a major issue are \emph{friction with collaboration tools} (22\%, C14), \emph{lack of dining options} (24\%, C18), and being \emph{blocked waiting on others} (28\%, C16).
\end{itemize}

\paragraph{Relationships between Challenges and Productivity}
\BEGINADDED The \emph{Delta} column of Table~\ref{tab:challenges} shows the difference between the percentage of respondents who reported that their productivity stayed ``about the same'' or increased (``more productive'', ``significantly more productive'') for those who did \emph{not} experience the challenge vs.\ those who did experience the challenge. \ENDADDED We make the following observations:

\begin{itemize}
\item We found that all of the challenges were associated with lower productivity. For all but two out of the 20 challenges, the difference was statistically significant. 
\item The challenges with the largest reduction in productivity are \emph{more distractions and interruptions} (\BEGINADDED --36.5 percentage points\ENDADDED, C11), \emph{lack of motivation} (\BEGINADDED --33.7\ENDADDED, C15), \emph{poor home work environment} (\BEGINADDED --32.6\ENDADDED, C17), \emph{less time to complete work} (\BEGINADDED --31.2\ENDADDED, C20), \emph{lack of a routine} (\BEGINADDED --25.1\ENDADDED, C12), \emph{difficulty communicating with colleagues} (\BEGINADDED --21.9\ENDADDED, C6), \BEGINADDED and \emph{less awareness of colleagues work} (--20.4, C4)\ENDADDED. 
\end{itemize}

\BEGINADDED
\paragraph{\RADDED{Subgroup analysis}} \RADDED{We compared the following subgroups to identify differential characteristics in the survey data.}
\begin{itemize} 
\item People managers vs.\ individual contributors. Compared to individual contributors, people managers were \emph{more} likely to indicate \emph{too many meetings} (+26.1 percentage points), \emph{fewer breaks} (+21.1), \emph{lack of childcare} (+15.2), \emph{less time to complete work} (+12.8), and \emph{poor work life balance} (+11.4) as challenges. 
They were \emph{less} likely to indicate \emph{lack of motivation} (--9.3 percentage points), and \emph{being blocked waiting on others} (--5.2) as challenges.
\item \RBEGINADDED Software engineers vs.\ program managers. Compared to program managers, software engineers were \emph{more} likely to indicate \emph{connectivity problems} (+12.1 percentage points), 
\emph{difficult to find dining options} (+8.3),
\emph{lack of a routine} (+7.5),
\emph{lack of motivation} (+7.2),
\emph{difficult to communicate with colleagues} (+6.6),
\emph{insufficient hardware, monitors or devices} (+6.3) as challenges.
They were \emph{less} likely to indicate 
\emph{too many meetings} (--26.1 percentage points),
\emph{fewer breaks} (--16.1),
\emph{less time to complete my work} (--9.4), and
\emph{lack of childcare} (--7.4)
as challenges. 
\end{itemize} 
\ENDADDED

\input{tables/challengeslasso.tex}

\paragraph{Lasso Analysis}
The previous analysis shows the relationship between individual challenges and change in productivity but does not take into account the presence of multiple challenges and the interaction effects between two challenges, which can be particularly important for productivity. Therefore, we ran a Lasso analysis to see which challenges  were most strongly associated with productivity change in a combined model, and to check whether interactions between the challenges matter.

Table \ref{tab:challengesLASSO} shows the \BEGINADDED marginal effects from the logit analysis for the Lasso-selected variables (details on the method are in Section \ref{ss:lassodetails}). \ENDADDED People are substantially less likely to report that their productivity is the same or increased when they say that  having \emph{more distractions and interruptions} (\BEGINADDED--40.9\%\ENDADDED) or \emph{lack of motivation} (\BEGINADDED--26.4\%\ENDADDED) were major issues; 
\emph{difficult to communicate with colleagues} (\BEGINADDED --13.1\%\ENDADDED), \emph{connectivity problems} (\BEGINADDED --11.9\%\ENDADDED), \emph{missing social interactions} (\BEGINADDED --11.2\%\ENDADDED), \BEGINADDED\emph{poor home work environment} (--8.0\%),\ENDADDED{} and \emph{less awareness of colleagues work} (\BEGINADDED --6.2\%\ENDADDED)
were also associated with a significantly lower probability of reporting unchanged or increased productivity.

The challenge \emph{less time spent to complete work} was also selected by the Lasso algorithm, though the coefficient is not significant. However, when combined with a \emph{lack of childcare}, the interaction between both challenges is associated with a substantial and significant lower probability (\BEGINADDED --34.0\%\ENDADDED) of reporting unchanged or increased productivity.

\section{Improvements \BEGINADDED(RQ4)\ENDADDED}
\label{sec:improvements}

In this section, we address the research question \emph{``What recommendations should be made to companies whose engineers may wish to work from home?''} (RQ4). To identify these improvements, we first analyzed the open-ended questions in \surveyone, and to identify the most-requested improvements, we used the responses to a closed question in \surveytwo.

\subsection{Survey 1: Improvements to the Work from Home Experience}
Respondents included several improvements that could be made to support their work from home experience.
We briefly describe the most frequently mentioned improvements below.
Survey 1 was sent within the first two weeks of employees working from home. Microsoft implemented many improvements throughout the pandemic to provide a better work from home experience to its employees, for example, employees facing school closures due to the pandemic were offered up to three months of paid parental leave~\cite{microsoft:childcare} as well as resources and activities to support their physical, emotional and financial well-being. %

\wfhcategory{Hardware.} Although many respondents noted they were able to bring some equipment home or purchase additional devices, the most frequent possible improvement noted by respondents in the first survey was related 
to hardware. Insufficient hardware was also noted as a key challenge (as discussed in Section~\ref{sec:challenges}) and was mentioned by Ralph et al. in the Pandemic Programming study~\cite{ralph2020pandemic}.  

\myquote{I think that this is a reminder that when employing individuals that need certain equipment both at an office and at home, that we come up with a way to fully equip both locations simultaneously.}{293}{R_1E703c3ifvaeSOm}

Many employees work with multiple monitors and sometimes multiple machines.  In contrast, when working from home, employees are sometimes limited to just a laptop or a desktop with a single monitor.  Large and/or multiple monitor setups have been found to improve productivity in information workers~\cite{czerwinski2003toward}. Indeed, the most frequently requested type of hardware we noted in the second survey was either larger or more monitors.  Developers also asked for more powerful workstations (especially those working on laptops at home) as well as peripherals such as mice, keyboards, and noise-cancellation headphones. 

\myquote{Employees should be provided with equipment to make the experience better for everyone: webcams, good noise-cancelling headsets, etc.}{346}{R_xsYCYMjGFuT2g9z}

Employees were allowed to take any non-confidential property off campus to use at home after informing their manager. 

\wfhcategory{Connectivity.} Another frequently requested improvement was better Internet connectivity or improved VPN access from home, and many that mentioned this as an improvement in Survey 1 added that it was their biggest challenge to address: (\myquotex{Network connectivity is the biggest pain}{XXX}{R_27qHZG3G4O8kBzq}.
Some respondents indicated that ensuring good connectivity at home was expensive.  Paying for upgraded home internet connectivity was also a key recommendation from the Pandemic Programming study~\cite{ralph2020pandemic}. 

\myquote{Perhaps subsidize higher speed internet connections, or speak with the broadband providers to get better service in our areas.  With multiple people at home, there's only so much bandwidth to be shared.}{74}{R_3oLL90vAb5X36TW}

\wfhcategory{Stipend/Budget for home office.} The third most frequently selected item was to have a stipend to purchase equipment for a home office, with just slightly more engineers requesting this as an improvement if they experienced lower productivity. Several tech companies, including Microsoft, 
have been offering such a stipend to their employees~\cite{shopify:stipend,playstation:stipend,facebook:stipend}.

\myquote{Provide employees a one time reasonable allowance to set up a home office such as sit/stand desk, ergonomic chair, allowance for monitors, budget for coffee/drinks/snacks.}{80}{R_3qTKULFIQXyuz2F}

\wfhcategory{Improvements to communication tools.} Once working from home, engineers were totally reliant on communication tools to collaborate with their colleagues and for meetings. Many noted specific improvements, including some that were engineering specific:

\myquote{Support whiteboard drawing, support multi desktop sharing from multiple people and sharing on split windows or on my local multiple screens.}{}{R_232kf2XMtgFVNVS}

\myquote{Add a bunch of developer specific features. A simple example is how do you go around the room in standup and know that everyone got to talk.}{XXX}{R_2wSR3WGvZOQhtHB}
\wfhcategory{Provide more ergonomic furniture.} Many participants noted in the first survey that their furniture at home was not as ergonomic as their furniture at work (e.g., no standup desk, small desk space, less ergonomic keyboards etc.).
To better support employees, Microsoft provides recommendations on how to setup physical workspaces in an ergonomic way and some stipends for the workspace.

Not being able to exercise as much was also an issue, as this respondent mentioned: 
\myquote{I also don't have the ability to stand up and work since my home desk doesn't move. This means that I'm sitting down even more every day, which also leads to back and neck pain and frustration.}{XXX}{R_1DTfsPqdppaBXoB}%
A couple of respondents in our sample suggested treadmill desks as a possible improvement for WFH to address less of exercise. 

For some respondents, working from home goes beyond furniture, as one participant noted: 
\myquote{If I am going to continue work from home, I need a new house/space from which I can work w/fewer distractions.}{XXX}{R_1rwgesezsRPpCOL}

\wfhcategory{Support for remote work post-pandemic.}  In the first survey, many suggested that full or partial remote work should be supported after the pandemic, and that they appreciated the opportunity to work from home and experienced a variety of benefits (as discussed in Section~\ref{sec:benefits}). 
In particular, several mentioned that working at home, at least some of the time, would help them be more productive:

\myquote{Allow more people to do this long-term after this current crisis ends; I feel I'm more productive for it, I'm contributing better to my team, and keeping more people home will help us meet our sustainability and environmental goals as a company.}{XXX}{R_wO8hPFy88l0S7jb}

Some provided concrete suggestions for supporting WFH long-term:

\myquote{Every team should treat their tech stack as if they had to work from home at least 2 days a week.  This way things such as VPN, workstations, deploy pipelines, local builds, etc are naturally able to support remote workers.  This will lead us to hiring better remote talent and allow us to institute remote working policies during health or environmental changes that hinder the ability of some workers to be in the office.  That's what we've done and it's made our org more productive during the WFH period.}{XXX}{R_2ClDHYit6qdzUJm}

\wfhcategory{Provide guidance for working from home.}  
Many noted that curated guidance for working from home would benefit not just their own work but also the work of their colleagues.  
For example, on how to use different communication platforms, one respondent suggested: 

\myquote{Get some primers out so that people can feel comfortable in the space and know the use cases it's for. Lack of this knowledge keeps translating to inefficient use of email threads.}{XXX}{R_1mruQc8Ny3qH6Wv}

Improving how knowledge is externalized is also more important when everyone is remote: 
\myquote{Encourage documentation as part of our culture. It's difficult to impossible to use libraries from within our org without directly talking to the repository owners. Now the only way to get information about these libraries is to send a message and hope they respond.}{XXX}{R_1joYf29ApLyGaAY} %

Some respondents also noted that it isn't just about improving tools and processes for WFH, there is also a need for organizational guidance regarding maintaining a positive work-life balance:
\myquote{Broad communication across the company to say "General work ends at 5:00 PM local unless business critical" as a way to force work/life balance now that work and life are in the same place.}{XXX}{R_3nqMMmi9S24lDpF}

Microsoft continuously provided guidance, tips, and resources for employees working from home during the COVID-19 outbreak.  In addition, communities were created to connect employees with colleagues around the world for tips and support on working from home.

\wfhcategory{Other improvements.} Other improvements that were suggested by respondents were to \emph{improve and encourage social interactions within teams};
\emph{be more understanding of WFH scenarios beyond the pandemic}; 
\emph{encourage people to be more responsive}; \emph{minimize the number of meetings}; and \emph{guidance for managers to manage WFH employees}. 

\subsection{Survey 2: Relation between Improvements and Productivity}

From the themes that emerged in Survey 1, we inferred a list of 12 characteristic improvements (S1..S12) that we included in Survey 2, in which respondents could select up to three improvements. The results are displayed in Table~\ref{tab:improvements}. The table shows the frequencies for how often an improvement was selected by all respondents (column "All"), by respondents who reported a decrease in productivity (column "Low"), and by respondents who reported an increase in productivity (column "High"). 
The improvements are sorted and numbered in descending order of frequency by all respondents.

We make the following observations:
\begin{itemize}

\item The most frequently selected improvement was to \emph{provide more/better hardware for home} (41.6\%), \emph{improve connectivity} (41.5\%), and \emph{provide a stipend for improving work from home environment} (40.8\%).

\item Several improvements were more frequently selected by respondents who experienced a \textbf{decrease} in their productivity:\footnotemark{}\ \emph{improve connectivity} (45.8\% vs.\ 35.5\%, S2), \emph{make improvements to communication tools} (36.9\% vs.\ 29.3\%, S4), \emph{provide guidance for successfully working from home} (23.7\% vs.\ 16.7\%, S7), and \emph{minimize the number of meetings} (7.6\% vs.\ 3.8\%, S11).

\footnotetext{\BEGINADDED We break down the improvements by productivity, not because we think that getting those improvements explains productivity, but because it shows how the needs of people who are already productive differ from those who do not. For example, if a company wants to focus on helping people who report difficulties remaining productive while working from home, then they should initially focus on improving connectivity and collaboration tools, not providing ergonomic furniture (since that was primarily requested by those who were already productive).\ENDADDED}

\item Several improvements were more frequently selected by respondents who experienced an \textbf{increase} in their productivity. 
The improvement \emph{provide ergonomic furniture} was more than twice as likely to be selected (40.7\% vs 17.2\%, S5). This may be because more productive respondents were satisfied with other potentially pressing needs. 
The improvement \emph{support remote work better during normal circumstances} was selected twice as frequently (30.7\% vs. 15.1\%, S6). This may be because these respondents' basic needs were met and they were focused on future needs and being able to continue to working from home. 

\end{itemize}

\input{tables/improvements}

\section{Discussion}
\label{sec:discussion}

Although the pandemic is an unusual---and hopefully uncommon---event in the lives of software engineers, the sudden work from home directive provides an opportunity to study what happens when engineers at a very large company are suddenly in a remote working condition with the rest of their team and the entire organization. Engineering work is similar to knowledge work in general, but engineers may require highly intense periods of focus work but also rely on tight collaboration to develop modern software. 

As such, the pandemic and the force to \wfh provides an interesting opportunity to understand more about developer productivity, but also to find guidance for developers that work remotely or for developers that collaborate with remote team members. \BEGINADDED As many companies are anticipating supporting much more remote work in the future, some even declaring they will be entirely remote, the findings from our study are important. However, our study is focused on the study of remote work during the pandemic.  We found many overlapping factors from those found in studies conducted before the pandemic, such as control over their work, time to complete work, focus time and work life balance. But some new factors emerged that were specific to the pandemic context, such as commute time and reduced health risks. \ENDADDED

Before discussing our findings further, we remind the reader that the context for this study is a large multi-national software company, and that our study focuses on engineers working in the US.

\subsection{The Yin and Yang of Working from Home} 

As we saw earlier \BEGINADDED in Tables~\ref{tab:benefits} and~\ref{tab:challenges}\ENDADDED, for some developers that previously worked in-office with their co-workers, certain factors that were described as a challenge by some, were described as a benefit by others. In the statistical analysis, we even found that for some factors the corresponding challenge was associated with statistically significant lower levels of productivity, while the corresponding benefit was associated with statistically significant higher levels of productivity. Examples are the ability to focus (B7, C11) and the home work environment (B12, C17). %

These dichotomous experiences are expected, which we see from extensive research around the world on people's experiences of lock-down or social distancing during the pandemic. Furthermore, divergent experiences are expected given the varied family life, living conditions/location, job characteristics, and personality characteristics of our studied population. For employers and managers, knowing that ``one size does not fit all'' is critically important for the future of software development work.

The main divergent factors were: 
 \begin{itemize}
      \item \textbf{Ability to focus.} The number and nature of interruptions and distractions varied considerably, with some reporting more focus time at home and higher levels of productivity, and others having less focus time, especially those facing interruptions with family members at home and lower productivity. But even for those that appreciated fewer ``randomizations'' from colleagues since \wfh, at the same time, they missed the knowledge and awareness they gleaned from these and other informal interactions.
      \item \textbf{Work autonomy and motivation.} Increased autonomy and control over tasks and timing increased motivation for some, but reduced motivation for others and also their reported productivity. 
     \item \textbf{Work environment.} Some appreciated the novelty of \wfh, having natural light and more comfort at home, which was associated with higher productivity, while others missed their office work environment with extra amenities such as the cafeteria and reported lower levels of productivity.
     \item \textbf{Meetings.} Some felt there were too many meetings since WFH, and they missed face-to-face social cues and whiteboards, but others liked the shorter meetings and the associated artifacts they could refer to later. \RADDED{We observed that people managers and product managers reported more frequently having too many meetings.}
     \item \textbf{Work-life balance.} Many appreciated having more time due to no commute and being able to use that for extra time with family or to do personal chores or for self care, but others also found it difficult to disconnect from work, worked too many hours and did not have healthy habits since \wfh. \RADDED{People managers reported more frequently a lack of work-life balance and benefited less from flexible work hours.}
     \item \textbf{Childcare needs.} Having children with a need for childcare led to some surprising dichotomous experiences. From our analysis of the data from %
     \surveytwo, we saw that employees with children who had no difficulty handling childcare less frequently reported a drop in productivity (20\%) than employees without children (30\%). However, for employees with children who had difficulties handling childcare, 40\% reported a drop in productivity. We also found that 
     employees who previously had children in school or childcare were MORE likely to indicate these major challenges: \emph{lack of childcare}, \emph{more distractions or interruptions}, \emph{less time to complete my work}, and were LESS likely to indicate \emph{lack of motivation} as a challenge. \RADDED{In the subgroup analysis, people managers reported more frequently the lack of childcare as a challenge.}
    \item \textbf{Social connections.} Having fewer social connections was reported as a challenge for many, but for others, a minority, they felt MORE connected to their team and appreciated online activities such as standups, social lunches, games, daily check-ins etc.
 \end{itemize}

\subsection{Triangulating the Impact of Working from Home} 
\label{sec:triangulation}

In this paper, we have presented insights based on surveys with self-reported data through the lens of individual productivity. This is just one of many possible analyses on how remote work affects productivity. To illustrate an alternative perspective, we show an analysis of productivity by mining software data collected through the company's engineering systems. 
For this section, we analyze trends across Microsoft to see how developer productivity changed during work from home. 
We compared the pull request counts during the pandemic (March/April 2020) with prior historical values during comparable periods of the fiscal year (March/April 2018 and March/April 2019). %

    As the Microsoft engineering workforce grew in numbers since 2018, we normalized the pull request counts by the number of engineers to control for the growth. 
    The number of pull requests \emph{opened} per developer during the pandemic increased compared to previous years: 4.4\% compared to 2019 and +3.4\% compared to 2018.
    Similarly, the number of pull requests \emph{closed} per developer also increased: +4.0\% compared to 2019 and +2.1\% compared to 2018.

Different parts of the world went into ``lockdown'' at different times in March and April 2020. Microsoft has development teams spread across the world, spanning all continents. 
In order to control for geographic differences, we further analyzed the pull request data for the three main Microsoft regions separately: \emph{Puget Sound} in North America; \emph{ASIA}, which includes China, India and Japan; and \emph{EMEA}, which includes UK, Netherlands, Germany, France and Scandinavian countries. We observed that there was no  discernible drop in the number of pull requests for all three regions, including when normalized by engineer count. 

Overall we observe that there is no clear or significant drop (at statistically significant levels) in terms of the number of pull requests and the pull requests per developer. This data analysis suggests that the pandemic has not significantly influenced productivity at the company level. While this particular analysis shows that productivity has been stable or has slightly improved on average, it is important to recognize that just focusing on the company level alone loses the nuance of how individual people are affected differently. This highlights the need to run a family of experiments that investigate work from home using different types of data and methodologies such as diary studies~\cite{butler20} and workplace analytics~\cite{workplace-analytics}. 

\subsection{From Pandemic to Future of Work}
\label{subsec:future-work}

The pandemic has been a major disruption and will change how engineers work in the future and beyond the pandemic.
Of course, this is not unique to software development and this disruption is visible in other professions and for many kinds of knowledge workers. 
Many companies, software companies in particular, have announced either a shift to full remote work, or to partial remote work~\footnote{See \url{https://www.forbes.com/sites/jackkelly/2020/05/19/after-announcing-twitters-permanent-work-from-home-policy-jack-dorsey-extends-same-courtesy-to-square-employees-this-could-change-the-way-people-work-where-they-live-and-how-much-theyll-be-paid/\#4ac1881b614b}} in a hybrid fashion where more developers may be allowed or encouraged to work from home several days a week.
Working remotely is already the norm for some companies, for example GitHub~\footnote{\url{https://github.com/clef/handbook/blob/master/Employment\%20Policies/Working\%20Remotely.md}},  Automattic~\cite{berkun2013year}, and many other highly successful open source systems have been designed, written and maintained by distributed developers, many of them volunteers~\cite{raymond1999cathedral}. Much can be learned from these existing success stories, but there is more to learn and many factors to consider in a future hybrid setting.  An organization the size of Microsoft that has  primarily relied on co-located work must now adapt and find new ways to work in this new hybrid world the pandemic has left behind~\footnote{For example, articles such as this one that discusses the advantages of remote work from an organization that was previously remote: \url{https://www.nytimes.com/2020/07/12/business/matt-mullenweg-automattic-corner-office.html}}.

Remote work may suit many developers and projects, but there may be other aspects of software development that are negatively affected and the system engineering output data may not show those limitations.
We already see signals from our survey that teams may face collaboration and communication challenges.
In our first survey, one of our respondents noted that working at home does not provide the same information about the pulse of work but it could be addressed by tools:
\myquote{Automate team trends and share across the team, it's difficult to determine the real focus the team is trying to solve without seeing people stressing behind their desks (or not).}{XXX}{R_1Om0SQTG1ltL99a}
And individuals that are part of a team, may be concerned that a lot of the work they do (such as helping others) may not be visible to the entire team, as one noted:
\myquote{My biggest fear is being ``out of sight, out of mind''}{XXX}{R_1IRAcPopyNfYbpG}

Some reported addressing team work challenges through daily stand-up calls over video, virtual coffee hours, and more impromptu meetings (which as we mentioned above, was greatly appreciated by some participants). 
But some development activities, such as long-term planning and creative aspects of development, may be affected differently as some early work indicates~\cite{burke2020}. 
These aspects of development work need to be studied in a longitudinal fashion, especially if work becomes hybrid for some developers. Managers should also be studied as they may face additional stress working from home and managing a team. For example, it may be harder for managers to give feedback (an important factor for developer satisfaction and productivity~\cite{storey2019theory}) and maintain awareness of the well-being and productivity of their team members.  
A shift to hybrid remote work will also have some societal implications---our survey respondents recognized this and appreciated the positive effect on the environment less commuting may lead to. 

\section {Related Work}
\label{sec:relatedwork}

\BEGINADDED
There has been extensive research that compares distributed/remote development work with  co-located work, as well as several other studies that have studied developer productivity and well-being during the pandemic.
We summarize the key findings from these prior works below and compare findings across these studies.
\ENDADDED

\BEGINADDED
\subsection{Working Remotely}
\ENDADDED

Remote work was adopted by many large technology organizations \BEGINADDED long before the pandemic \ENDADDED because of the advantages that working remotely provides to employees. 
Remote work provides workers with the opportunity to engage with a globally distributed team introducing a wide range of perspectives to the project.
Another advantage is the flexibility of \emph{how} to work. Specifically, workers have the autonomy over when to engage and disengage with colleagues, providing unique opportunities for deeper concentrated work~\cite{ford2019remote}. Focused time to work is often challenging when colleagues face unscheduled interruptions~\cite{haynes2007office}. Remote work also provides the flexibility of \emph{where} to work, granting workers the ability to work from many parts of the world---which if well supported, can lead to distributed teams being just as effective as co-located teams~\cite{bird2009does}.

Despite the benefits with remote work, there are also several challenges that remote work presents for workers.
For example, the ability to build trust with colleagues while working remotely is critical for collaboration~\cite{bos2002effects} but can be harder to achieve. Close proximity and in-person work provides opportunities for unplanned interactions in-office
that build trust. In contrast, interactions in remote settings must be intentional or they will affect the building of social capital across distributed teams~\cite{haines2013here}. In remote settings, there is a need for more devoted time, resources, communication channels, and events to foster relationships.
Although remote work implies the ability to work from any location, working remotely has often been synonymous with \wfh--which has its own set of challenges. Some of the challenges with \wfh include supporting family members who may be sharing the same working space. For example, Heisman's interviews with remote workers at \gh identifies how they have been able to take advantage of flexible work hours with use of support groups to support their children~\cite{heisman2020remoteparents}. 

It is important to note that there were a few multinational technology companies who supported a ``remote-first''~\cite{mazzina2017remotefirst} work environment before the pandemic. Some of these companies have shared their best practices to support other organizations.  For example, in 2017, \so shared a blogpost about how their organization has created a successful remote work environment. \BEGINADDED In this post, they proclaim that the most important aspect that has contributed to their success is assigning an employee to be the main point of contact to respond to all remote work-related questions~\cite{pardue2017remotework}. \ENDADDED Their article further describes how effective it has been for someone in the organizations' leadership to advocate on behalf of remote workers. 

Similarly, GitLab, an all-remote DevOps technology company, released an inaugural remote work report to reveal the state of distributed work and explore the future of remote work~\cite{gitlab2020remotereport}. This timely report was released only days after King County employees were instructed to work from home (see Figure~\ref{fig:study-timeline}). The report shares research conducted by a third-party company that provides insights from over 3,000 remote workers across four countries in a variety of industries and roles. Some of the key takeaways from this report are \emph{all-remote work is surging}, \emph{remote work can foster a better sense of work-life harmony}, \emph{allowing remote work provides a hiring advantage}, and that \emph{``remote$\neq$alone''}, meaning that remote work does not have to mean workers are isolated.%
In response to \covid, GitLab produced a ``Remote Work Playbook''~\cite{gitlab2020playbook} where they describe strategic tactics to help support their more than 1,200 remote workers across 67 countries feel more supported. It also describes what other now remote companies can do for their newly transitioned remote workers. The playbook outlines guidelines on how to align values with expectations, how to manage remote teams, how to identify tools for effective communication, and how to encourage a healthy remote work lifestyle.

\subsection{Studies of Developer Productivity during the COVID-19 Pandemic}

\BEGINADDED
In addition to previously remote companies sharing their insights about effective remote work, several studies have looked at the impact of the pandemic on developers now working from home.  We summarize some of these studies below and briefly compare our findings. 
\ENDADDED

\subsubsection{Pandemic Programming: Developer Experiences and How Companies Can Help}

Ralph et al.~\cite{ralph2020pandemic} conducted an online questionnaire with over 2,000 responses from developers around the world (the largest proportion of 22.7\% were from Germany, followed by 16.4\% from Russia, 12.2\% from Brazil, and 4.4\% from US). They aimed to understand how working at home during the COVID-19 pandemic affected developer well-being and productivity. The survey was run at the end of March 2020 and participants were primarily recruited through social media channels frequented by developers. 

They found that developers' productivity and their well-being have suffered since \wfh, and that well-being and productivity are closely related. Dealing with the pandemic and home office ergonomics affected well-being and productivity and that women/parents/people with disabilities may be disproportionately affected.  
Their study leads to several recommendations how companies can support their employees:  pay for home internet, help with home equipment, pay attention to employee emotional well-being and assure them that their reduced productivity is expected and will not negatively affect their job.
\BEGINADDED
In a different study, Machado et al. studied the impact of the pandemic on developers in Brazil and likewise found negative implications on gender equality and on women in particular~\cite{Machado21}. 
\ENDADDED

\subsubsection{\gh Study in the Early Days of the Pandemic}

Forsgren et al. conducted an analysis of developer activity on projects hosted on \gh in the early days of COVID-19~\cite{forsgren2020covid}. They considered both open source and private project data. They found the following key insights when comparing the first three months of 2020 to the same time period in 2019: 
\begin{itemize}
    \item Developer activity (pushes, pull requests, code review and commented issues) was mainly similar to or slightly increased compared to the previous year.
    \item There was some disruption in the early days of work from home for enterprise projects but this quickly stabilized.
    \item Developer work days were longer by up to an hour per day with more work on weekdays and on weekends. They suggest this could indicate a risk of burnout.  
    \item Collaboration had increased on open source projects (in terms of number of users and projects). 
\end{itemize}
Their ongoing study from an engineering system performance point of view indicates that developers have stayed productive throughout the pandemic.  Their findings align with the findings we found from analyzing system data at Microsoft (see Section~\ref{sec:triangulation}). 

\subsubsection{The Baidu Pandemic Study}
Bao et al. studied the effect of the pandemic and \wfh at one of the largest IT companies in China, Baidu~\cite{bao2020does}. 
They conducted a quantitative analysis of 139 developers’ daily activities (over 138 working days).
They found that \wfh is associated with positive and negative changes in developer productivity in terms of the number of builds, commits and code reviews.
They also considered the influence of different programming languages, project size/age/type, and considered individual developers.  

They found \wfh was associated with negative changes for  large projects and has different effects for different developers. Their data suggests that developer productivity may be more stable working from home than working onsite (less variation in their levels of productivity). 

They also considered data from individual developers working from home and before work from home.
They found that for the majority (approx. 85\%), their productivity is about the same, but different for others (some are more productive, some are less). 
They asked developers to share feedback on their WFH productivity. The benefits that more productive developers reported were: working from home is exciting and energizing; developers can focus with fewer disturbances;
WFH decreases transportation costs and saves time; WFH increases flexibility of when to work and improves work-life balance. 
The challenges that developers with lower productivity included: more home demands; a need for self discipline; and decreased collaboration with others.
For the developers that found no difference with WFH, they experienced no barriers to completing their work, they could keep track of their schedule using online scheduling tools, and they found conferencing tools were powerful and effective for screen sharing.  

\BEGINADDED

\subsubsection{From Working from Home during the Pandemic to Working from Anywhere: A Case Study}

Smite et al. conducted a case study with a large international company with offices in Sweden, USA and the UK. They analyzed system data and conducted interviews to study how \wfh during the pandemic impacted developer productivity, satisfaction and collaboration~\cite{Smite21}.  They found from their system data that work activity continued without significant interruptions, but they reported changes in their daily routines. They also report different benefits and challenges speaking to the Tale of Two cities theme, such as less pairing, more loneliness, higher burnout, more communication friction and more scheduled meetings, but they reported more focus time and better work-life balance. They conclude their paper with recommendations for the future of work from anywhere to improve work culture but also to balance individual and team productivity. 

\subsubsection{A Longitudinal Study of Developers During the Pandemic}

Russo et al. followed 200 developers across the world working in similar lockdown conditions to understand the impact of the lockdown on their work activity, as well as to understand any changes in their self-reported productivity and well-being over the course of several weeks during the pandemic~\cite{russo2021daily, russo2021}. They found that developers spent relatively the same amount of time on different activities, but they spent less time in meetings and on breaks. They found no significant relations between productivity, well-being and working activities, suggesting that \wfh by itself does not present a significant challenge for organizations or developers, but introverts were potentially more impacted by the lockdown. 

\subsubsection{Other Pandemic Studies at Microsoft}
At the same time that this study was conducted at Microsoft, several other studies (using a variety of methods and focusing on different populations) were also ongoing in parallel. 
An extensive report by Teevan et al.~\cite{teevan2021} summarizes the findings from this and other studies. 
For example, one notable study of developers at Microsoft that was done in parallel to our study, by Butler et al., describes the use of a diary study to investigate not just challenges but also the gratitude felt by developers over the course of the first 10 weeks of the WFH directive~\cite{butler2021}. Some challenges reported by developers in this diary study included too many meetings, feeling overworked, and challenges managing their mental and physical health.  However, these developers also reported feeling gratitude for their families, their job, increased work flexibility and their team.  These authors report how insights from the diaries shape recommendations to improve developer work from home experiences. 

We also conducted a followup survey study at Microsoft where we considered in more depth the impact of work from home on \textbf{team productivity}~\cite{Miller2021}. Similarly, we found that for some developers, team productivity had gone up, while other developers reported that their team productivity was lower.  Some of the important factors that were associated with these changes include ability to brainstorm with colleagues, difficulties communicating with team members, and their satisfaction with social activities. Another follow-on study to the surveys we describe in this paper was conducted by Wang et al.~\cite{Wang21} to study the impact of returning to the office in China when the country started to recover from the pandemic.  They found workers preferred a hybrid style of working as many of the workers they studied were comfortable working at home after the pandemic and were used to using communication technology to do so.

In summary, the work from home studies conducted at Microsoft and across the world agree on a tale of two cities theme, where for some developers they did very well, but others did not, while many adapted over time. The studies led to important insights and recommendations that can shape developer work into a hopefully post-pandemic world.

\ENDADDED

\section{Conclusion}
\label{sec:conclusions}
 
 The COVID-19 pandemic has been and continues to be a worldwide human and economic disaster, with many repercussions that are already evident, but with other effects that we can't even yet imagine. 
 One thing is clear, a return to business as it was before is unlikely, and many predict that the future of development work is likely to be either fully remote, or for many, some form of hybrid work.
 Thus, it is critical to understand what has worked well and what has not gone well with remote work. We recognize that our study is only a start in understanding the implications of the pandemic on the software developer. 
In particular, new models of hybrid work are likely to lead to new challenges and benefits over remote work. 
 
 Our study reveals a ``tale of two cities''---even in a company that has support in place for its developers and remote work---and delivers not just quantitative insights \BEGINADDED (as we saw from several studies) \ENDADDED on how certain factors may be associated with higher and lower productivity, but also deeper insights into the narratives from these differential experiences. 
 The improvements our study participants recommended shine some light on how organizations (and managers) may support their developers, and we hope that the lessons learned from our study and other studies of development work during the pandemic will help others recognize and react to the disruptive changes we see unfolding in our industry.

\paragraph{Acknowledgments}
We thank the many respondents who answered our surveys and shared their experiences about working from home with us. We thank Victor Bahl, Peter Bergen, Surajit Chaudhuri, Jacek Czerwonka, Nicole Forsgren, Sam Guckenheimer, Brent Hecht, Donald Kossmann, Courtney Miller, Brendan Murphy, Madan Musuvathi, Paige Rodeghero, Jaime Teevan, and Scott Wadsworth for the great discussions about work from home and their support of this research. We also thank the entire Future of Remote Work v-team for the inspiring meetings. We thank Caroline Davis, Susan Hastings Tiscornia, Tanya Platt, and Kelly Sieben for the  privacy and ethics review of the surveys. We thank the anonymous reviewers and Cassandra Petrachenko for suggestions on how to improve this paper.

\bibliographystyle{IEEEtran}
\bibliography{main}

\appendix

\section{Codebook}
\label{appendix:codebook}

\newcommand{\bcreference}[1]{}
\newcommand{\codetitle}[1]{\hangindent=\dimexpr 0.5cm+\tabcolsep\emph{#1:}}
\newcommand{\themetitle}[1]{\noindent\begin{tabular}{p{\dimexpr \linewidth-2\tabcolsep}}%
\midrule \textbf{#1} \\ \midrule%
\end{tabular}}

{\footnotesize%
\parskip=1mm
\parindent=\tabcolsep%

\input{tables/codebook}
\noindent\begin{tabular}{p{\dimexpr \linewidth-2\tabcolsep}}%
\midrule\\
\end{tabular}%
}%

\input{tables/codecount}

\end{document}

%% file: aliases.tex
\definecolor{darkgreen}{rgb}{0.05,0.5,0.05}

\newcommand{\myquote}[3]{\begin{quote}{\small\faComment}~\emph{#1} (\lookupGet{#3})\end{quote}}
\newcommand{\myquotewithcodez}[4]{\begin{quote}{\small\faComment}~\emph{#1} (\lookupGet{#3})\newline#4\end{quote}}
\newcommand{\myquotex}[3]{\emph{``#1''} (\lookupGet{#3})\xspace}

\newcommand{\mycodez}[1]{\framebox{\emph{#1}\vphantom{g}}\xspace}

\newcommand{\wfh}{working from home\xspace}
\newcommand{\gh}{GitHub\xspace}
\newcommand{\so}{Stack Overflow\xspace}

\newcommand{\covid}{COVID-19\xspace}

\newcommand{\surveyoneresponses}{1,369\xspace}
\newcommand{\surveytworesponses}{2,265\xspace}
\newcommand{\surveytotalresponses}{3,634\xspace}

\newcommand{\numberCodes}{32\xspace}

\newcommand{\wfhcategory}[1]{\smallskip\textbf{#1}}

\newcommand{\surveyone}{Survey~1\xspace}
\newcommand{\surveytwo}{Survey~2\xspace}

\RequirePackage{color}
\definecolor{BLUE}{rgb}{0,0,1} %
\definecolor{BLACK}{rgb}{0,0,0}
\definecolor{HIGHLIGHT}{rgb}{0,0,0} 
\definecolor{REVISION}{rgb}{0,0,0} 

\newcommand{\ADDED}[1]{{\protect\color{HIGHLIGHT}#1}}

\newcommand{\BEGINADDED}{\protect\color{HIGHLIGHT}}

\newcommand{\RADDED}[1]{{\protect\color{REVISION}#1}}

\newcommand{\RBEGINADDED}{\protect\color{REVISION}}

\newcommand{\ENDADDED}{\protect\color{black}}

%% file: participants.tex
\lookupPut{R_2QhYxxrOzIJ8xcE}{P1}
\lookupPut{R_2yec50up2YE3MFg}{P2}
\lookupPut{R_2SdgeMfsYXuiNez}{P3}
\lookupPut{R_2uqyJTbPN1G8uNL}{P4}
\lookupPut{R_BEZYRo3WhyMopQR}{P5}
\lookupPut{R_1d0uJyznXvNvmDa}{P6}
\lookupPut{R_1OPFk7XW80g52V1}{P7}
\lookupPut{R_w5jdQfEOrGZQF0t}{P8}
\lookupPut{R_31ASeSO1VkXlXJ8}{P9}
\lookupPut{R_2dGc6Iztrx5kgGg}{P10}
\lookupPut{R_2uybGocFlttRSZb}{P11}
\lookupPut{R_3rUHyHzC9I0E5PZ}{P12}
\lookupPut{R_1LTjhYsr6YlisbY}{P13}
\lookupPut{R_1E061czKYYqisgp}{P14}
\lookupPut{R_3p8ZBsVqroAOPRp}{P15}
\lookupPut{R_ypPI0WNCipWRVxT}{P16}
\lookupPut{R_2zeEUFZaCZ1YyrW}{P17}
\lookupPut{R_qwQ0z1lnV6QvlQZ}{P18}
\lookupPut{R_24ut71AH3jmzjjB}{P19}
\lookupPut{R_24iLqbAJOv4OT2Q}{P20}
\lookupPut{R_263kXSgt11Q6XME}{P21}
\lookupPut{R_3KZuZTqH2xrRWrz}{P22}
\lookupPut{R_3kOezPysJPNV1FF}{P23}
\lookupPut{R_2SCCa8OoLuJgqDp}{P24}
\lookupPut{R_1NCTFXkeDgDc6og}{P25}
\lookupPut{R_2qlqurlcghoHphk}{P26}
\lookupPut{R_PCfMq75Hw44ENzP}{P27}
\lookupPut{R_27U06WlU6igY2Bn}{P28}
\lookupPut{R_Qht6lHTRma8Jly9}{P29}
\lookupPut{R_2X69xzk9boloIKb}{P30}
\lookupPut{R_2QxkOqAVCsSIiE5}{P31}
\lookupPut{R_2az6235kOVuP6Nh}{P32}
\lookupPut{R_2ClDHYit6qdzUJm}{P33}
\lookupPut{R_3HgX0AYGgmRWl2e}{P34}
\lookupPut{R_cSL3ZlNXD8jNW2B}{P35}
\lookupPut{R_vwVLqA7RiwCMD3X}{P36}
\lookupPut{R_2y2RYhzd2Ogysvl}{P37}
\lookupPut{R_ZyLjhxur9bd029b}{P38}
\lookupPut{R_10GcerD7kXaLzrv}{P39}
\lookupPut{R_Q6yJTfPnCRgtBCN}{P40}
\lookupPut{R_1gvVhR1eeEpZDir}{P41}
\lookupPut{R_246FLoQPhWvkxgM}{P42}
\lookupPut{R_2t5uYNqejqNNlqw}{P43}
\lookupPut{R_27qHZG3G4O8kBzq}{P44}
\lookupPut{R_yJwmzVifabHq0sF}{P45}
\lookupPut{R_roNDzIsneX6BIAN}{P46}
\lookupPut{R_1F4WnzNfVTC6ZQz}{P47}
\lookupPut{R_2dsTV5Z8nG9DmW7}{P48}
\lookupPut{R_2zpX2l8aEkGMTMv}{P49}
\lookupPut{R_3psfVQOzvvQ0pKI}{P50}
\lookupPut{R_3lG5IMc8mO1q82f}{P51}
\lookupPut{R_1NlfW9AzD8Rttxp}{P52}
\lookupPut{R_1Lev62noFWmf2N2}{P53}
\lookupPut{R_9MqHB6OX0ft0bbH}{P54}
\lookupPut{R_3p54sLnFp1FC3m4}{P55}
\lookupPut{R_1KiG8EEU5Bl6oSV}{P56}
\lookupPut{R_3OjlX4UtMe4SNwJ}{P57}
\lookupPut{R_8p44FZdAgP4Xmxj}{P58}
\lookupPut{R_vwQbPDsRR4O3y6J}{P59}
\lookupPut{R_1dMNkrzmrhNuLmx}{P60}
\lookupPut{R_2uQE8iGBPhrAsBc}{P61}
\lookupPut{R_3MlNczDKngkAUhr}{P62}
\lookupPut{R_1Qt1rrOQPEmjPkE}{P63}
\lookupPut{R_sMqFVMumSKyRAnn}{P64}
\lookupPut{R_07DnymPaSvLvpHb}{P65}
\lookupPut{R_1HhQdS7M4ZuOZ3v}{P66}
\lookupPut{R_2fx4zGNUB5aAzT3}{P67}
\lookupPut{R_zcZ8yEzaYsTLJhD}{P68}
\lookupPut{R_1mK4roVUn5bDrEZ}{P69}
\lookupPut{R_wTAZ1uQxo5xyXYt}{P70}
\lookupPut{R_ritNjFhf7eAVDwt}{P71}
\lookupPut{R_1K2yfUfcFB1wjLn}{P72}
\lookupPut{R_1LzFvjfrSUNcEnU}{P73}
\lookupPut{R_86sKEzl1KJHhdvP}{P74}
\lookupPut{R_2eW6bK9FYsGSmyH}{P75}
\lookupPut{R_2dhyHPfnESToWJe}{P76}
\lookupPut{R_1hHmcHnftEat9NR}{P77}
\lookupPut{R_2as27CDxgGZAJ6Q}{P78}
\lookupPut{R_3oLL90vAb5X36TW}{P79}
\lookupPut{R_1gOvu7NBkFhTkrc}{P80}
\lookupPut{R_BtsrgMRSw2nyAfv}{P81}
\lookupPut{R_3sgoxm1m0OXjXXR}{P82}
\lookupPut{R_3Kw5QGeoVdwKE28}{P83}
\lookupPut{R_2zciC6BTPcFhSwB}{P84}
\lookupPut{R_28ZArZkX9sfWKci}{P85}
\lookupPut{R_3qTKULFIQXyuz2F}{P86}
\lookupPut{R_3HoZnh3BUg8oOTY}{P87}
\lookupPut{R_3fuWSF2zZovfH4B}{P88}
\lookupPut{R_C8QFvMHRgRpm3aV}{P89}
\lookupPut{R_2XmMf7DWwDvTOMV}{P90}
\lookupPut{R_qWxdknK126JbTMZ}{P91}
\lookupPut{R_1hXPR5DTWSHiTli}{P92}
\lookupPut{R_eQzl5OJltP01b3P}{P93}
\lookupPut{R_cIqEmqSaj7UuFup}{P94}
\lookupPut{R_VPtaYKAcrxcWONX}{P95}
\lookupPut{R_2duc4sOpGoyWwQt}{P96}
\lookupPut{R_URsvwVym57gXPBD}{P97}
\lookupPut{R_3kh7dAbvTFvUODs}{P98}
\lookupPut{R_1NFJfRWRpVQqrjB}{P99}
\lookupPut{R_33mAgQTiIeeKI3p}{P100}
\lookupPut{R_38n8etKfdB1gf29}{P101}
\lookupPut{R_3plu6LTn9ufuBLR}{P102}
\lookupPut{R_PMMjc41KvBY2Pwl}{P103}
\lookupPut{R_1FEcpXZJ8McnSJm}{P104}
\lookupPut{R_3KB1QpJWvHc9S9b}{P105}
\lookupPut{R_2VyDRYfns44pNL0}{P106}
\lookupPut{R_UGimlI0xLi5LFwR}{P107}
\lookupPut{R_zcfYOjt4DKP33q1}{P108}
\lookupPut{R_3F3DvSuwbpgocYe}{P109}
\lookupPut{R_1i86kuKSZCwm3A2}{P110}
\lookupPut{R_2PcJpH0O6NEOSmy}{P111}
\lookupPut{R_3hA6hMdbJyUEt0f}{P112}
\lookupPut{R_zTHtRgiaTeG883T}{P113}
\lookupPut{R_XNA2xtrluHD1InD}{P114}
\lookupPut{R_1odQgDXWbeeM7wg}{P115}
\lookupPut{R_2wuI155wwAOdoPo}{P116}
\lookupPut{R_VLsxnwAeN8N5uwh}{P117}
\lookupPut{R_3Rfnac86j1HufbD}{P118}
\lookupPut{R_2bOu5DBEyMjKk52}{P119}
\lookupPut{R_2qFbI0PVK7OT9pe}{P120}
\lookupPut{R_cMggqn03fhHsZ2h}{P121}
\lookupPut{R_2ztGzJ9hjrThn2v}{P122}
\lookupPut{R_C7uYO4QDc4jCAsp}{P123}
\lookupPut{R_2PjvFMzOL8SJfZm}{P124}
\lookupPut{R_1kZeD32GoRPLf93}{P125}
\lookupPut{R_1LG7J5tuVFMfQI2}{P126}
\lookupPut{R_W7LyE4vCEPjrpNT}{P127}
\lookupPut{R_pc8zC0D72BdCRvr}{P128}
\lookupPut{R_2Yb5PREKlyLwCkd}{P129}
\lookupPut{R_cMFeXkjzNsCU2LD}{P130}
\lookupPut{R_2QXnFU2xaIjATzA}{P131}
\lookupPut{R_2VR4F4LfZK5oeT8}{P132}
\lookupPut{R_1OIFFmE5ZbP8QKg}{P133}
\lookupPut{R_3COEJEGjrN27Akd}{P134}
\lookupPut{R_1nNLMm7Tpwl7uRl}{P135}
\lookupPut{R_2AEjD9TC6oY8jse}{P136}
\lookupPut{R_RlQ40sZ1vWqgqNb}{P137}
\lookupPut{R_2CDGro15oOlzY7W}{P138}
\lookupPut{R_3Gf7Up40MGWE6qq}{P139}
\lookupPut{R_23W9feuiPcOPyDm}{P140}
\lookupPut{R_1mJCxiNxrZj0GCZ}{P141}
\lookupPut{R_2bQ5hpsf0cLwttA}{P142}
\lookupPut{R_2fvODXiyftq6nis}{P143}
\lookupPut{R_263dtopdzmNfW7T}{P144}
\lookupPut{R_Q4D6syPn394SeqZ}{P145}
\lookupPut{R_3DERDufB5YtfC5b}{P146}
\lookupPut{R_1Netav5ATYNhgLY}{P147}
\lookupPut{R_aeHsQVP2zxLZthT}{P148}
\lookupPut{R_2UhAKSESSyjrx26}{P149}
\lookupPut{R_3lJrmIQhD1TBlIx}{P150}
\lookupPut{R_saRI451J7Kn7uKt}{P151}
\lookupPut{R_xAFUa3TqDIPflSh}{P152}
\lookupPut{R_bHNiOEbhnEhMtG1}{P153}
\lookupPut{R_22qRRvdY7X057Cu}{P154}
\lookupPut{R_3h53A1oNOlvt0uI}{P155}
\lookupPut{R_XA11UlWMxsiPHTH}{P156}
\lookupPut{R_2tL0BH0t9M8yTUI}{P157}
\lookupPut{R_yOqMom5Jmq75LI5}{P158}
\lookupPut{R_32VEa92xmxnxu4E}{P159}
\lookupPut{R_1CDcs7hNkfBAzGL}{P160}
\lookupPut{R_3iQgCcULw8KZOS8}{P161}
\lookupPut{R_8dg16ECJ4p2to1X}{P162}
\lookupPut{R_snAKw4628EsDy9P}{P163}
\lookupPut{R_VJxIwLovTsVTFlv}{P164}
\lookupPut{R_3D5AB6024euQ4sl}{P165}
\lookupPut{R_3iEjaee3U9a2tps}{P166}
\lookupPut{R_1mEO1siQSQxO2JE}{P167}
\lookupPut{R_1r6tIJdLDXZ53NZ}{P168}
\lookupPut{R_3kp8Jdv3xF8sLlW}{P169}
\lookupPut{R_1pseHJa8wXOGLGJ}{P170}
\lookupPut{R_2nLRFGLrjsPmbgB}{P171}
\lookupPut{R_25zpZRbo2i1dXGP}{P172}
\lookupPut{R_2zSF7deUqILgPxP}{P173}
\lookupPut{R_urzu4fDjMutW8aB}{P174}
\lookupPut{R_abmG1f5sUFDzGU1}{P175}
\lookupPut{R_1rJTavlyGEPu1wP}{P176}
\lookupPut{R_1jrmaS4lZtnG8Uq}{P177}
\lookupPut{R_uk890vApAdg3Wbn}{P178}
\lookupPut{R_2y8yHQ7WbRTmxnb}{P179}
\lookupPut{R_3qPvBTddlFdezQ7}{P180}
\lookupPut{R_1CC7G8Szz27ZMqY}{P181}
\lookupPut{R_24CYwbutnarT7Mb}{P182}
\lookupPut{R_b7NB23FuGVMWQBr}{P183}
\lookupPut{R_2VEYc6ssookniGu}{P184}
\lookupPut{R_3ffcuPqYR1gyBtl}{P185}
\lookupPut{R_2uQyLbyDO80ltib}{P186}
\lookupPut{R_1dt2EiEzrA69Cvw}{P187}
\lookupPut{R_2SCzNJqz4gfxD1P}{P188}
\lookupPut{R_3D6kOSpOvEo4QKe}{P189}
\lookupPut{R_1ePOSZQEwSbLm5R}{P190}
\lookupPut{R_1rjP5XiStS6mKMD}{P191}
\lookupPut{R_2uVEQgDMhAGfAel}{P192}
\lookupPut{R_3DjWbJXzzMzFUK7}{P193}
\lookupPut{R_3D8eyLr1ZcdxPTt}{P194}
\lookupPut{R_2Y9XvduDVpracQo}{P195}
\lookupPut{R_3CPWNBTQLXAUVTl}{P196}
\lookupPut{R_1LcFQJSZFfpj5zn}{P197}
\lookupPut{R_1mxRzhVR2bS4EAI}{P198}
\lookupPut{R_2QWWPK7EE69CpLv}{P199}
\lookupPut{R_3ewrdYhCPyOH4NT}{P200}
\lookupPut{R_3NDoq1PIRR2fMMf}{P201}
\lookupPut{R_10NlJL7TJWl4o6y}{P202}
\lookupPut{R_1mUGOvVroKygr3F}{P203}
\lookupPut{R_2qC1FpXxbN9yKOq}{P204}
\lookupPut{R_24CMItWb8oUY30U}{P205}
\lookupPut{R_1Dp94PMtYUKwoT8}{P206}
\lookupPut{R_2Wv8wMxg2jSRYnr}{P207}
\lookupPut{R_2SvpTpItziTfm0R}{P208}
\lookupPut{R_3qU8Ant97kZDa6m}{P209}
\lookupPut{R_1Qs94WBWvxJyLaT}{P210}
\lookupPut{R_27amzn4O1xlgTjv}{P211}
\lookupPut{R_1joYf29ApLyGaAY}{P212}
\lookupPut{R_6A5Y6DfVPAU8Vzj}{P213}
\lookupPut{R_8q2mDiTKVvoMf3r}{P214}
\lookupPut{R_2dKZf5IG6crztdq}{P215}
\lookupPut{R_ripeYoeh8wOfQ1X}{P216}
\lookupPut{R_1DY0mn6VxiO6Qxu}{P217}
\lookupPut{R_UH39zSVd1du1Dk5}{P218}
\lookupPut{R_3QQ9OVCzTb7TKRd}{P219}
\lookupPut{R_31nhmZaOJG5E3C4}{P220}
\lookupPut{R_3PnPgxkqeEEggfI}{P221}
\lookupPut{R_2bJI5GxKT31IaMF}{P222}
\lookupPut{R_1rD0XIqTLYEcnpq}{P223}
\lookupPut{R_27eQD9fY67lBkdT}{P224}
\lookupPut{R_31MC5a99v7NIc4s}{P225}
\lookupPut{R_2PdiQpYWl09Ly9J}{P226}
\lookupPut{R_117liAwRyqWYeSR}{P227}
\lookupPut{R_23eoNVYdPRKpkQv}{P228}
\lookupPut{R_3iRociSQUC7WFJr}{P229}
\lookupPut{R_2vY6mPne6j1mRmn}{P230}
\lookupPut{R_0lypTF3WmPQHpct}{P231}
\lookupPut{R_1qdPBS18m8ON2Rk}{P232}
\lookupPut{R_22JhjLQglqxy3Fo}{P233}
\lookupPut{R_3gRTjONOe4ip6ho}{P234}
\lookupPut{R_1pzmVREaSpUixK1}{P235}
\lookupPut{R_33kEEBhA73F3vBX}{P236}
\lookupPut{R_27TdJ9DU6himlly}{P237}
\lookupPut{R_sHluAWJNuB3M8Gl}{P238}
\lookupPut{R_1JUCX5aT3ibnk9w}{P239}
\lookupPut{R_2DSvmG6QmMfudS8}{P240}
\lookupPut{R_1K2YKX8oIIat5mU}{P241}
\lookupPut{R_1JL14inAYypcPCh}{P242}
\lookupPut{R_2tnGSQkXgy4aU6P}{P243}
\lookupPut{R_esnxM2V2NtAMa3L}{P244}
\lookupPut{R_10VqcFDroE9voYy}{P245}
\lookupPut{R_3qeyY0MxNVK7ZkU}{P246}
\lookupPut{R_2D1YtNFY6xENPG4}{P247}
\lookupPut{R_1cTgKieFHUT8x9h}{P248}
\lookupPut{R_3Dno8nq0CAdhZOc}{P249}
\lookupPut{R_1hA2rimN3kIyFpG}{P250}
\lookupPut{R_3g5HUAJP4eZInTz}{P251}
\lookupPut{R_wY08ghCZL7Xpk7D}{P252}
\lookupPut{R_plOJWNvkhExTXJ7}{P253}
\lookupPut{R_bCn11CoJmM9yQlb}{P254}
\lookupPut{R_1N47ZKlLj0dFsu9}{P255}
\lookupPut{R_1gwHQkfD7tSs5zh}{P256}
\lookupPut{R_2VyVa2bn4dKQ8G9}{P257}
\lookupPut{R_2SernBYVQAWUIR7}{P258}
\lookupPut{R_1oneuCw6zSnOeaH}{P259}
\lookupPut{R_AuIvVAaEq7A1HEZ}{P260}
\lookupPut{R_1d3MbxEeoXzteNA}{P261}
\lookupPut{R_9pN21mNpaVJy0jT}{P262}
\lookupPut{R_3rG5j3xXL8nsguN}{P263}
\lookupPut{R_1hSFppfk9kmnQBC}{P264}
\lookupPut{R_2z7rJaNfrOsDzoS}{P265}
\lookupPut{R_3LiOoDmpDLw0RPx}{P266}
\lookupPut{R_1rqdpRwbgJ6nsYT}{P267}
\lookupPut{R_3KpQYBDJqZRrjam}{P268}
\lookupPut{R_3sAo493oY3jcnSW}{P269}
\lookupPut{R_XmI3LDGR4iT1KcF}{P270}
\lookupPut{R_1dKPKfAvVKnHNsj}{P271}
\lookupPut{R_0qwl9lBJlBNAxod}{P272}
\lookupPut{R_2dhV70dWvH4rRei}{P273}
\lookupPut{R_2wtqI7xtRZPWLCl}{P274}
\lookupPut{R_2di2r4Vya2dhBzp}{P275}
\lookupPut{R_1LOkOCtqEX4WVlo}{P276}
\lookupPut{R_1hBhy4czqX2tmHK}{P277}
\lookupPut{R_qURqxoG0p0OlC0x}{P278}
\lookupPut{R_0D6fzJNzkTvZz0Z}{P279}
\lookupPut{R_24rfY33H5RuwygC}{P280}
\lookupPut{R_1jjPiaJLD0OPsq1}{P281}
\lookupPut{R_25BMmykC4FUHiE1}{P282}
\lookupPut{R_1ocoJj2znV7h3z0}{P283}
\lookupPut{R_cSnE1iGUypXdgqZ}{P284}
\lookupPut{R_2OUNc64O6VypHc8}{P285}
\lookupPut{R_2thwMO7G0FfTLVC}{P286}
\lookupPut{R_Rf3SeSMg8oyHeCZ}{P287}
\lookupPut{R_xhCWE8qDaMq6smZ}{P288}
\lookupPut{R_31sBkX3q0PZV6o6}{P289}
\lookupPut{R_BFL28hyc3IPT8iZ}{P290}
\lookupPut{R_2SfdTJgydwoyvSJ}{P291}
\lookupPut{R_2vjlRAIjMyFgrA1}{P292}
\lookupPut{R_1gHMbhc7GQNGjK7}{P293}
\lookupPut{R_ehV75JV12f7S1eF}{P294}
\lookupPut{R_yjB6Nru2HbLXwmB}{P295}
\lookupPut{R_1CpXdz7QHGGTe72}{P296}
\lookupPut{R_2BeOzQUO9eaeiso}{P297}
\lookupPut{R_2BhWAVBiG0l5D3Z}{P298}
\lookupPut{R_AgH6tvlbjnzAgw1}{P299}
\lookupPut{R_Okj81j1io5kZ62B}{P300}
\lookupPut{R_3lFjuW6VwdRmBfG}{P301}
\lookupPut{R_sLJWp3Ym37BGcGR}{P302}
\lookupPut{R_2cbUihVC3Dv6kiV}{P303}
\lookupPut{R_2CoZISB8FFoInKV}{P304}
\lookupPut{R_3ELOJOiP5IdYfY5}{P305}
\lookupPut{R_25NMc04EQLBYTQL}{P306}
\lookupPut{R_29c3of0cEW9r2AZ}{P307}
\lookupPut{R_2V9JI8O05IDgI9F}{P308}
\lookupPut{R_3HHb7BNkil4pwJH}{P309}
\lookupPut{R_3fCqI2NKXllCB4t}{P310}
\lookupPut{R_2gCCb1gEzSxlumZ}{P311}
\lookupPut{R_2trPSUFUj2NvcId}{P312}
\lookupPut{R_1dAzjdfk48upEn2}{P313}
\lookupPut{R_2SGEFmgobQjQ8PX}{P314}
\lookupPut{R_3ELxbvWYi1uRZzJ}{P315}
\lookupPut{R_1BVs4ilMQOANnUO}{P316}
\lookupPut{R_3nvBZSH99z6v7xf}{P317}
\lookupPut{R_XO1hFRbuXuv53DH}{P318}
\lookupPut{R_25Tk6AIykepP9VW}{P319}
\lookupPut{R_3MEZiVXADnefQf5}{P320}
\lookupPut{R_6QBhGWrTysGFt0B}{P321}
\lookupPut{R_2v8oZBfQGrDgkrf}{P322}
\lookupPut{R_2upSs5ctw4atEMX}{P323}
\lookupPut{R_tFisJqI5ZXUxsBP}{P324}
\lookupPut{R_2aFDcTv5D89AvCd}{P325}
\lookupPut{R_32KsnV2i9rU3P0L}{P326}
\lookupPut{R_9mKzu6ZXWKPwbQZ}{P327}
\lookupPut{R_2Pbc5xMxN89j19f}{P328}
\lookupPut{R_1oAtyXs5Lzd91a8}{P329}
\lookupPut{R_31QJb6Uk5LeRIi2}{P330}
\lookupPut{R_2tJzdjKNaKDDItC}{P331}
\lookupPut{R_2tL22NE58as1wUP}{P332}
\lookupPut{R_1immlUzAEXPS5Yc}{P333}
\lookupPut{R_p9mBUYkp3nGna8h}{P334}
\lookupPut{R_2B4oog2pDyJsuB6}{P335}
\lookupPut{R_1QgSTIUrqkTFkYm}{P336}
\lookupPut{R_390LxvoAGXyiLgB}{P337}
\lookupPut{R_C9QURDTfT96z7AR}{P338}
\lookupPut{R_1pRKZdKywGjyVGy}{P339}
\lookupPut{R_1ihbXeOSH7WNXAI}{P340}
\lookupPut{R_3JFdVZtVHPd0jsZ}{P341}
\lookupPut{R_2fDHmT8sYn0PHxq}{P342}
\lookupPut{R_7aE5kUKe5As4mQx}{P343}
\lookupPut{R_sTetq07mzpLTm9j}{P344}
\lookupPut{R_UDygtpcxmqF95yp}{P345}
\lookupPut{R_31bdGbOoTC0O36O}{P346}
\lookupPut{R_2dnnA0Oy9SC0ldj}{P347}
\lookupPut{R_3MhdlJnJaDddr8S}{P348}
\lookupPut{R_30cSbqs2aT7B585}{P349}
\lookupPut{R_2DS6eqxDIwR8KhM}{P350}
\lookupPut{R_zTIMb1o6jhiMvE5}{P351}
\lookupPut{R_2uxMzYIyWhQpSaU}{P352}
\lookupPut{R_3EiKrDrB3aHWr2g}{P353}
\lookupPut{R_Z1XP7NY2CpQCjvP}{P354}
\lookupPut{R_DMKQ7QWeojQeqrf}{P355}
\lookupPut{R_2cdfsLIOtUVF6km}{P356}
\lookupPut{R_yW9ld2e4jJsUpEZ}{P357}
\lookupPut{R_1QnKULWzGkYEPL5}{P358}
\lookupPut{R_3KIRGZHKFOt5wmG}{P359}
\lookupPut{R_0GMVjj8AGUu8ZKV}{P360}
\lookupPut{R_3g1MuqjxxoePvbl}{P361}
\lookupPut{R_sIukv6LxHBaYNlT}{P362}
\lookupPut{R_2YkT5joiClRR52a}{P363}
\lookupPut{R_2Cyu7JUuv4c6hly}{P364}
\lookupPut{R_zdpoNPsLURodyNj}{P365}
\lookupPut{R_yWUss6P9aGUfLsB}{P366}
\lookupPut{R_1i5tt0NzsJZ9hyh}{P367}
\lookupPut{R_30wdqRilTw1dRSF}{P368}
\lookupPut{R_2eW1wJTRFgZMbBG}{P369}
\lookupPut{R_VOVEaLSI52TKE3n}{P370}
\lookupPut{R_6hxPRq3I1dVhE8V}{P371}
\lookupPut{R_2BleNqqDd1VQ3ME}{P372}
\lookupPut{R_2ifCltwPGa4wy3L}{P373}
\lookupPut{R_3ltBuJgHw4FZYfT}{P374}
\lookupPut{R_2QVQOFa3uD1gYJw}{P375}
\lookupPut{R_3JDazXkeTrY34A8}{P376}
\lookupPut{R_2ZZyxQC7L7C0HaF}{P377}
\lookupPut{R_qDCQSQgSWmYj65X}{P378}
\lookupPut{R_3rIktuQ5WhtENnX}{P379}
\lookupPut{R_12D4dW7ANxTnIwB}{P380}
\lookupPut{R_22nzYgD0TdzHk1F}{P381}
\lookupPut{R_db6hKQyZRBCBunf}{P382}
\lookupPut{R_2OVbH4VVXhVkdX0}{P383}
\lookupPut{R_32PHtzXJggT42hZ}{P384}
\lookupPut{R_1LhTAsbJihdf8Vd}{P385}
\lookupPut{R_1o0v0z3KgrKGlFb}{P386}
\lookupPut{R_1QmRfcbf3QQgcI4}{P387}
\lookupPut{R_1Nea8HJ46usu0b9}{P388}
\lookupPut{R_3R2FBCbw96hICaV}{P389}
\lookupPut{R_1ovdkDTXpl7CgEB}{P390}
\lookupPut{R_1d51TyOa75B4UBK}{P391}
\lookupPut{R_1Ka4gqNe3nnRF4V}{P392}
\lookupPut{R_3hAO6dmmrSjTU7V}{P393}
\lookupPut{R_2qlr2coKJQddzwy}{P394}
\lookupPut{R_3hychKhlHBYfWuD}{P395}
\lookupPut{R_3GClcYwGt18Kk7X}{P396}
\lookupPut{R_by2a2nQVhN8g6bf}{P397}
\lookupPut{R_2rBpaJY8Oc2mSvv}{P398}
\lookupPut{R_DTrwG3cL9X9Y1CF}{P399}
\lookupPut{R_27yzjDTh6KYXvJR}{P400}
\lookupPut{R_30l608tVczGNLEQ}{P401}
\lookupPut{R_1FtBWeUuzNsFO0h}{P402}
\lookupPut{R_1OPIgbkj0lHb05E}{P403}
\lookupPut{R_3Hi6Cr1zpHvejOk}{P404}
\lookupPut{R_3J5jX4zozcJJEt6}{P405}
\lookupPut{R_23W8Z1OuAUi14Nk}{P406}
\lookupPut{R_2U4oU45gD68VZl6}{P407}
\lookupPut{R_2OPjeQIw1PTNISY}{P408}
\lookupPut{R_UR8gjYMf9bd1hND}{P409}
\lookupPut{R_1FmlPRJC38FQ8uL}{P410}
\lookupPut{R_12LuLJOKoc5ewdR}{P411}
\lookupPut{R_3QFpsxUR3RCA99h}{P412}
\lookupPut{R_tGuPlwwTEDiExuF}{P413}
\lookupPut{R_3RyJmgDyS7R4emh}{P414}
\lookupPut{R_eR1j35RyvigWBOh}{P415}
\lookupPut{R_3psDMrJfJB9IAVh}{P416}
\lookupPut{R_1hVV1zWldPVMHZ7}{P417}
\lookupPut{R_3ELpQVFXlFPLVXg}{P418}
\lookupPut{R_3Pzd3hVZK6iyan0}{P419}
\lookupPut{R_3kbxn47kHr9A0m7}{P420}
\lookupPut{R_2q98lUdyWJi4Pfe}{P421}
\lookupPut{R_yx3TcRRZLpkJyF3}{P422}
\lookupPut{R_27ythWmrfg5QY40}{P423}
\lookupPut{R_3nSGOKTqj7r6YzL}{P424}
\lookupPut{R_3ESvPsAjM1QijCv}{P425}
\lookupPut{R_28NcKPlPyu60TNH}{P426}
\lookupPut{R_1dmhM2tvePMfU2q}{P427}
\lookupPut{R_2zGeOMbxSS9sTC1}{P428}
\lookupPut{R_2m1h93SIYXFZNux}{P429}
\lookupPut{R_2czrcGXNa8Riwhf}{P430}
\lookupPut{R_rl2U8G4Ouqgncfn}{P431}
\lookupPut{R_3PXBfwPjK3UqSei}{P432}
\lookupPut{R_3rOQE9XhwoaBCIk}{P433}
\lookupPut{R_cYgLpNB5pGtQOwV}{P434}
\lookupPut{R_3QDY0gllmhb8kDN}{P435}
\lookupPut{R_1MJWfIh3Hl2QRwZ}{P436}
\lookupPut{R_BPPRrqHqMI8DSQV}{P437}
\lookupPut{R_2fAKy9qtcFqI6kD}{P438}
\lookupPut{R_2OU1NYjO9YPXxcH}{P439}
\lookupPut{R_3L29joBsVOssQ9r}{P440}
\lookupPut{R_2bZf6qYybIbEo7L}{P441}
\lookupPut{R_WvCqhVGQNqXOFnr}{P442}
\lookupPut{R_2usaiaXJioBiZJK}{P443}
\lookupPut{R_1M5pigAOBAlOjTB}{P444}
\lookupPut{R_2dvkxLJ9ceA2JrC}{P445}
\lookupPut{R_W2LfQ6gsxz5vxYJ}{P446}
\lookupPut{R_3k4VwA2uuxfgiiX}{P447}
\lookupPut{R_2ZErMz5vmgeLzqs}{P448}
\lookupPut{R_5i4oa0BUB1lyhjP}{P449}
\lookupPut{R_ueS14VAQ9I5EHZL}{P450}
\lookupPut{R_qUSOqaaWDOTB8at}{P451}
\lookupPut{R_3novrSIFGwJtNTA}{P452}
\lookupPut{R_8CFrvjTc9SI03Wp}{P453}
\lookupPut{R_2qypJ4BWECDUyOi}{P454}
\lookupPut{R_279yYMIOwv2eyw2}{P455}
\lookupPut{R_2TRb2wmnMQ9yoQu}{P456}
\lookupPut{R_23nluYZnNDgx5mN}{P457}
\lookupPut{R_1MSsJ24Y3fxZvAo}{P458}
\lookupPut{R_2ymFG1W8hhxg0Xo}{P459}
\lookupPut{R_2dzuw8l4ku46xJe}{P460}
\lookupPut{R_24D9BbdJt6w7PzR}{P461}
\lookupPut{R_2fJ44PventcqXgB}{P462}
\lookupPut{R_1pVwxXd1A5lON3d}{P463}
\lookupPut{R_3iWpT4gkfZbR1L8}{P464}
\lookupPut{R_4IvLlGbO0vet2Ct}{P465}
\lookupPut{R_T73ey7QX76wk3PH}{P466}
\lookupPut{R_1BQkUcvubr8kBya}{P467}
\lookupPut{R_2rJybzBo8CjAzDL}{P468}
\lookupPut{R_27JBYI8bI0oSHNM}{P469}
\lookupPut{R_2AYXkgg7oW0zHcA}{P470}
\lookupPut{R_w18XWZrl5QwGqFH}{P471}
\lookupPut{R_2v05hMtz83iP8QS}{P472}
\lookupPut{R_2TCaIVIlsesERcC}{P473}
\lookupPut{R_a4TDGiMtUrm1SSd}{P474}
\lookupPut{R_2E00uAwyx4RVwY4}{P475}
\lookupPut{R_2z6JgYxbVcxQ9DD}{P476}
\lookupPut{R_vYOHiHIPqHbM9z3}{P477}
\lookupPut{R_1roALdYLC1cY7hE}{P478}
\lookupPut{R_2b1uGMmm4vFGi3Q}{P479}
\lookupPut{R_30o9E3jls9Fmxmj}{P480}
\lookupPut{R_2DY6iyOxdTzvKQS}{P481}
\lookupPut{R_2PveDuIW1xSXPnB}{P482}
\lookupPut{R_1kMoGk1UPGYuybk}{P483}
\lookupPut{R_3QWcrQqPwCbt7s9}{P484}
\lookupPut{R_3qk6MIjwqVni3Cu}{P485}
\lookupPut{R_3e7X9ukI4Chv84l}{P486}
\lookupPut{R_3kG17vE4fk6zAy0}{P487}
\lookupPut{R_1PTb9SE9PLMm1EW}{P488}
\lookupPut{R_22GhCSvmLzw3Vkm}{P489}
\lookupPut{R_20VjqtNTNbQWlBp}{P490}
\lookupPut{R_10ZB31w9IIczbJC}{P491}
\lookupPut{R_1IQbI4y4sucQj42}{P492}
\lookupPut{R_20SeMyFnr02Baic}{P493}
\lookupPut{R_Pv96O5pA1gVsld7}{P494}
\lookupPut{R_2Vl46TaYa5qutt0}{P495}
\lookupPut{R_0xlwfvBocPIgdKV}{P496}
\lookupPut{R_12b4dVX9K4oUdXH}{P497}
\lookupPut{R_2954wfjgbmeP5kV}{P498}
\lookupPut{R_tFpL1pXbTs76YuZ}{P499}
\lookupPut{R_3k0adONtvYQNx11}{P500}
\lookupPut{R_XnccZZlni4vmmdP}{P501}
\lookupPut{R_Zh3CADAyXQpnVTz}{P502}
\lookupPut{R_1FQSYPHqUV6czsE}{P503}
\lookupPut{R_C9yAmQlgmCwlD33}{P504}
\lookupPut{R_1Nh3t5RH93w6hnZ}{P505}
\lookupPut{R_bQSXpPnY0BoFfnb}{P506}
\lookupPut{R_eY8UoNT1qNZBA5z}{P507}
\lookupPut{R_ZltABV8DO4sv5kZ}{P508}
\lookupPut{R_3CBnFzCj0T3LunY}{P509}
\lookupPut{R_3hfGUUrmwT0TsUK}{P510}
\lookupPut{R_1HqpWVJhn1KXx3f}{P511}
\lookupPut{R_2oFb6G4rTsGu6Vr}{P512}
\lookupPut{R_xG7x0O0JKY5Njwt}{P513}
\lookupPut{R_1Itm4JHlk5EcVtU}{P514}
\lookupPut{R_3HZz8PM9kTSqJ6l}{P515}
\lookupPut{R_1dd9t1VhdMxIpEH}{P516}
\lookupPut{R_3lSlo8kkjR53pS0}{P517}
\lookupPut{R_D03QcjHQF89BkD7}{P518}
\lookupPut{R_2tsC0zXifP0l6aR}{P519}
\lookupPut{R_vCbIR7P0fKcRsZz}{P520}
\lookupPut{R_2RR4MwdJApLo3eq}{P521}
\lookupPut{R_2EsLiW5WkbTOOoH}{P522}
\lookupPut{R_2AZI7Pz6IeC9Ig1}{P523}
\lookupPut{R_1d7grWka0tUzeLz}{P524}
\lookupPut{R_0x5pZHzhB2hEvUl}{P525}
\lookupPut{R_10C9CfcR2DABuYI}{P526}
\lookupPut{R_aUZaozmIx5mChCF}{P527}
\lookupPut{R_BtUfLnBV2ZsCTzH}{P528}
\lookupPut{R_3kfQjXS8yJog7uN}{P529}
\lookupPut{R_1I9Hf4cgnFtOttj}{P530}
\lookupPut{R_2tgmuEtLpdUmeDt}{P531}
\lookupPut{R_8jZjruF7JSWQyEV}{P532}
\lookupPut{R_1NhpvHlFOunAzIs}{P533}
\lookupPut{R_2PjJNjWJ7JJSj2e}{P534}
\lookupPut{R_2wEK7iouSP9ii6S}{P535}
\lookupPut{R_zeAklvOWtIwUD7z}{P536}
\lookupPut{R_TvYYjR6xRs82CWt}{P537}
\lookupPut{R_3EKbN4XMLRdhOPJ}{P538}
\lookupPut{R_3CNJDMVRDlaDUeF}{P539}
\lookupPut{R_djnYn3UndlAx0l3}{P540}
\lookupPut{R_3PHMr03sWpGJc6V}{P541}
\lookupPut{R_3HFqJ0eMtFKqSml}{P542}
\lookupPut{R_1qXvHDYa6663MYB}{P543}
\lookupPut{R_3qF1J5lGenS4LSV}{P544}
\lookupPut{R_1dnrWm47wl7UWhF}{P545}
\lookupPut{R_2VBFh65XflPD1k9}{P546}
\lookupPut{R_1DOWjFQHYpsaacG}{P547}
\lookupPut{R_3KDB8lQ1O4XSOgY}{P548}
\lookupPut{R_1gbGR1fHLOmuEHZ}{P549}
\lookupPut{R_2tsQTevU9GehGd5}{P550}
\lookupPut{R_3GqPVd9urP2FPAc}{P551}
\lookupPut{R_3EbfzmvK96BSYJI}{P552}
\lookupPut{R_2ZTxbVHtqY7pq71}{P553}
\lookupPut{R_3Ras30IhMXftfun}{P554}
\lookupPut{R_6PtHdi4udsXn9PX}{P555}
\lookupPut{R_OkgRGtotY66sBLX}{P556}
\lookupPut{R_AMrqGQ7Rc4gUFax}{P557}
\lookupPut{R_3PACVRYskz9rhsW}{P558}
\lookupPut{R_bNNgV8AOSoKIHMB}{P559}
\lookupPut{R_p5Y1rroZXBMMOHv}{P560}
\lookupPut{R_yJTEq6Te1BjRmgh}{P561}
\lookupPut{R_1N5rvPo072Pay7K}{P562}
\lookupPut{R_VOkExUuSzre79Kx}{P563}
\lookupPut{R_2qggkHKxll4vnUW}{P564}
\lookupPut{R_cUP7DcTQ4UqUiu5}{P565}
\lookupPut{R_0celH9oEiGnAkaB}{P566}
\lookupPut{R_1yP7oNZIui88rzb}{P567}
\lookupPut{R_PCb1vmXM8gPXhMR}{P568}
\lookupPut{R_1gcMXNZ2PJNvmh9}{P569}
\lookupPut{R_2rG6ZTZJicVoDgR}{P570}
\lookupPut{R_1MLgmRQGuAwBfIA}{P571}
\lookupPut{R_9mnvmUbGrQr47Lz}{P572}
\lookupPut{R_2Wv4y4xXTOnhGsc}{P573}
\lookupPut{R_OpUFK5jQArToB6p}{P574}
\lookupPut{R_Anx7w6jf34JV3vr}{P575}
\lookupPut{R_1itk2BsKI53Ied0}{P576}
\lookupPut{R_3MGKnTxdsKRzxWe}{P577}
\lookupPut{R_1CfPmmlU1Wiz6Vb}{P578}
\lookupPut{R_3HUEXT8us3Ow24o}{P579}
\lookupPut{R_1C1LsfTu7JrBT2y}{P580}
\lookupPut{R_bwJykqUw1JolJGp}{P581}
\lookupPut{R_31WpfXR6Br9gyqR}{P582}
\lookupPut{R_2CTk1AI0oGJSp7R}{P583}
\lookupPut{R_2dHjDshgz276hVn}{P584}
\lookupPut{R_ZBJL4gKzoj1f92p}{P585}
\lookupPut{R_2wNAIR7xxH4S4O6}{P586}
\lookupPut{R_2ANo3EuRJzyWGes}{P587}
\lookupPut{R_Rl5836USj8kHA9b}{P588}
\lookupPut{R_24AF5iBmzuFpeS6}{P589}
\lookupPut{R_1LJvUaJJceq0Hmh}{P590}
\lookupPut{R_2WI1ZXxBaFo7hgN}{P591}
\lookupPut{R_3RruZEU2ofciHO2}{P592}
\lookupPut{R_2RUQZMv2QoUGbdA}{P593}
\lookupPut{R_2rTMAeNF13wtOiz}{P594}
\lookupPut{R_3suETDPEnjdEaIF}{P595}
\lookupPut{R_22LhsPdaRlZlbmB}{P596}
\lookupPut{R_2teSTCglf0l7Dfg}{P597}
\lookupPut{R_stVhqrceEm9Oisx}{P598}
\lookupPut{R_3I3SPVGtMpCnRQq}{P599}
\lookupPut{R_oZfbVsTy2X9tDBn}{P600}
\lookupPut{R_3fJXgb2TQfqnCV3}{P601}
\lookupPut{R_OBYcRjLsj4qfmjn}{P602}
\lookupPut{R_1LZ0JToA3n6Cstg}{P603}
\lookupPut{R_2zBZBQTmUF7zCgM}{P604}
\lookupPut{R_265CSKGVMAK5k69}{P605}
\lookupPut{R_1f9gTgSJVj93ZtL}{P606}
\lookupPut{R_1P4MtlTAEWu3AOV}{P607}
\lookupPut{R_1dpkctpR4nkX4FB}{P608}
\lookupPut{R_1H6k5xVBSigRTqk}{P609}
\lookupPut{R_b7QBbRFVgPJLNG9}{P610}
\lookupPut{R_3CT4zha0L6MiTEu}{P611}
\lookupPut{R_2fqQXLolqkGceg8}{P612}
\lookupPut{R_Bzdi7g4YUGn4p45}{P613}
\lookupPut{R_1M3w0xqhVMBQniI}{P614}
\lookupPut{R_29tf6mNgwpyIbbw}{P615}
\lookupPut{R_3EEJnW4aldQE9Rc}{P616}
\lookupPut{R_2YRhr7ZndVNEfRZ}{P617}
\lookupPut{R_wXFz8W4saUbpvr3}{P618}
\lookupPut{R_31XI2rebAnejIqC}{P619}
\lookupPut{R_8Aq7wBTnJCxRpDP}{P620}
\lookupPut{R_2AGd1HKvERqpOmc}{P621}
\lookupPut{R_2cchrVAI7b1kYtq}{P622}
\lookupPut{R_2tDaYKYTdaB5kHj}{P623}
\lookupPut{R_sjy4SJRi9AoMYfv}{P624}
\lookupPut{R_1DI8MxIDOikKOoj}{P625}
\lookupPut{R_sRwqVdLeUGMTsch}{P626}
\lookupPut{R_3ivTpEsnqkL3bg9}{P627}
\lookupPut{R_2ygbzRXLwFhkcex}{P628}
\lookupPut{R_3FRJMuk8CVduA34}{P629}
\lookupPut{R_1Q9HWVFMUTZ8VKg}{P630}
\lookupPut{R_QgBoz895c2Z1FQJ}{P631}
\lookupPut{R_XGHiobDwzEXtWcp}{P632}
\lookupPut{R_ym9DdCldebQ9TNv}{P633}
\lookupPut{R_6Rm45Q9J4hpqZi1}{P634}
\lookupPut{R_2VdxwUcL9kFfEn4}{P635}
\lookupPut{R_AjO6ZXY7JoF969z}{P636}
\lookupPut{R_2To9VkwldrkuuvC}{P637}
\lookupPut{R_PM08lxtxOQ4tY2J}{P638}
\lookupPut{R_1riTvwf3vhSsE2s}{P639}
\lookupPut{R_32OPCUw1zw3cRZK}{P640}
\lookupPut{R_30iXEwFp9FvYy3o}{P641}
\lookupPut{R_2Eb32x980Tk1Bis}{P642}
\lookupPut{R_3NCX9VpQpluY3OF}{P643}
\lookupPut{R_29nt19EDs6laNRH}{P644}
\lookupPut{R_1pEjzYeozgisP63}{P645}
\lookupPut{R_xu3KJDkdlK1w49z}{P646}
\lookupPut{R_30jf8xiVF9Od4Cz}{P647}
\lookupPut{R_22ukI9pBecuZhm2}{P648}
\lookupPut{R_1ls21923Py8PC51}{P649}
\lookupPut{R_33DuzxVMDP8VQTX}{P650}
\lookupPut{R_3qPRDBGtfCWnCEo}{P651}
\lookupPut{R_1Kd9mUZsLGiXSLw}{P652}
\lookupPut{R_2YaJrj6qqXfEm0Y}{P653}
\lookupPut{R_1LAqSANA3wHNpPW}{P654}
\lookupPut{R_3wOcURxYquERoit}{P655}
\lookupPut{R_3isqFXM5NEOrlsv}{P656}
\lookupPut{R_3CHMNXUCoopMKov}{P657}
\lookupPut{R_25AQfv50oY9o8UX}{P658}
\lookupPut{R_3j7fQGfQfnUnlgd}{P659}
\lookupPut{R_3QQX0HzGgPEtz7o}{P660}
\lookupPut{R_3kHB9YjdWWE7V0d}{P661}
\lookupPut{R_11ZauEIdsVy91br}{P662}
\lookupPut{R_1PbLrOBFq83uor9}{P663}
\lookupPut{R_2Ce5xqGXzXTuRKl}{P664}
\lookupPut{R_3nOfIpD4vkMBygu}{P665}
\lookupPut{R_2S0ozbTYQMJp37a}{P666}
\lookupPut{R_3hfHJULgw1QhVE7}{P667}
\lookupPut{R_3ixlwGuIC1Ftkcc}{P668}
\lookupPut{R_1mruQc8Ny3qH6Wv}{P669}
\lookupPut{R_3D6DKjjxvjiVj2P}{P670}
\lookupPut{R_vc9rfLMi6ljfm25}{P671}
\lookupPut{R_1gUcbBH4CIjt2lE}{P672}
\lookupPut{R_UYozoyzobSGAyml}{P673}
\lookupPut{R_2wvn1VIOvuxufXW}{P674}
\lookupPut{R_1pAKVLN8WASgam0}{P675}
\lookupPut{R_3nAFJsJthnZ24Sk}{P676}
\lookupPut{R_1Dv6pjFbJ5UFWp6}{P677}
\lookupPut{R_10vmKitSIhXK10f}{P678}
\lookupPut{R_1CHGAHXk5Sh8DKa}{P679}
\lookupPut{R_vvIXB1U0UPTv0dP}{P680}
\lookupPut{R_1meDUhvlmb9zhrA}{P681}
\lookupPut{R_2EmAVDs6AGudmp0}{P682}
\lookupPut{R_eMbfltCT1IZzr0d}{P683}
\lookupPut{R_3kb6OpljnzUJIWA}{P684}
\lookupPut{R_3O1YsXOlNzDVumS}{P685}
\lookupPut{R_u9rxVmoI6DVGOZ3}{P686}
\lookupPut{R_22xAQxZn7ghoCJt}{P687}
\lookupPut{R_1dBt8B8QK76rTZm}{P688}
\lookupPut{R_3EA94mBy6gEvjIT}{P689}
\lookupPut{R_WpQLOp0JeJnERJD}{P690}
\lookupPut{R_sROzubcZSHDdNgB}{P691}
\lookupPut{R_279hEXAtMvGXOdD}{P692}
\lookupPut{R_1KvZCo8pKMAuWvu}{P693}
\lookupPut{R_1dKX9WvYqTNhWz4}{P694}
\lookupPut{R_3VMZl7YbsTQ9OcF}{P695}
\lookupPut{R_3nqMMmi9S24lDpF}{P696}
\lookupPut{R_1hAqSYdfMd6cdsS}{P697}
\lookupPut{R_2P7trrWAf7UeFk9}{P698}
\lookupPut{R_1N4xItp0k1iiZl3}{P699}
\lookupPut{R_2fWkB3lMgqUTb33}{P700}
\lookupPut{R_1EYxqerKjeH39Ht}{P701}
\lookupPut{R_3ZRFbS2BGS3HkBj}{P702}
\lookupPut{R_2c6kYGOjLOI5WI3}{P703}
\lookupPut{R_1P6j8GV4gXcOi2H}{P704}
\lookupPut{R_1rC7iDCXfliHR9b}{P705}
\lookupPut{R_3qfzVV2HgKXiQD0}{P706}
\lookupPut{R_3hEpkb0s53zVpfL}{P707}
\lookupPut{R_3iqjjrsS0ofKDhy}{P708}
\lookupPut{R_11i3KWzXnEkpQQr}{P709}
\lookupPut{R_2YlE8QdS0LgrDA2}{P710}
\lookupPut{R_eL02pn4ZqHcNWYF}{P711}
\lookupPut{R_2TpYINBFZU1T9TF}{P712}
\lookupPut{R_3Ok9gqVoixzAdWM}{P713}
\lookupPut{R_1IDUxwQL4YSLkOW}{P714}
\lookupPut{R_3EYXB8IoDxsgG3P}{P715}
\lookupPut{R_2Vl7cYM8yNA1LcE}{P716}
\lookupPut{R_6XaFRih7Wzii5d7}{P717}
\lookupPut{R_UMjz2EJabwubyzD}{P718}
\lookupPut{R_3CI7GaqTtyHL5cs}{P719}
\lookupPut{R_1Dv6qZ298lJDcXf}{P720}
\lookupPut{R_2AFgmQiaBzye2F8}{P721}
\lookupPut{R_3h3JwEjG6SdElaz}{P722}
\lookupPut{R_214W2Dimh6Xtk5i}{P723}
\lookupPut{R_3HoVC1bKzUkIY3z}{P724}
\lookupPut{R_2B4ZWxkbCrHkiyE}{P725}
\lookupPut{R_W3sV5veBz8G2gql}{P726}
\lookupPut{R_3NOEQoQZ7KCWF9M}{P727}
\lookupPut{R_9oeSQ7FL9HuuGad}{P728}
\lookupPut{R_1Tif7259koaHj1v}{P729}
\lookupPut{R_1H06SamhMBcgVsk}{P730}
\lookupPut{R_30eKhsUaNflxmQD}{P731}
\lookupPut{R_bPMpTuA5COBvgl3}{P732}
\lookupPut{R_3g1kpqO0k1V3kFp}{P733}
\lookupPut{R_9uhMIz5Vxgs4POF}{P734}
\lookupPut{R_3iWplonXLKSmLto}{P735}
\lookupPut{R_2rvgJ2FaoUXOCB9}{P736}
\lookupPut{R_2VK31rvFCJ87ejI}{P737}
\lookupPut{R_pgCImwaEmRHJlgl}{P738}
\lookupPut{R_24HOtELfrTovMpG}{P739}
\lookupPut{R_3nfxpAmi1KJVNgM}{P740}
\lookupPut{R_YVOfflZMPJ0GjOV}{P741}
\lookupPut{R_2b2EYDHYZmVIBF9}{P742}
\lookupPut{R_3MPzZn0jzGTgeO8}{P743}
\lookupPut{R_21pmD25bm66eQtc}{P744}
\lookupPut{R_xzxVQlHpLbQiMh3}{P745}
\lookupPut{R_11j9HNj4cOM1pfa}{P746}
\lookupPut{R_2CTxBxhmOcrUnzz}{P747}
\lookupPut{R_3gUdJSj5SeM68gr}{P748}
\lookupPut{R_2zYez9aW5eSco6F}{P749}
\lookupPut{R_DnNCgyRXmr2fTyh}{P750}
\lookupPut{R_2bN0UPLkr72H2bQ}{P751}
\lookupPut{R_pF7z7MUGyFrllYJ}{P752}
\lookupPut{R_AdFkpWIkoo73snn}{P753}
\lookupPut{R_xr83MWoPYpZ5gd3}{P754}
\lookupPut{R_1gjD9RHUhGdPlCF}{P755}
\lookupPut{R_2rO6yEIAqeliRj3}{P756}
\lookupPut{R_2rFDSXbAzg1epUt}{P757}
\lookupPut{R_Ry2Dr9E4Y4AfNL3}{P758}
\lookupPut{R_z8AQfywjMIDnTbP}{P759}
\lookupPut{R_3G0tgIfLYOM9qdV}{P760}
\lookupPut{R_1CrooKGLrpGP8i2}{P761}
\lookupPut{R_9tLid2Mh2VSXtn3}{P762}
\lookupPut{R_2ZWuDgnd1WeUZaq}{P763}
\lookupPut{R_3FUS1dupyrbU06D}{P764}
\lookupPut{R_1dayzupM8DdFI30}{P765}
\lookupPut{R_2QtW7EMSYQ4ibGL}{P766}
\lookupPut{R_22of40DU48EyRwL}{P767}
\lookupPut{R_2q7wfOyKCTzOWky}{P768}
\lookupPut{R_3qeULwkpRaXDGnG}{P769}
\lookupPut{R_2D5j0gKRMp6J7JY}{P770}
\lookupPut{R_27dKpru7hX2y0XL}{P771}
\lookupPut{R_yl8ThKcllgBUVdn}{P772}
\lookupPut{R_0ofCH705i6Xpxfz}{P773}
\lookupPut{R_3PZCztraJWdpA9H}{P774}
\lookupPut{R_2fB84x9S8vYsBpl}{P775}
\lookupPut{R_308EEvJcekurO9J}{P776}
\lookupPut{R_2rNmmPZI9A7bQMZ}{P777}
\lookupPut{R_p4s2OrAM4DGjz7b}{P778}
\lookupPut{R_C1aAI5PduYIEjtv}{P779}
\lookupPut{R_1JCjQayokmOtVXW}{P780}
\lookupPut{R_BPOp1yK5vgc1VLP}{P781}
\lookupPut{R_1rlhphEJQ8jgnzc}{P782}
\lookupPut{R_2f1Kke10utIIOy8}{P783}
\lookupPut{R_25Z9lDGd7W0dD7b}{P784}
\lookupPut{R_31a9vJePRUEHxaO}{P785}
\lookupPut{R_1eJhh93KSbcnHoj}{P786}
\lookupPut{R_00nUjvZT9bdnDTb}{P787}
\lookupPut{R_3EuIQBWzQYIOs1z}{P788}
\lookupPut{R_1mRFyEPsICaOYGh}{P789}
\lookupPut{R_1oAWUcUoESLHnF9}{P790}
\lookupPut{R_2EsqYDXocagFMUo}{P791}
\lookupPut{R_2vhQtvNHaX9dzhs}{P792}
\lookupPut{R_2ZP4GXQenGFgIJC}{P793}
\lookupPut{R_30hFG7OJVno5faG}{P794}
\lookupPut{R_325eCwTchvw6y7A}{P795}
\lookupPut{R_1Cl7f3O11lkx4DU}{P796}
\lookupPut{R_33evIFp16rMb7Ty}{P797}
\lookupPut{R_3CHjAFDxUSQ0heE}{P798}
\lookupPut{R_2aRzWxnhQBuYBLa}{P799}
\lookupPut{R_3h9qPAqOssMWcbJ}{P800}
\lookupPut{R_73656UZsWujNJ1n}{P801}
\lookupPut{R_2bJGOzf3cfzFmBa}{P802}
\lookupPut{R_3CBduVNtLmNJI94}{P803}
\lookupPut{R_1NbmBLtwVa6K1Ma}{P804}
\lookupPut{R_3HF7zegKAqZURZQ}{P805}
\lookupPut{R_2uDnIM9YywmBW4v}{P806}
\lookupPut{R_2EFSL5evu6lRlu5}{P807}
\lookupPut{R_3CCPlSLF7DRwxMS}{P808}
\lookupPut{R_3PQSY6t7mZtMMPv}{P809}
\lookupPut{R_3MfDhkAuJbgweYv}{P810}
\lookupPut{R_31nq8JSqBtKaUkh}{P811}
\lookupPut{R_3gUhfc8xUgQUD0Y}{P812}
\lookupPut{R_2AMbOOYs9nmeH1X}{P813}
\lookupPut{R_1CdLbIhO3FgeWqW}{P814}
\lookupPut{R_1fdZWNkMnIwK6vb}{P815}
\lookupPut{R_3OlFnpbYfdNeH9S}{P816}
\lookupPut{R_sO25ymu0VtLjXr3}{P817}
\lookupPut{R_2QJRXiVowNdo5bq}{P818}
\lookupPut{R_1jrEbMkTMU00RNq}{P819}
\lookupPut{R_10Bl1NUr8hk0wiX}{P820}
\lookupPut{R_2qBG1wCJcHR86K1}{P821}
\lookupPut{R_1oaMZOj8jJvNKZR}{P822}
\lookupPut{R_1IWd8SOlwM8sNAQ}{P823}
\lookupPut{R_2lgLgldc55XB453}{P824}
\lookupPut{R_3rNxF4zVzGuXVHU}{P825}
\lookupPut{R_vdBS1cD3Ok8MlHP}{P826}
\lookupPut{R_9F7rPJ5ceoT2Os1}{P827}
\lookupPut{R_1hBkA6wna6XAaJQ}{P828}
\lookupPut{R_1LYHv2CkDgADXXR}{P829}
\lookupPut{R_yqDOUYlkpCfZFWV}{P830}
\lookupPut{R_1CrP7DLZkETQ989}{P831}
\lookupPut{R_3MgPUxMfIOdkwkx}{P832}
\lookupPut{R_03AB2ycRjm2WuFX}{P833}
\lookupPut{R_p9ICYhP39q9KMxj}{P834}
\lookupPut{R_3nBiWCBhA2MlFnP}{P835}
\lookupPut{R_2SwVKX0OkfJwlWc}{P836}
\lookupPut{R_3QYT0GXf77MNduc}{P837}
\lookupPut{R_2xXB9ZgYNer5aFt}{P838}
\lookupPut{R_3s4v0Ih4O8myhWj}{P839}
\lookupPut{R_8CCfJWho4K4URVf}{P840}
\lookupPut{R_PvxSoqC0YSJfkHv}{P841}
\lookupPut{R_2Tpea7vT7HIUXJd}{P842}
\lookupPut{R_1HjnFH1gpmrJWyu}{P843}
\lookupPut{R_06AFofuZpPTYKIx}{P844}
\lookupPut{R_1hycnKfGlFjRB8E}{P845}
\lookupPut{R_1mUv5KSuazAjQrp}{P846}
\lookupPut{R_2aXmIkfEWO8AgQC}{P847}
\lookupPut{R_A1lJxujHc5K13DX}{P848}
\lookupPut{R_qQGKrAjwYPyw8LL}{P849}
\lookupPut{R_1LSDuxXsz3ItZXw}{P850}
\lookupPut{R_2ckF4Lb6YvnLAGl}{P851}
\lookupPut{R_2WCKxekSq7zgYrT}{P852}
\lookupPut{R_3fTbZ67wbiLkyL5}{P853}
\lookupPut{R_3qJvzd2BwIQULx1}{P854}
\lookupPut{R_2bTZb1MtpMn7USl}{P855}
\lookupPut{R_3PX3jlGQ7vv6mCM}{P856}
\lookupPut{R_2wvllC2KZio8vUV}{P857}
\lookupPut{R_3kHVGnvsJTXBXcR}{P858}
\lookupPut{R_e4YmPjYMzuydUwp}{P859}
\lookupPut{R_2PvzVDfPcpMmJxr}{P860}
\lookupPut{R_1lhEbvOpFRZFpIA}{P861}
\lookupPut{R_3g5SaT7rK8JdSve}{P862}
\lookupPut{R_27W0pCuewtJaCEY}{P863}
\lookupPut{R_3iHVUHGVMCdqPOf}{P864}
\lookupPut{R_2AZO0AA9C6aZvHQ}{P865}
\lookupPut{R_12SEcdADTOtbl1o}{P866}
\lookupPut{R_Uo0MGDk1CyXVmqB}{P867}
\lookupPut{R_9prBDqLI2xVhlUB}{P868}
\lookupPut{R_0ibDpwiEdtjvrxL}{P869}
\lookupPut{R_1daFOHN8ZqSoKFe}{P870}
\lookupPut{R_RlzreebAE4ieWyt}{P871}
\lookupPut{R_1jJa2eohV8tLhSC}{P872}
\lookupPut{R_3ewcxFO2dtFamUN}{P873}
\lookupPut{R_BrGiA2RqPnLhCY9}{P874}
\lookupPut{R_2y7QbtKpFUbqe5R}{P875}
\lookupPut{R_1PTUejux9q4aciL}{P876}
\lookupPut{R_acaZEVAatRpWg3n}{P877}
\lookupPut{R_2dWmL00PEiCDj7n}{P878}
\lookupPut{R_22rWPmwmShQjRqI}{P879}
\lookupPut{R_27lLGeBkmhUdBGZ}{P880}
\lookupPut{R_3PUog910Ut2BOgP}{P881}
\lookupPut{R_1Fm7FlagiFqN5a9}{P882}
\lookupPut{R_3DoPgbYBMtH8QG0}{P883}
\lookupPut{R_3GdJzjY9LwY90ri}{P884}
\lookupPut{R_2wSR3WGvZOQhtHB}{P885}
\lookupPut{R_1rrxiSVJVOySSfj}{P886}
\lookupPut{R_1CJbeOXrszX43zT}{P887}
\lookupPut{R_DxeiSyylib98vgB}{P888}
\lookupPut{R_C2AHSKDkYIJAbJL}{P889}
\lookupPut{R_1QGGzsOxCK255va}{P890}
\lookupPut{R_1Om0SQTG1ltL99a}{P891}
\lookupPut{R_1FKGfD04ju5u8fO}{P892}
\lookupPut{R_2THLgTy4dKgggh6}{P893}
\lookupPut{R_1rxKrSPFfJZFtMN}{P894}
\lookupPut{R_29mNGcorrGQqAGM}{P895}
\lookupPut{R_2VlOSW7nf9Hm2nT}{P896}
\lookupPut{R_2A0VrvzPiSRxOMX}{P897}
\lookupPut{R_2YfwW8XAwVdZmAx}{P898}
\lookupPut{R_1OV2K377879LWL8}{P899}
\lookupPut{R_1CHfiFfZDOrkWVq}{P900}
\lookupPut{R_1OUL5JwXWqySvEh}{P901}
\lookupPut{R_1LUJHwZiKmqmomG}{P902}
\lookupPut{R_2TvpPwm2uzNyQvQ}{P903}
\lookupPut{R_2eUMuYhNF80MUIo}{P904}
\lookupPut{R_2RRlZFlcJdaFR3F}{P905}
\lookupPut{R_1Iza3yCtjvUmQS6}{P906}
\lookupPut{R_2c7rcZZZTiJpKWB}{P907}
\lookupPut{R_25Eez8sloLGliJq}{P908}
\lookupPut{R_3DoxMUjrnZFCa96}{P909}
\lookupPut{R_uylkgEmw2EEuFMt}{P910}
\lookupPut{R_2zO8MVD1mg8mZcs}{P911}
\lookupPut{R_5w4csxELm8S4EmZ}{P912}
\lookupPut{R_2UW7fAsvCPfjecr}{P913}
\lookupPut{R_31Mdq2LEbt9ClE1}{P914}
\lookupPut{R_qCMMTsmqWMrNN05}{P915}
\lookupPut{R_3O1ZTQqIbTgEEbM}{P916}
\lookupPut{R_XNffiSnpZWQPq3D}{P917}
\lookupPut{R_1ojYhV8Q9elDGre}{P918}
\lookupPut{R_1jZdw49Ixa8u2ZD}{P919}
\lookupPut{R_3HFCoGE7YOmbjhT}{P920}
\lookupPut{R_1mfnyVFAzMiyxvj}{P921}
\lookupPut{R_2uEe4youONZBwAk}{P922}
\lookupPut{R_3dDDqBMmq1fl3iz}{P923}
\lookupPut{R_2XhxkyVRGok3PiB}{P924}
\lookupPut{R_Cmhs6uTjdEuxnLr}{P925}
\lookupPut{R_2CkBmbzn06g2JvQ}{P926}
\lookupPut{R_3J3qu9GzYM2KAaf}{P927}
\lookupPut{R_3stVxOaPuXcUgJP}{P928}
\lookupPut{R_2aP5gbH6HHk4pDU}{P929}
\lookupPut{R_Y4bK9fRAQNF03kt}{P930}
\lookupPut{R_1OPsqHkxnIfwxtG}{P931}
\lookupPut{R_1IRAcPopyNfYbpG}{P932}
\lookupPut{R_YXOo2vYnBi4zGxz}{P933}
\lookupPut{R_1dKpzADDEmGGrep}{P934}
\lookupPut{R_2eRnnSGsUWUznTj}{P935}
\lookupPut{R_vkpwL4X8TgyLMKB}{P936}
\lookupPut{R_1dcXHNN6ZbeURQy}{P937}
\lookupPut{R_125U78s7Qs21PMi}{P938}
\lookupPut{R_2S948KHQlApGBai}{P939}
\lookupPut{R_3dMPaDDcZNYexYy}{P940}
\lookupPut{R_2WD802oitAQCLSE}{P941}
\lookupPut{R_1rBO7bUPL86joek}{P942}
\lookupPut{R_2QgtTEeUw4U0zdu}{P943}
\lookupPut{R_2P5mTY1vgoCWLm2}{P944}
\lookupPut{R_3DvVjEBhum8lDsG}{P945}
\lookupPut{R_1Hq5rXTEcu3UGBD}{P946}
\lookupPut{R_3EnTrkN4cjb06S1}{P947}
\lookupPut{R_2znALRsVHIsAAxj}{P948}
\lookupPut{R_2ZE1ZhDaoMoRS3b}{P949}
\lookupPut{R_3m4QuhahdssTPpx}{P950}
\lookupPut{R_3dX1qJKCXmFwzCt}{P951}
\lookupPut{R_r1oTo1NyKMJoEGl}{P952}
\lookupPut{R_dajYFyh6llyjUyJ}{P953}
\lookupPut{R_2a8j5yT3nJXhquX}{P954}
\lookupPut{R_3isPdLN0ZH46ytE}{P955}
\lookupPut{R_3PzQP5QSlICIVtd}{P956}
\lookupPut{R_yl5sCr6JfzyDSbD}{P957}
\lookupPut{R_1Du814bCaPzaf53}{P958}
\lookupPut{R_WpVF1tErtbCFxqF}{P959}
\lookupPut{R_sjbhGX7JoEWMkqR}{P960}
\lookupPut{R_1Ow3iuowq6TEvx3}{P961}
\lookupPut{R_1IiSh180CAtWfQ4}{P962}
\lookupPut{R_2ZOlP7NrqBgKc4r}{P963}
\lookupPut{R_AtcPVzUEo7JCyK5}{P964}
\lookupPut{R_3JFyq5b9TgpaNon}{P965}
\lookupPut{R_1GWjw2XfeenFibj}{P966}
\lookupPut{R_sbaSPj0rplBFuEx}{P967}
\lookupPut{R_3dHa6AMNobCkre7}{P968}
\lookupPut{R_3j1FR6g7peYe0o2}{P969}
\lookupPut{R_3D54pnaA09wS4le}{P970}
\lookupPut{R_3QM3nE4RRPKuphL}{P971}
\lookupPut{R_XjGWRl15lL3Nmzn}{P972}
\lookupPut{R_XN6uTfZ8XrMgMZb}{P973}
\lookupPut{R_Dcs5AAgxCwSYX05}{P974}
\lookupPut{R_1gT15NbPosujoPF}{P975}
\lookupPut{R_vo67G4xGmCndzpf}{P976}
\lookupPut{R_spPHcXhsOy1Mjuh}{P977}
\lookupPut{R_215qJXDdeQvfWwE}{P978}
\lookupPut{R_2wQt0njZeExIB0F}{P979}
\lookupPut{R_2cAQTy06OWQ0twk}{P980}
\lookupPut{R_25NOPP7l7kmc7Zl}{P981}
\lookupPut{R_1hEw0gPKS9XHVHc}{P982}
\lookupPut{R_22Y2FYymUpOe1PA}{P983}
\lookupPut{R_eD8TNDaNxjWTnZ7}{P984}
\lookupPut{R_2cqNsESluNYmRHS}{P985}
\lookupPut{R_07nwQMFy55bAFzz}{P986}
\lookupPut{R_2COpBSBUW8g5sQt}{P987}
\lookupPut{R_1LtewUMRC81lrHC}{P988}
\lookupPut{R_sc9VqLt101o1E41}{P989}
\lookupPut{R_3ERmga8Qvu4YGdC}{P990}
\lookupPut{R_1GU4gcwUVWoh0kg}{P991}
\lookupPut{R_xAEeaa7dxfGjz45}{P992}
\lookupPut{R_7TVMPUENwN8LrUd}{P993}
\lookupPut{R_1E703c3ifvaeSOm}{P994}
\lookupPut{R_3Hv72SmHJGFU9Nz}{P995}
\lookupPut{R_11au5A1mvqoSp1f}{P996}
\lookupPut{R_1IQ6bkf28KENtGF}{P997}
\lookupPut{R_2TNv9thEkDTvur4}{P998}
\lookupPut{R_3CGUVw40PSbFYZ4}{P999}
\lookupPut{R_OMXiedx7N52TqLv}{P1000}
\lookupPut{R_32Q39KOhYUBFvj0}{P1001}
\lookupPut{R_QoBdiCebl8YO0uZ}{P1002}
\lookupPut{R_3M3tovcZxhRMBtF}{P1003}
\lookupPut{R_1g6COFm1UUnO98q}{P1004}
\lookupPut{R_DB7ayVrJll93dHX}{P1005}
\lookupPut{R_3PTbl1r7Q7pjgYS}{P1006}
\lookupPut{R_2VsY7Tr2NIQZMNG}{P1007}
\lookupPut{R_vlwS2SfhXvhMP73}{P1008}
\lookupPut{R_1d1fYmo9dQUtODx}{P1009}
\lookupPut{R_8evIyAQLpzWipnb}{P1010}
\lookupPut{R_278MmyB3x80zUol}{P1011}
\lookupPut{R_t0tbPiiZ4b2vs2Z}{P1012}
\lookupPut{R_3ER3V3aSCaZkbS7}{P1013}
\lookupPut{R_RfXwdogbLGztrUJ}{P1014}
\lookupPut{R_3qwYFY2BLU36w7Q}{P1015}
\lookupPut{R_25BdiNhowgFFQZI}{P1016}
\lookupPut{R_pTaEUWW7Uxid2GB}{P1017}
\lookupPut{R_2CZKhGtLbLOOz3K}{P1018}
\lookupPut{R_2vYMMoRPTuq7hD8}{P1019}
\lookupPut{R_OuNkRKRIS5g7jmF}{P1020}
\lookupPut{R_237icOE3srAbSmV}{P1021}
\lookupPut{R_22xL2NBKSRrlyqA}{P1022}
\lookupPut{R_2uIEjOVsQjWeFQe}{P1023}
\lookupPut{R_3KNrcROUuhkIrUE}{P1024}
\lookupPut{R_zfoLqmNA06rlfRn}{P1025}
\lookupPut{R_3HIg7WV8IBGXrTY}{P1026}
\lookupPut{R_DwKI2SEYW5TpPl7}{P1027}
\lookupPut{R_27BSf7htpAju5bm}{P1028}
\lookupPut{R_3luBXRbJkZKzqfT}{P1029}
\lookupPut{R_1GQhT9rEoQW1jow}{P1030}
\lookupPut{R_11jypE5Uu1ZgdLe}{P1031}
\lookupPut{R_1IGyjeZo7vOOMwG}{P1032}
\lookupPut{R_3OjGku31S2j5qIh}{P1033}
\lookupPut{R_232eLNU8osGoLh9}{P1034}
\lookupPut{R_2Ph6FXaGooj1yRF}{P1035}
\lookupPut{R_1MXIQNHEhW4DBIw}{P1036}
\lookupPut{R_1KlkhWK0C4CZtOH}{P1037}
\lookupPut{R_2zSDg1HkPCpBb1e}{P1038}
\lookupPut{R_bE4u95QdGvA6VGN}{P1039}
\lookupPut{R_1o1kgjika901VJ9}{P1040}
\lookupPut{R_3ho8UpPxEyL8Mq5}{P1041}
\lookupPut{R_OBAvpKnEIGe023f}{P1042}
\lookupPut{R_eaE0qC0VdjHAayt}{P1043}
\lookupPut{R_27Vhr4hTjwhmlnF}{P1044}
\lookupPut{R_1LSE9QiwnuxNphN}{P1045}
\lookupPut{R_1rvMKsakew7AE0K}{P1046}
\lookupPut{R_33wdKDr6FFzrAIV}{P1047}
\lookupPut{R_1DTfsPqdppaBXoB}{P1048}
\lookupPut{R_30izw8NJ9m0plFk}{P1049}
\lookupPut{R_dcA1qgi7sBDCowN}{P1050}
\lookupPut{R_33lL7GSZRZXmpiO}{P1051}
\lookupPut{R_3iDWN85DTkLJhWv}{P1052}
\lookupPut{R_3hGCF4qvkplcpaY}{P1053}
\lookupPut{R_24794aHozvAp59l}{P1054}
\lookupPut{R_2dLakyJtpq0ovgU}{P1055}
\lookupPut{R_3qInQEtj4h6UrWd}{P1056}
\lookupPut{R_2PqbnBg6MNzPj8s}{P1057}
\lookupPut{R_1i20KKdEZG7KLMh}{P1058}
\lookupPut{R_3HoALxxI9ppwRTY}{P1059}
\lookupPut{R_3POfjbvEkP9SAIE}{P1060}
\lookupPut{R_2QRGqWTlj85FnVZ}{P1061}
\lookupPut{R_1ISCza5V0SUHhz0}{P1062}
\lookupPut{R_3Mb8woQNQQFnDF2}{P1063}
\lookupPut{R_sMOfMHNWNyEmEVP}{P1064}
\lookupPut{R_1j6UA413Mhtskfi}{P1065}
\lookupPut{R_9AcIthtG4mDt26J}{P1066}
\lookupPut{R_85FbcOOvK1Hrpyp}{P1067}
\lookupPut{R_2s6AAoPAG1u0Dos}{P1068}
\lookupPut{R_3KNMZhUSPPeFtFQ}{P1069}
\lookupPut{R_R37fylNX4ZnzYdz}{P1070}
\lookupPut{R_w767RZQ6ysat4nn}{P1071}
\lookupPut{R_pDATl3Oa6FYxSPD}{P1072}
\lookupPut{R_1dnb63XxxO8BjxO}{P1073}
\lookupPut{R_Od6BiCbo1Op7OX7}{P1074}
\lookupPut{R_3qg1ct2CxJZYEer}{P1075}
\lookupPut{R_3KSD4QQOpJuKLFT}{P1076}
\lookupPut{R_1H2AWFyQ8rZrM0l}{P1077}
\lookupPut{R_3s4TSpZcHdTFATy}{P1078}
\lookupPut{R_3O6ctJ6mTQnngI9}{P1079}
\lookupPut{R_1QlRLgdtVrACOp6}{P1080}
\lookupPut{R_25RQ3QNofD16QAk}{P1081}
\lookupPut{R_AcFLsTD2Fe0NwVH}{P1082}
\lookupPut{R_3Jk1t04cgx7oUTu}{P1083}
\lookupPut{R_aaAjjJ9U6YL0vct}{P1084}
\lookupPut{R_Z90e2aI6pnKieDn}{P1085}
\lookupPut{R_1QxO9gb1AcnOUGd}{P1086}
\lookupPut{R_10VtCJEiSRXXTHZ}{P1087}
\lookupPut{R_2ckoU49UxJYp8db}{P1088}
\lookupPut{R_10U7CbliunQVx5e}{P1089}
\lookupPut{R_2f24DLJcqApEojq}{P1090}
\lookupPut{R_1mRedoti2uxOZGK}{P1091}
\lookupPut{R_2ztGhD6P1PqR1hs}{P1092}
\lookupPut{R_1F8tfWmdRoSDK3a}{P1093}
\lookupPut{R_3HjFjcDmvHb4EbX}{P1094}
\lookupPut{R_BPT9CLM5sUJIUHT}{P1095}
\lookupPut{R_Z1MFatR1amuYRXj}{P1096}
\lookupPut{R_eQ0bgOv0y1KCDfz}{P1097}
\lookupPut{R_0CAoCrYym8qY1Oh}{P1098}
\lookupPut{R_0lhyPUVyc2I4zmN}{P1099}
\lookupPut{R_2RV8khuYOMHKKAI}{P1100}
\lookupPut{R_3HjAuuTMZpn3aUs}{P1101}
\lookupPut{R_3Dba1Ft8BNecltg}{P1102}
\lookupPut{R_yF27pRRnYo7yZvr}{P1103}
\lookupPut{R_2akpGm0i1o1xDfV}{P1104}
\lookupPut{R_3m2eGxdUehv9F1w}{P1105}
\lookupPut{R_daSrbIbQ902Lt8B}{P1106}
\lookupPut{R_22J0iRoU6jcN5rk}{P1107}
\lookupPut{R_3g19rwnBRLFuNvP}{P1108}
\lookupPut{R_4UVKPDWnianYNTb}{P1109}
\lookupPut{R_ZxEamq6b0E7sxod}{P1110}
\lookupPut{R_9Z7pu5IM2YMF02Z}{P1111}
\lookupPut{R_1n9NSitdQcdlpwo}{P1112}
\lookupPut{R_1kFa0iWYgJnHXU5}{P1113}
\lookupPut{R_3MDqqENMnT0nlww}{P1114}
\lookupPut{R_3JsgTkafrR6RNBW}{P1115}
\lookupPut{R_3JqhRhaacnYZv6h}{P1116}
\lookupPut{R_1mqjd5iLTfct2e2}{P1117}
\lookupPut{R_3dDJJcGQIZWo4Pd}{P1118}
\lookupPut{R_1NF8WUXoZJ5JZMU}{P1119}
\lookupPut{R_2zo9xcxBg3NXdk0}{P1120}
\lookupPut{R_BrIlP2cBr8ngxA5}{P1121}
\lookupPut{R_yZQgmsRfiXjv9Tz}{P1122}
\lookupPut{R_2uPAGayQlnljsz8}{P1123}
\lookupPut{R_2PyVe6erzujwqhK}{P1124}
\lookupPut{R_25Sdt6WlC2xKuwn}{P1125}
\lookupPut{R_Av2V4qgyOBi82I1}{P1126}
\lookupPut{R_dg0qwkWIWkdqEEx}{P1127}
\lookupPut{R_1nW5blHGmrQaG0V}{P1128}
\lookupPut{R_3HSIFsxtpZutfWB}{P1129}
\lookupPut{R_1IbFKrn4rDifPDa}{P1130}
\lookupPut{R_x6rcs5dZfu0Xd6x}{P1131}
\lookupPut{R_3Mbtz7U6UNq2ae5}{P1132}
\lookupPut{R_1eXlrTVSBmuQ7Lu}{P1133}
\lookupPut{R_1FwvDXvIPeJ9rng}{P1134}
\lookupPut{R_2S5ZF8UDlrSU8hC}{P1135}
\lookupPut{R_25zJ5TgUKqcODi5}{P1136}
\lookupPut{R_3G6Q1zcjLRosqdP}{P1137}
\lookupPut{R_2xS53tZ0OgurcqL}{P1138}
\lookupPut{R_6ogDR4YGwQ1Dxzr}{P1139}
\lookupPut{R_3EQe6C8WJjKPFvv}{P1140}
\lookupPut{R_3F4G1gQRnV4Y4bi}{P1141}
\lookupPut{R_xyjGVJQnyWY3vr3}{P1142}
\lookupPut{R_29dm1ksA2UL5b5l}{P1143}
\lookupPut{R_216OZ4ukfvXLxop}{P1144}
\lookupPut{R_2VPbxV0kQKTYRBS}{P1145}
\lookupPut{R_eJ83LD2l42o4iY1}{P1146}
\lookupPut{R_27scbayPakhTjeD}{P1147}
\lookupPut{R_PXnHpM4Y910fVg5}{P1148}
\lookupPut{R_XmONWtdQM5MdXzj}{P1149}
\lookupPut{R_2uqxKlvJLuxdQWe}{P1150}
\lookupPut{R_3oYZg03kinfXGbV}{P1151}
\lookupPut{R_38WbMe4aPfUNU2Z}{P1152}
\lookupPut{R_123ATmAv8zG1zaM}{P1153}
\lookupPut{R_1k0Oxced3Nuobgo}{P1154}
\lookupPut{R_3kzDZnAK1NA1oSk}{P1155}
\lookupPut{R_3gO4i4OvQ8nAoIp}{P1156}
\lookupPut{R_2Slsk34Di5tizNg}{P1157}
\lookupPut{R_3qIlD9GA1gkkTpv}{P1158}
\lookupPut{R_1H1VHPvqx9mFIqo}{P1159}
\lookupPut{R_yE1PugW7KWHidGN}{P1160}
\lookupPut{R_31H6S5VxBfooa14}{P1161}
\lookupPut{R_O7FevUmoaWD1PNv}{P1162}
\lookupPut{R_1gvVWO1a5zbHjKE}{P1163}
\lookupPut{R_3PFXRGAKnRO5X5i}{P1164}
\lookupPut{R_2uxo27Zu61Lmwnd}{P1165}
\lookupPut{R_1K259nev257f9PQ}{P1166}
\lookupPut{R_3oN6yjVE4NJ4MG1}{P1167}
\lookupPut{R_2xIOuZAdDWuGodJ}{P1168}
\lookupPut{R_2wM2trNgmLd0u7B}{P1169}
\lookupPut{R_zUNu6MmpxOVa9iN}{P1170}
\lookupPut{R_71UXNx084wrtqtX}{P1171}
\lookupPut{R_A7nDCT5URu76tpL}{P1172}
\lookupPut{R_1irJ2NoRN14onFb}{P1173}
\lookupPut{R_xsYCYMjGFuT2g9z}{P1174}
\lookupPut{R_00uLDyyfjsXnztD}{P1175}
\lookupPut{R_W1Wi9I6tkE6K9R7}{P1176}
\lookupPut{R_cLPUGqbsDdCvNlf}{P1177}
\lookupPut{R_2P1F8fAnrfPr4va}{P1178}
\lookupPut{R_pc56WeJlyQCRBjb}{P1179}
\lookupPut{R_1SRUcxQ4WoNwn6h}{P1180}
\lookupPut{R_pmauUrmnJZjJUQh}{P1181}
\lookupPut{R_0vqU2zhDJ3xuxuF}{P1182}
\lookupPut{R_2ydNFrgPm6ErV4B}{P1183}
\lookupPut{R_vP3t4EORrS7pwnT}{P1184}
\lookupPut{R_3Li0IXHM6bq05hw}{P1185}
\lookupPut{R_1jx0ynuEnwSBQ6N}{P1186}
\lookupPut{R_p4UBAO9QATka3dL}{P1187}
\lookupPut{R_1d9V7P68DE1EW6S}{P1188}
\lookupPut{R_3QXnbOChER8ptUI}{P1189}
\lookupPut{R_VX5SfXugYSk8EaB}{P1190}
\lookupPut{R_yD3CFKO2Ig21MHL}{P1191}
\lookupPut{R_2ZTwjFqZ1phuRiq}{P1192}
\lookupPut{R_3sztjdKmtQuvYjX}{P1193}
\lookupPut{R_vDI2a6huvHjrgHf}{P1194}
\lookupPut{R_Q6Vhg6dIKaGZ0el}{P1195}
\lookupPut{R_1CjUKHHPaZUS8SW}{P1196}
\lookupPut{R_3QMqxgj94rvkx7O}{P1197}
\lookupPut{R_zSHmXjVCcghSU4V}{P1198}
\lookupPut{R_Zy0TSyIG8K2irWF}{P1199}
\lookupPut{R_308hjlT84H4ZYoN}{P1200}
\lookupPut{R_1Eh3AhviiirxIIc}{P1201}
\lookupPut{R_24rer2jiUHhbo7I}{P1202}
\lookupPut{R_1iaOAgsLf8pTvX2}{P1203}
\lookupPut{R_3O8ksVDFsvqsm8w}{P1204}
\lookupPut{R_SE877JOEPuo8EmZ}{P1205}
\lookupPut{R_dg97eFqGI4QenQJ}{P1206}
\lookupPut{R_1OZDhDZ6AozybVF}{P1207}
\lookupPut{R_3HFDYqfP0n7pZSB}{P1208}
\lookupPut{R_1EchiP3COAVjAWQ}{P1209}
\lookupPut{R_DMMNmzfioVDuJqh}{P1210}
\lookupPut{R_sAlzID58RV5vOxP}{P1211}
\lookupPut{R_26alTYnj7XMD1dU}{P1212}
\lookupPut{R_3g7KK0mCSrsRnCg}{P1213}
\lookupPut{R_2SxxP74N4g2AgMt}{P1214}
\lookupPut{R_1Q9KkDAJLmuGl0O}{P1215}
\lookupPut{R_Rsqi57297Zhs3n3}{P1216}
\lookupPut{R_3ew42vwSIh8zkHo}{P1217}
\lookupPut{R_1IhgJenn0zsd2il}{P1218}
\lookupPut{R_2YD2vzL2ohBOZ4P}{P1219}
\lookupPut{R_3CPi3XvrpfTqcYA}{P1220}
\lookupPut{R_1Ol4qvvOLzyslMH}{P1221}
\lookupPut{R_3EYC7CTfe2Fz2BK}{P1222}
\lookupPut{R_ZvIPh3BviqkWdJD}{P1223}
\lookupPut{R_1rwgesezsRPpCOL}{P1224}
\lookupPut{R_2VBJ7vt7ZngF7xK}{P1225}
\lookupPut{R_T0cKOyvmcrzdrrj}{P1226}
\lookupPut{R_vDNwBlb0PaIdRbr}{P1227}
\lookupPut{R_2Cm0ZJ6pobg03lX}{P1228}
\lookupPut{R_5haHLV3DPGzKqpX}{P1229}
\lookupPut{R_3e9r8HwZ53nS5z1}{P1230}
\lookupPut{R_3lYsnuKMwIePh5a}{P1231}
\lookupPut{R_vqmRBkp8RJYSIcF}{P1232}
\lookupPut{R_1i2U3Oa5z0qd3D3}{P1233}
\lookupPut{R_3EYzeYnE7FfBTZg}{P1234}
\lookupPut{R_1mxefQYMgrb1Gjk}{P1235}
\lookupPut{R_3rYv78QDkbNsAiw}{P1236}
\lookupPut{R_Yb5tCHeyTfTEMYp}{P1237}
\lookupPut{R_1P8jQb3srOJyMsN}{P1238}
\lookupPut{R_22P17f4CFJrl357}{P1239}
\lookupPut{R_2VeIm4ho5bvrsis}{P1240}
\lookupPut{R_3PjXIewmbtzOxAu}{P1241}
\lookupPut{R_3s0aBWbysqTPwEg}{P1242}
\lookupPut{R_1dHxoAZI3wWGowQ}{P1243}
\lookupPut{R_1K7PPERvh7lb2Xh}{P1244}
\lookupPut{R_3OqxUEbpl72qeVK}{P1245}
\lookupPut{R_2yab0yIXYvctOMW}{P1246}
\lookupPut{R_3qD5koRJkcNDyH6}{P1247}
\lookupPut{R_10Bgt4uqLZTpKS2}{P1248}
\lookupPut{R_1oaO7FOSOtbkg0A}{P1249}
\lookupPut{R_2rYrCJb1dO9GDCv}{P1250}
\lookupPut{R_20ZIQoVrjk54eFt}{P1251}
\lookupPut{R_9yLXk2wX0MOThGp}{P1252}
\lookupPut{R_1Eg2jXOcPfEhXLl}{P1253}
\lookupPut{R_PTe8vfVjbZ7shB7}{P1254}
\lookupPut{R_3ktZ9MPIJvkJnju}{P1255}
\lookupPut{R_PG1pPcLqdH0uwZX}{P1256}
\lookupPut{R_3kkbxbuLur0evUN}{P1257}
\lookupPut{R_BQEMf9RBECk4yJz}{P1258}
\lookupPut{R_sLKskQPFiHwpg9H}{P1259}
\lookupPut{R_3irieQeNpjMJsT5}{P1260}
\lookupPut{R_AguPFuHfef4kwdb}{P1261}
\lookupPut{R_11jtLfSDYeRftaC}{P1262}
\lookupPut{R_1o23gvJUzZhuRnW}{P1263}
\lookupPut{R_PwioOeZ7vwHWYKZ}{P1264}
\lookupPut{R_3RscSpCHH7ci6Cl}{P1265}
\lookupPut{R_pmHG0JZ9Su8UE3n}{P1266}
\lookupPut{R_1LnzXFUNwgWkXWu}{P1267}
\lookupPut{R_ezFRpAsq9GLJzGN}{P1268}
\lookupPut{R_1f6NoOYKjXY9Nm1}{P1269}
\lookupPut{R_2YVBfNzPWagHmgv}{P1270}
\lookupPut{R_2y7pW492voDuy9k}{P1271}
\lookupPut{R_cuLmdqDEaNexMD7}{P1272}
\lookupPut{R_BKPkODZ1FAQ6rE5}{P1273}
\lookupPut{R_1rCbqFwFtUiSYNA}{P1274}
\lookupPut{R_YW8K14lf2MrgDDP}{P1275}
\lookupPut{R_1P5UOJ6id6W61sc}{P1276}
\lookupPut{R_232kf2XMtgFVNVS}{P1277}
\lookupPut{R_2c2B8gdDkKilYPc}{P1278}
\lookupPut{R_2VaaXdILYcIUEAx}{P1279}
\lookupPut{R_3kgEaEWOwjNxZg5}{P1280}
\lookupPut{R_2s0awqEYTZM7wQL}{P1281}
\lookupPut{R_25ZXcPJpMZvJNhL}{P1282}
\lookupPut{R_28T0txiwRQwsKpC}{P1283}
\lookupPut{R_2Cp5u5KjPJTWkMZ}{P1284}
\lookupPut{R_28zwrmS68vVCYZX}{P1285}
\lookupPut{R_2bPf7ozaUvsWv41}{P1286}
\lookupPut{R_1QZWNDMeBPuGL9n}{P1287}
\lookupPut{R_1hy8XuJRwe5tPXy}{P1288}
\lookupPut{R_sRu5g2TlKbcQLVD}{P1289}
\lookupPut{R_p6hiac6AwwmNkGt}{P1290}
\lookupPut{R_WfD6zaKupTmEq2d}{P1291}
\lookupPut{R_2q93BeGDTftj4So}{P1292}
\lookupPut{R_2dBenQkCdhtFFV4}{P1293}
\lookupPut{R_1ihbQfyJNTZIU5r}{P1294}
\lookupPut{R_1rNTkHVKd2Qoykd}{P1295}
\lookupPut{R_2OTLXjuksCOQOUr}{P1296}
\lookupPut{R_wO8hPFy88l0S7jb}{P1297}
\lookupPut{R_AgOmVN8nRkHR3cR}{P1298}
\lookupPut{R_p3tec5dAHQcXWmd}{P1299}
\lookupPut{R_2f2TcE01QVuNBfU}{P1300}
\lookupPut{R_2qCB4D8GEiWRRSr}{P1301}
\lookupPut{R_1OO9ne1JO7bzDkA}{P1302}
\lookupPut{R_2Ej9QwC4drqTtM8}{P1303}
\lookupPut{R_XvT3nypUD7SpBqp}{P1304}
\lookupPut{R_2CwCpPrITUNA2wX}{P1305}
\lookupPut{R_z7og8ya636z4CnT}{P1306}
\lookupPut{R_3HuxNkhagsvwqco}{P1307}
\lookupPut{R_2VeDNTgNcJEDrCs}{P1308}
\lookupPut{R_p48npNWz6Q3MvMB}{P1309}
\lookupPut{R_3ijXkf2ZsNDsSd7}{P1310}
\lookupPut{R_tWIux8zQH0y0Ns5}{P1311}
\lookupPut{R_SKb4WW7ELTJlmp3}{P1312}
\lookupPut{R_2V4w3c89vCZG4uB}{P1313}
\lookupPut{R_31t4wZcQkDPP4K9}{P1314}
\lookupPut{R_21AqtrOFvXBtOyR}{P1315}
\lookupPut{R_PZmoY1jxkoXJB8B}{P1316}
\lookupPut{R_2f1jpqiSJ1ZKUIL}{P1317}
\lookupPut{R_1rP3pIIs06EtOMF}{P1318}
\lookupPut{R_1nUf920iu7yE44X}{P1319}
\lookupPut{R_doqPhmABnWFcuAh}{P1320}
\lookupPut{R_qz5Ylui0T2dMkX7}{P1321}
\lookupPut{R_24GXPznQcG0ycvM}{P1322}
\lookupPut{R_24cJCA3wKQkvSJo}{P1323}
\lookupPut{R_3s0PYik3LkWGEb6}{P1324}
\lookupPut{R_3p8QAlp1lhpPXE3}{P1325}
\lookupPut{R_Ox9SGpyL2UWSf9n}{P1326}
\lookupPut{R_1hGvqlVhNsPLnob}{P1327}
\lookupPut{R_2bZUiWpv4Sq2R8r}{P1328}
\lookupPut{R_6xISrKB429R9SG5}{P1329}
\lookupPut{R_2WGR9ZsdxumegbY}{P1330}
\lookupPut{R_2rx2GerDxaGhG33}{P1331}
\lookupPut{R_sZOqmUSyXNklMuR}{P1332}
\lookupPut{R_SPGXlXcaTkMMcet}{P1333}
\lookupPut{R_sBwzwhxT1S13bX3}{P1334}
\lookupPut{R_3DhFxWmt8nP1MNb}{P1335}
\lookupPut{R_RLCCM6AquxKYzPH}{P1336}
\lookupPut{R_2R8QL40WAHVBGHS}{P1337}
\lookupPut{R_3FVuK8u8mltYdll}{P1338}
\lookupPut{R_TvdFf0hXPAUQhiN}{P1339}
\lookupPut{R_u1vCd5L2MVzvvAB}{P1340}
\lookupPut{R_1jHQkkShMUoeADg}{P1341}
\lookupPut{R_1Nb6JhWgFuV6cSI}{P1342}
\lookupPut{R_2Cm0KQeRGuoW8Xv}{P1343}
\lookupPut{R_2VpwyhDCqMqXhTD}{P1344}
\lookupPut{R_2t5RnMyc7nNuowv}{P1345}
\lookupPut{R_12DV3RkiL3yOrHm}{P1346}
\lookupPut{R_29fSCXZEyfvDJ56}{P1347}
\lookupPut{R_bOh1lOyXVdI0TvP}{P1348}
\lookupPut{R_1mrVP9EO7fHPJ52}{P1349}
\lookupPut{R_3qQGP8Mmo4awoNY}{P1350}
\lookupPut{R_2DTuj0F3FYdRaih}{P1351}
\lookupPut{R_3MrPRL5VmZjiQPu}{P1352}
\lookupPut{R_1o88RXWarZoFqtT}{P1353}
\lookupPut{R_x4w7HqaG9qddI5z}{P1354}
\lookupPut{R_1E6y5r069quaTeJ}{P1355}
\lookupPut{R_2CDniMkpmXX1A5F}{P1356}
\lookupPut{R_3qIhyK2HV1DPdVV}{P1357}
\lookupPut{R_3rIApURAVnu1z8E}{P1358}
\lookupPut{R_d41aLgqUKselzb3}{P1359}
\lookupPut{R_RXMtU3UsO0w4xLH}{P1360}
\lookupPut{R_9uFmEFZ1foENHPP}{P1361}
\lookupPut{R_2uX320lu95Og3oQ}{P1362}
\lookupPut{R_ernTUoYLS39CnBv}{P1363}
\lookupPut{R_3lQR3xKIzgcyPE9}{P1364}
\lookupPut{R_OPPQTwVr2oAfj1v}{P1365}
\lookupPut{R_27OuMxGU3IOrqDU}{P1366}
\lookupPut{R_269nIeTeWhTVxSj}{P1367}
\lookupPut{R_3NE3qPGMXc6I9MX}{P1368}
\lookupPut{R_3NCiPFla3egA1KT}{P1369}

%% file: figures/timeline.tex
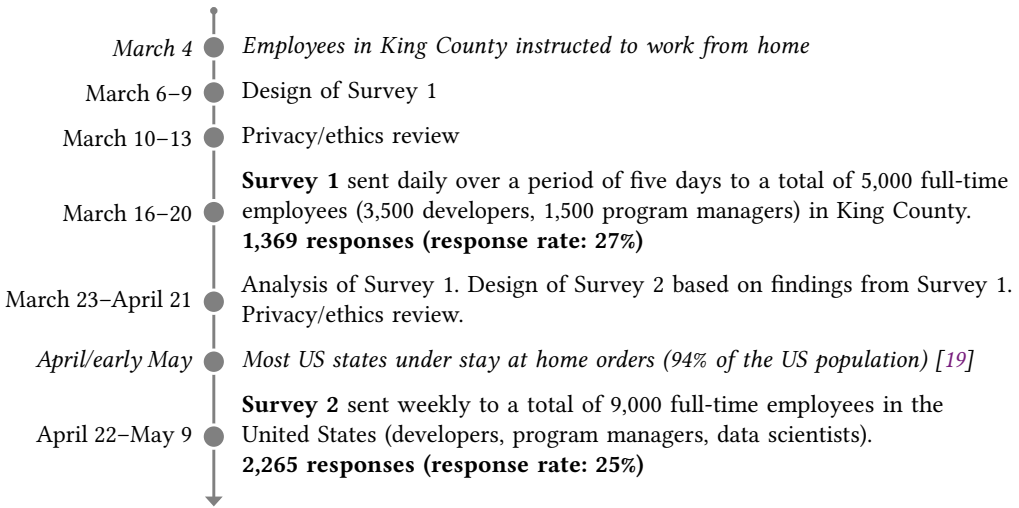
\begin{figure}[!htb]
\centering

\resizebox{\linewidth}{!}{%
\begin{tikzpicture}[
  node distance = 1mm and 3mm,
  start chain = A going below,
  dot/.style = {circle, draw=white, very thick, fill=gray, minimum size=3mm},
  box/.style = {rectangle, text width=110mm, inner xsep=4mm, inner ysep=1mm,
                on chain},
  ]
\begin{scope}[every node/.append style={box}]
\node { \emph{Employees in King County instructed to work from home}} ; 
\node { Design of Survey 1  } ;  %
\node { Privacy/ethics review } ; 
\node { \textbf{Survey 1} sent daily over a period of five days to a total of 5,000 full-time employees (3,500 developers, 1,500 program managers) in King County. \newline \textbf{\surveyoneresponses responses (response rate: 27\%)} } ; 
\node { Analysis of Survey 1. Design of Survey 2 based on findings from Survey~1. Privacy/ethics review. } ;
\node { \emph{Most US states under stay at home orders (94\% of the US population)~\cite{stay-at-home-orders}} } ;
\node { \textbf{Survey 2} sent weekly to a total of 9,000 full-time employees in the United States (developers, program managers, data scientists). \newline \textbf{\surveytworesponses responses (response rate: 25\%)}} ;
\end{scope}
\draw[very thick, gray, {Circle[length=3pt]}-{Triangle[length=4pt)]},
      shorten <=-3mm, shorten >=-3mm]   %
      (A-1.north west) -- (A-7.south west);
\foreach \i [ count=\j] in {\emph{March 4},March 6--9,March 10--13,March 16--20,March 23--April 21,\emph{April/early May},April 22--May 9}
    \node[dot,label=left:\i] at (A-\j.west) {};
\end{tikzpicture}
}%
\caption{Timeline for our multi-survey study}
\label{fig:study-timeline}
\end{figure}

%% file: tables/productivity.tex
\begin{table}[t]\centering

\newlength{\myboxheight}
\settoheight{\myboxheight}{1234567890\%}

\def\mybarchart#1{
\resizebox {#1} {\myboxheight} {%
\begin{tikzpicture}[]
\definecolor{clr1}{RGB}{99,99,99}
\definecolor{clr2}{RGB}{240,240,240}
\begin{axis}[
      axis background/.style={fill=gray!10, draw=gray!50},
      axis line style={draw=none},
      tick style={draw=none},
      ytick=\empty,
      xtick=\empty,
      ymin=0, ymax=1, %
      xmin=0, xmax=1]
\addplot [
      ybar interval=.5,
      fill=black,
      draw=none,
]
	coordinates {(1,1) (1,1)}; 
\addplot [
      ybar interval=.5,
      fill=black,
      draw=none,
]
	coordinates {(1,1) (0,1)}; 
\end{axis}%
\end{tikzpicture}%
}%
}

\caption{Changes in self-reported productivity based on the responses to the question \emph{``Compared to working in office, how has your productivity changed?''} (Q13, Q13*) The responses to Survey 2 are further broken down by week: April 22--25 (W1), April 26--May 2 (W2), and May 3--9 (W3).}
\label{tab:productivity}

\begin{tabular}{rp{2cm}p{2cm}rrr}
\toprule
& \multicolumn{1}{l}{Survey 1} & \multicolumn{1}{l}{Survey 2} & W1 & W2 & W3 \\
\cmidrule(r){1-1}\cmidrule(lr){2-2}\cmidrule(l){3-6}
Significantly more productive & \mybarchart{8pt} 8\% & \mybarchart{11pt} 11\% & 13\% & 10\% & 10\% \\ 
More productive & \mybarchart{22pt} 22\% & \mybarchart{26pt} 26\% & 23\% & 28\% & 26\% \\ 
\midrule
About the same & \mybarchart{32pt} 32\% & \mybarchart{32pt} 32\% & 31\% & 30\% & 34\% \\ 
\midrule
Less productive & \mybarchart{32pt} 32\% & \mybarchart{26pt} 26\% & 26\% & 26\% & 24\% \\ 
Significantly less productive & \mybarchart{6pt} 6\% & \mybarchart{6pt} 6\% & 7\% & 6\% & 6\% \\ 
\bottomrule
\end{tabular}

\end{table}

%% file: tables/mybarcharts.tex
\newlength{\myDistLength}
\newlength{\myDistHeight}

\newlength{\myDistHeightX}
\settoheight{\myDistHeightX}{Distribution}  

\definecolor{clgclr1}{RGB}{240,240,240}
\definecolor{clgclr2}{RGB}{189,189,189}
\definecolor{clgclr3}{RGB}{99,99,99}

\definecolor{benclr1}{RGB}{247,247,247}
\definecolor{benclr2}{RGB}{204,204,204}
\definecolor{benclr3}{RGB}{150,150,150}
\definecolor{benclr4}{RGB}{82,82,82}

\def\mybarchartbox#1{
\resizebox {\myDistHeightX} {\myDistHeightX} {%
\begin{tikzpicture}[]
\begin{axis}[
      axis background/.style={fill=none, draw=none}, %
      axis line style={draw=none},
      tick style={draw=none},
      ytick=\empty,
      xtick=\empty,
      ymin=0, ymax=1, %
      xmin=0, xmax=1]
\addplot [
      ybar interval=.5,
      fill=#1,
      draw=none,
]
	coordinates {(1,1) (1,1)}; 
\addplot [
      ybar interval=.5,
      fill=#1,
      draw=none,
]
	coordinates {(1,1) (0,1)}; 
\end{axis}%
\end{tikzpicture}%
}%
}

\def\mybarcharttwoboxes#1#2#3#4#5{%
\resizebox {2\myDistHeightX} {\myDistHeightX} {%
\begin{tikzpicture}[]
\definecolor{clr1}{RGB}{240,240,240}
\definecolor{clr2}{RGB}{189,189,189}
\definecolor{clr3}{RGB}{99,99,99}
\begin{axis}[
      axis background/.style={fill=none, draw=none}, %
      axis line style={draw=none},
      tick style={draw=none},
      ytick=\empty,
      xtick=\empty,
      ymin=0, ymax=1, %
      xmin=0, xmax=1]
\addplot [
      ybar interval=.5,
      fill=#5,
      draw=none,
]
	coordinates {(#1+#2+#3,1) (#1+#2,1)}; 
\addplot [
      ybar interval=.5,
      fill=#4,
      draw=none,
]
	coordinates {(#1+#2,1) (#1,1)}; 
\end{axis}%
\end{tikzpicture}%
}%
}

\def\mybarchartthreeboxes#1#2#3#4#5#6{%
\resizebox {3\myDistHeightX} {\myDistHeightX} {%
\begin{tikzpicture}[]
\definecolor{clr1}{RGB}{240,240,240}
\definecolor{clr2}{RGB}{189,189,189}
\definecolor{clr3}{RGB}{99,99,99}
\begin{axis}[
      axis background/.style={fill=none, draw=none}, %
      axis line style={draw=none},
      tick style={draw=none},
      ytick=\empty,
      xtick=\empty,
      ymin=0, ymax=1, %
      xmin=0, xmax=1]
\addplot [
      ybar interval=.5,
      fill=#6,
      draw=none,
]
	coordinates {(#1+#2+#3,1) (#1+#2,1)}; 
\addplot [
      ybar interval=.5,
      fill=#5,
      draw=none,
]
	coordinates {(#1+#2,1) (#1,1)}; 
\addplot [
      ybar interval=.5,
      fill=#4,
      draw=none,
]
	coordinates {(#1,1) (0,1)}; 
\end{axis}%
\end{tikzpicture}%
}%
}

%% file: tables/benefits.tex
\begin{table}\centering

\settowidth{\myDistLength}{Distribution}  

\def\mybarchart#1#2#3#4{
\resizebox {\the\myDistLength} {7.5pt} {%
\begin{tikzpicture}[]
\definecolor{clr1}{RGB}{247,247,247}
\definecolor{clr2}{RGB}{204,204,204}
\definecolor{clr3}{RGB}{150,150,150}
\definecolor{clr4}{RGB}{82,82,82}
\begin{axis}[
      axis background/.style={fill=none, draw=none}, %
      axis line style={draw=none},
      tick style={draw=none},
      ytick=\empty,
      xtick=\empty,
      ymin=0, ymax=1, %
      xmin=0, xmax=1]
\addplot [
      ybar interval=.5,
      fill=clr4,
      draw=none,
]
	coordinates {(#1+#2+#3+#4,1) (#1+#2+#3,1)}; 
\addplot [
      ybar interval=.5,
      fill=clr3,
      draw=none,
]
	coordinates {(#1+#2+#3,1) (#1+#2,1)}; 
\addplot [
      ybar interval=.5,
      fill=clr2,
      draw=none,
]
	coordinates {(#1+#2,1) (#1,1)}; 
\addplot [
      ybar interval=.5,
      fill=clr1,
      draw=none,
]
	coordinates {(#1,1) (0,1)}; 
\end{axis}%
\end{tikzpicture}%
}%
}

\newcommand{\oldvalue}[1]{\BEGINADDED}\newcommand{\oldvaluep}[1]{}

\caption{Benefits in Survey 2. The column \emph{Distribution} refers to the distribution of responses that (from left to right) do not experience a benefit (light gray\protect\mybarchartbox{benclr1}), experience a benefit and consider the benefit as unimportant (gray\protect\mybarchartbox{benclr2}), important (dark gray\protect\mybarchartbox{benclr3}) or very important (darker gray\protect\mybarchartbox{benclr4}). The column \emph{Prevalence} indicates the percentage of respondents who experienced the benefit ( \protect\mybarchartthreeboxes{0.34}{0.33}{0.33}{benclr2}{benclr3}{benclr4} ) while the column \emph{Importance} describes the percentage of participants who indicated this benefit to be important or very important ( percentage of \protect\mybarcharttwoboxes{0}{0.5}{0.5}{benclr3}{benclr4} with respect to \protect\mybarchartthreeboxes{0.34}{0.33}{0.33}{benclr2}{benclr3}{benclr4} ). The column \BEGINADDED\emph{Productivity Unchanged or Increased} reports the percentage of respondents who reported that productivity stayed ``about the same'' or increased (``more productive'', ``significantly more productive'') for respondents who \emph{did not} experience the benefit vs. respondents who \emph{did} experience the benefit; \ENDADDED statistically significant differences ($p<.01$, with Benjamini-Hochberg correction~\cite{benjamini1995controlling}) are indicated with an asterisk (*). The benefits are sorted and numbered in descending order by column \emph{Prevalence}.}
\label{tab:benefits}

\resizebox{\textwidth}{!}{%
\begin{tabular}{@{}rllcccccl@{}}
\toprule
& & & & & \multicolumn{4}{c}{\BEGINADDED Productivity Unchanged or Increased} \\
\cmidrule{6-9}
& Benefit & Distribution & Prevalence & Importance & \BEGINADDED \stackanchor{Benefit \textbf{not}}{experienced} & \BEGINADDED \stackanchor{Benefit}{experienced} & \BEGINADDED \stackanchor{Delta}{\small(pct points)} & \\
\midrule
B1 & Less time on commute & \mybarchart{0.0405888}{0.176182}{0.307315}{0.475914} & 96\% & 82\% & \BEGINADDED 63.3\% & \BEGINADDED 68.8\% & \BEGINADDED +5.5 &  \\ 
B2 & Spending less money & \mybarchart{0.163758}{0.284116}{0.3217}{0.230425} & 84\% & 66\% & \BEGINADDED 56.2\% & \BEGINADDED 70.9\% & \BEGINADDED +14.7 & (*) \\ 
B3 & Flexible work hours & \mybarchart{0.190775}{0.142857}{0.361845}{0.304523} & 81\% & 82\% & \BEGINADDED 57.9\% & \BEGINADDED 71.1\% & \BEGINADDED +13.2 & (*) \\ 
B4 & Closer to family & \mybarchart{0.194892}{0.12052}{0.380376}{0.304211} & 81\% & 85\% & \BEGINADDED 61.5\% & \BEGINADDED 70.4\% & \BEGINADDED +8.9 & (*) \\ 
B5 & More comfortable clothing & \mybarchart{0.197038}{0.416517}{0.236984}{0.149461} & 80\% & 48\% & \BEGINADDED 58.9\% & \BEGINADDED 70.9\% & \BEGINADDED +12.0 & (*) \\ 
B6 & Reduced health risks & \mybarchart{0.275395}{0.0884876}{0.256433}{0.379684} & 72\% & 88\% & \BEGINADDED 63.5\% & \BEGINADDED 70.5\% & \BEGINADDED +7.0 & (*) \\ 
B7 & Better Focus Time & \mybarchart{0.375619}{0.0458839}{0.316689}{0.261808} & 62\% & 93\% & \BEGINADDED 42.1\% & \BEGINADDED 84.7\% & \BEGINADDED +42.6 & (*) \\ 
B8 & Less distractions or interruptions & \mybarchart{0.446789}{0.0740907}{0.28379}{0.19533} & 55\% & 87\% & \BEGINADDED 49.2\% & \BEGINADDED 84.6\% & \BEGINADDED +35.4 & (*) \\ 
B9 & More time to complete work & \mybarchart{0.477498}{0.105311}{0.266877}{0.150315} & 52\% & 80\% & \BEGINADDED 55.4\% & \BEGINADDED 80.6\% & \BEGINADDED +25.2 & (*) \\ 
B10 & More breaks & \mybarchart{0.480432}{0.186685}{0.25596}{0.0769231} & 52\% & 64\% & \BEGINADDED 69.3\% & \BEGINADDED 68.0\% & \BEGINADDED --1.3 &  \\ 
B11 & Better Work Life Balance & \mybarchart{0.496861}{0.0255605}{0.241704}{0.235874} & 50\% & 95\% & \BEGINADDED 57.9\% & \BEGINADDED 79.1\% & \BEGINADDED +21.2 & (*) \\ 
B12 & Better Work Environment & \mybarchart{0.524175}{0.0899232}{0.222775}{0.163127} & 48\% & 81\% & \BEGINADDED 52.4\% & \BEGINADDED 86.5\% & \BEGINADDED +34.1 & (*) \\ 
B13 & More Efficient Meetings & \mybarchart{0.543694}{0.0581081}{0.26982}{0.128378} & 46\% & 87\% & \BEGINADDED 59.5\% & \BEGINADDED 79.5\% & \BEGINADDED +20.0 & (*) \\ 
B14 & More Control Over Work & \mybarchart{0.628764}{0.0485393}{0.216629}{0.106067} & 37\% & 87\% & \BEGINADDED 60.7\% & \BEGINADDED 82.1\% & \BEGINADDED +21.4 & (*) \\ 
B15 & More physical activity & \mybarchart{0.660242}{0.0291349}{0.167638}{0.142985} & 34\% & 91\% & \BEGINADDED 62.4\% & \BEGINADDED 80.3\% & \BEGINADDED +17.9 & (*) \\ 
\bottomrule
\end{tabular}
}%

\end{table}

%% file: tables/benefitslasso.tex
\begin{table}%
\centering
\caption{Results from the Lasso logit analysis. The dependent variable was whether a participant reported that productivity stayed the same or increased. The explanatory variables are the direct effects and interactions for experienced benefits that were considered as important or very important. \BEGINADDED The logit coefficients are converted to marginal effects for interpretability. The marginal effect is the difference between changing the variable from 0 (absent) to 1 (present). The overall mean is the percentage of participants who reported that productivity stayed the same or increased.}
\label{tab:benefitsLASSO}

\begin{tabular}{lr@{}lc}
\toprule
 Benefit  & \multicolumn{2}{l}{\BEGINADDED\stackanchor{Marginal}{Effect}} & Std.\ Error \\
\midrule
Better focus time (B7)    & \BEGINADDED +0.189&$^{***}$ &\BEGINADDED (0.026) \\ 
Less distractions or interruptions (B8) & \BEGINADDED +0.106&$^{***}$ &\BEGINADDED (0.033) \\ 
Better work environment (B12) & \BEGINADDED +0.036& & \BEGINADDED (0.059) \\[4pt] 
Less time on commute (B1)  \\
\hspace{15pt}\emph{and} Better work environment (B12)  & \BEGINADDED +0.157&$^{***}$ &\BEGINADDED (0.055) \\[4pt]
Less time on commute (B1)  \\
\hspace{15pt}\emph{and} More time to complete work (B9) & \BEGINADDED +0.122&$^{*\ADDED{**}}$ &\BEGINADDED (0.023) \\[4pt] 
\BEGINADDED Less distractions or interruptions (B8)  \\
\BEGINADDED \hspace{15pt}\emph{and} Closer to family (B4) & \BEGINADDED +0.089&$^{\ADDED{**}}$ & \BEGINADDED (0.035)\\[4pt]
\midrule
\BEGINADDED
Overall Mean & \BEGINADDED 0.685\\
\midrule
Observations & 2,104\\ 
\BEGINADDED Pseudo-R$^2$ & \BEGINADDED 0.238\\
\midrule
\multicolumn{1}{l}{\textit{Note:} $^{*}$p$<$0.1; $^{**}$p$<$0.05; $^{***}$p$<$0.01} \\ 
\bottomrule
\end{tabular}
\end{table}

%% file: tables/challenges.tex
\begin{table}\centering

\settowidth{\myDistLength}{Distribution}  
\settoheight{\myDistHeight}{Distribution}  

\def\mybarchart#1#2#3{%
\resizebox {\myDistLength} {\myDistHeight} {%
\begin{tikzpicture}[]
\definecolor{clr1}{RGB}{240,240,240}
\definecolor{clr2}{RGB}{189,189,189}
\definecolor{clr3}{RGB}{99,99,99}
\begin{axis}[
      axis background/.style={fill=none, draw=none}, %
      axis line style={draw=none},
      tick style={draw=none},
      ytick=\empty,
      xtick=\empty,
      ymin=0, ymax=1, %
      xmin=0, xmax=1]
\addplot [
      ybar interval=.5,
      fill=clr3,
      draw=none,
]
	coordinates {(#1+#2+#3,1) (#1+#2,1)}; 
\addplot [
      ybar interval=.5,
      fill=clr2,
      draw=none,
]
	coordinates {(#1+#2,1) (#1,1)}; 
\addplot [
      ybar interval=.5,
      fill=clr1,
      draw=none,
]
	coordinates {(#1,1) (0,1)}; 
\end{axis}%
\end{tikzpicture}%
}%
}

\caption{Challenges in Survey 2. The column \emph{Distribution} shows the distribution of responses that do not experience a challenge (light gray\protect\mybarchartbox{clgclr1} ), experience this challenge as a minor issue (gray\protect\mybarchartbox{clgclr2} ), and experience this challenge as a major issue (dark gray\protect\mybarchartbox{clgclr3} ). The following column \emph{Prevalence} indicates the percentage of respondents that experienced the challenge ( \protect\mybarcharttwoboxes{0}{0.5}{0.5}{clgclr2}{clgclr3} ), while the column \emph{Impact} describes the percentage of participants that indicated this challenge presented a major issue (percentage of \protect\mybarchartbox{clgclr3} with respect to \protect\mybarcharttwoboxes{0}{0.5}{0.5}{clgclr2}{clgclr3} ). The column \BEGINADDED\emph{Productivity Unchanged or Increased} reports the percentage of respondents who reported that productivity stayed ``about the same'' or increased (``more productive'', ``significantly more productive'') for respondents who \emph{did not} experience the challenge vs. respondents who \emph{did} experience the challenge; \ENDADDED statistically significant differences ($p<.01$, with Benjamini-Hochberg correction~\cite{benjamini1995controlling}) are indicated with an asterisk (*).  The challenges are sorted and numbered in descending order by the \emph{Prevalence} column.}
\label{tab:challenges}

\resizebox{\textwidth}{!}{%
\begin{tabular}{@{}rllcccccl@{}}
\toprule
& & & & & \multicolumn{4}{c}{\BEGINADDED Productivity Unchanged or Increased} \\
\cmidrule{6-9}
& Challenge & Distribution & Prevalence & Impact & \BEGINADDED \stackanchor{Challenge \textbf{not}}{experienced} & \BEGINADDED \stackanchor{Challenge}{experienced} & \BEGINADDED \stackanchor{Delta}{\small(pct points)} & \\
\midrule

C1 & Missing social interactions & \mybarchart{0.172939}{0.422043}{0.405018} & 83\% & 49\% & \BEGINADDED 82.4\% & \BEGINADDED 65.7\% & \BEGINADDED --16.7 & (*) \\ 
C2 & Lack of work life boundary & \mybarchart{0.221076}{0.407623}{0.3713} & 78\% & 48\% & \BEGINADDED 83.0\% & \BEGINADDED 64.5\% & \BEGINADDED --18.5 & (*) \\ 
C3 & Poor ergonomics & \mybarchart{0.297422}{0.340383}{0.362194} & 70\% & 52\% & \BEGINADDED 80.4\% & \BEGINADDED 64.4\% & \BEGINADDED --16.0 & (*) \\ 
C4 & Less awareness of colleagues  work & \mybarchart{0.346223}{0.419065}{0.234712} & 65\% & 36\% & \BEGINADDED 81.9\% & \BEGINADDED 61.6\% & \BEGINADDED --20.4 & (*) \\ 
C5 & Less physical activity & \mybarchart{0.350247}{0.315671}{0.334082} & 65\% & 51\% & \BEGINADDED 78.5\% & \BEGINADDED 63.3\% & \BEGINADDED --15.2 & (*) \\ 
C6 & Difficult to communicate with colleagues & \mybarchart{0.42973}{0.375676}{0.194595} & 57\% & 34\% & \BEGINADDED 81.1\% & \BEGINADDED 59.2\% & \BEGINADDED --21.9 & (*) \\ 
C7 & Insufficient hardware & \mybarchart{0.434802}{0.339478}{0.225719} & 57\% & 40\% & \BEGINADDED 77.1\% & \BEGINADDED 62.0\% & \BEGINADDED --15.1 & (*) \\ 
C8 & Connectivity problems & \mybarchart{0.463885}{0.350381}{0.185734} & 54\% & 35\% & \BEGINADDED 75.9\% & \BEGINADDED 62.1\% & \BEGINADDED --13.8 & (*) \\ 
C9 & Poor work life balance & \mybarchart{0.489869}{0.279154}{0.230977} & 51\% & 45\% & \BEGINADDED 78.5\% & \BEGINADDED 59.0\% & \BEGINADDED --19.5 & (*) \\ 
C10 & Too many meetings & \mybarchart{0.492335}{0.287647}{0.220018} & 51\% & 43\% & \BEGINADDED 71.2\% & \BEGINADDED 66.1\% & \BEGINADDED --5.0 &  \\ 
C11 & More distractions or interruptions & \mybarchart{0.505628}{0.312472}{0.1819} & 49\% & 37\% & \BEGINADDED 86.6\% & \BEGINADDED 50.0\% & \BEGINADDED --36.5 & (*) \\ 
C12 & Lack of a routine & \mybarchart{0.52518}{0.296763}{0.178058} & 47\% & 38\% & \BEGINADDED 80.5\% & \BEGINADDED 55.4\% & \BEGINADDED --25.1 & (*) \\ 
C13 & Fewer breaks & \mybarchart{0.561064}{0.285264}{0.153673} & 44\% & 35\% & \BEGINADDED 69.9\% & \BEGINADDED 66.9\% & \BEGINADDED --3.0 &  \\ 
C14 & Friction with collaboration tools & \mybarchart{0.563177}{0.338899}{0.0979242} & 44\% & 22\% & \BEGINADDED 75.2\% & \BEGINADDED 60.0\% & \BEGINADDED --15.2 & (*) \\ 
C15 & Lack of motivation & \mybarchart{0.582318}{0.266125}{0.151556} & 42\% & 36\% & \BEGINADDED 82.6\% & \BEGINADDED 48.9\% & \BEGINADDED --33.7 & (*) \\ 
C16 & Blocked waiting on others & \mybarchart{0.596752}{0.290483}{0.112765} & 40\% & 28\% & \BEGINADDED 75.0\% & \BEGINADDED 59.1\% & \BEGINADDED --15.9 & (*) \\ 
C17 & Poor home work environment & \mybarchart{0.602882}{0.263845}{0.133273} & 40\% & 34\% & \BEGINADDED 81.5\% & \BEGINADDED 48.9\% & \BEGINADDED --32.6 & (*) \\ 
C18 & Lack of dining options & \mybarchart{0.666967}{0.251575}{0.0814581} & 33\% & 24\% & \BEGINADDED 72.6\% & \BEGINADDED 60.2\% & \BEGINADDED --12.4 & (*) \\ 
C19 & Lack of childcare & \mybarchart{0.730873}{0.112511}{0.156616} & 27\% & 58\% & \BEGINADDED 72.5\% & \BEGINADDED 57.8\% & \BEGINADDED --14.7 & (*) \\ 
C20 & Less time to complete work & \mybarchart{0.756208}{0.156659}{0.0871332} & 24\% & 36\% & \BEGINADDED 76.2\% & \BEGINADDED 45.1\% & \BEGINADDED --31.2 & (*) \\ 

\bottomrule
\end{tabular}
} %

\end{table}

%% file: tables/challengeslasso.tex
\begin{table}\centering

\settowidth{\myDistLength}{Distribution}  

\def\mybarchart#1#2#3{
\resizebox {\the\myDistLength} {7.5pt} {%
\begin{tikzpicture}[]
\definecolor{clr1}{RGB}{240,240,240}
\definecolor{clr2}{RGB}{189,189,189}
\definecolor{clr3}{RGB}{99,99,99}
\begin{axis}[
      axis background/.style={fill=gray!10, draw=gray!50},
      axis line style={draw=none},
      tick style={draw=none},
      ytick=\empty,
      xtick=\empty,
      ymin=0, ymax=1, %
      xmin=0, xmax=1]
\addplot [
      ybar interval=.5,
      fill=clr3,
      draw=none,
]
	coordinates {(#1+#2+#3,1) (#1+#2,1)}; 
\addplot [
      ybar interval=.5,
      fill=clr2,
      draw=none,
]
	coordinates {(#1+#2,1) (#1,1)}; 
\addplot [
      ybar interval=.5,
      fill=clr1,
      draw=none,
]
	coordinates {(#1,1) (0,1)}; 
\end{axis}%
\end{tikzpicture}%
}%
}

\caption{Results from the Lasso logit analysis. The dependent variable was whether a participant reported that productivity stayed the same or increased. The explanatory variables are the direct effects and interactions for major challenges. \BEGINADDED The logit coefficients are converted to marginal effects for interpretability. The marginal effect is the difference between changing the variable from 0 (absent) to 1 (present). The overall mean is the percentage of participants who reported that productivity stayed the same or increased.}%
\label{tab:challengesLASSO}

\begin{tabular}{lr@{}lc}
\toprule
 Challenge & \multicolumn{2}{c}{\BEGINADDED\stackanchor{Marginal}{Effect}} & Std.\ Error \\
\midrule
More distractions or interruptions (C11) & \BEGINADDED $-$0.409&$^{***}$& \BEGINADDED (0.033) \\ 
Lack of motivation (C15) & \BEGINADDED $-$0.264&$^{***}$ &\BEGINADDED (0.037) \\ 
Difficult to communicate with colleagues (C6) & \BEGINADDED $-$0.131&$^{***}$ &\BEGINADDED (0.034) \\ 
Connectivity problems (C8)& \BEGINADDED $-$0.119&$^{***}$ &\BEGINADDED (0.032) \\ 
Missing social interactions (C1) & \BEGINADDED $-$0.112&$^{***}$ &\BEGINADDED (0.025) \\ 
\BEGINADDED Poor home work environment (C17)& \BEGINADDED  $-$0.080&$^{\ADDED{**}}$& \BEGINADDED (0.039)\\
Less awareness of colleagues work (C4)& \BEGINADDED $-$0.062&$^{**}$ &\BEGINADDED (0.030) \\[4pt]
Less time to complete work (C20) & \BEGINADDED $-$0.081&$^{}$&\BEGINADDED (0.060)\\[4pt]
Lack of childcare (C19) \\
\hspace{15pt} \emph{and} Less time to complete work (C20) &\BEGINADDED $-$0.340&$^{***}$ &\BEGINADDED (0.099) \\ 
\midrule 
\BEGINADDED
Overall Mean & \BEGINADDED 0.685\\
\midrule
Observations & 2,106 \\ 
\BEGINADDED Pseudo-R$^{2}$ & \BEGINADDED 0.243 \\ 
\midrule 
\multicolumn{1}{l}{\textit{Note:} $^{*}$p$<$0.1; $^{**}$p$<$0.05; $^{***}$p$<$0.01} \\ 
\bottomrule
\end{tabular}

 \end{table}

%% file: tables/improvements.tex
\begin{table}\centering

\settowidth{\myDistLength}{Distribution}  

\def\mybarchart#1#2#3{
\resizebox {\the\myDistLength} {7.5pt} {%
\begin{tikzpicture}[]
\definecolor{clr1}{RGB}{99,99,99}
\definecolor{clr2}{RGB}{240,240,240}
\begin{axis}[
      axis background/.style={fill=gray!10, draw=gray!50},
      axis line style={draw=none},
      tick style={draw=none},
      ytick=\empty,
      xtick=\empty,
      ymin=0, ymax=1, %
      xmin=0, xmax=1]
\addplot [
      ybar interval=.5,
      fill=clr2,
      draw=none,
]
	coordinates {(#1+#2,1) (#1,1)}; 
\addplot [
      ybar interval=.5,
      fill=clr1,
      draw=none,
]
	coordinates {(#1,1) (0,1)}; 
\end{axis}%
\end{tikzpicture}%
}%
}

\caption{Improvements in Survey 2. Participants could select up to three items. Column ``All'' indicates the frequency the improvement was suggested among all respondents, ``Low'' the frequency among respondents who reported a decrease in productivity, and ``High'' the frequency among respondents who reported an increase in productivity. Differences between the frequency for ``Low'' and ``High'' that are statistically significant with $p<.01$ after Benjamini-Hochberg correction~\cite{benjamini1995controlling} are labeled with an asterisk (*).}
\label{tab:improvements}

\resizebox{\textwidth}{!}{%
\begin{tabular}{@{}r@{\phantom{X}}lrrr@{\phantom{X}}l@{}}
\toprule
& Improvement & All & Low & High \\
\midrule
S1 & Provide more/better hardware for home (more screens, more powerful laptop, etc) & 41.6\% & 39.1\% & 42.7\% &  \\ 
S2 & Improve connectivity (fewer VPN drops, reimburse for faster internet) & 41.5\% & 45.8\% & 35.5\% & (*) \\ 
S3 & Provide a stipend for improving work from home environment & 40.8\% & 42.4\% & 39.4\% &  \\ 
S4 & Make improvements to communication tools & 33.1\% & 36.9\% & 29.3\% & (*) \\ 
S5 & Provide ergonomic furniture & 30.1\% & 17.2\% & 40.7\% & (*) \\ 
S6 & Support remote work better during normal circumstances & 22.3\% & 15.1\% & 30.7\% & (*) \\ 
S7 & Provide guidance for successfully working from home (e.g., online meeting ettiquette) & 20.8\% & 23.7\% & 16.7\% & (*) \\ 
S8 & Improve and encourage team socialization & 16.8\% & 16.9\% & 15.8\% &  \\ 
S9 & Be more understanding of WFH scenarios beyond COVID-19 & 14.6\% & 15.7\% & 15.0\% &  \\ 
S10 & Encourage people to be more responsive & 9.8\% & 8.9\% & 12.0\% &  \\ 
S11 & Minimize the number of meetings & 5.8\% & 7.6\% & 3.8\% & (*) \\ 
S12 & Give guidance to management on how to manage WFH employees & 5.6\% & 5.9\% & 5.6\% &  \\ 
\bottomrule
\end{tabular}
} %
\end{table}

%% file: tables/codebook.tex
We identified \numberCodes codes in the following six themes.

\themetitle{Beyond Work} \par \codetitle{Ecological Impact} The impact of working from home on ecological factors (e.g. affecting the environment).  The most common is less pollution due to not commuting. \par
\codetitle{Family, Children, and  Pets} Factors related to pets, children, and family. This includes the proximity to them, interruptions from them, lack of childcare, and needing to help children who are doing school remotely from home. \bcreference{B4, C19} \par
\codetitle{Food} The impact of working from home on meals and snacks.  This includes quality and quantitiy of food, access to food, diversity of food consumed, and the need to or opportunity to cook for one's self. \bcreference{C18} \par
\codetitle{House Work} The impact of working from home on home-related tasks or activities such as laundry, paying bills, picking up packages, chores, and maintenance. \par
\codetitle{Money} The impact on money and spending.  This may include spending less money due to not eating out or commuting as well as spending more money on groceries, setting up a home office, or upgrading internet. \bcreference{B2} \par
\themetitle{Collaboration} \par \codetitle{Blocks} Comments about being blocked from making progress due to waiting on others to relay information, make decisions, or complete pre-requisite tasks. \bcreference{C16} \par
\codetitle{Collaboration} Aspects of coordination or collaboration that are not explicitly about communication.  This also includes general statements about collaboration such as "Collaboration is worse" or "It's hard to be creative with people". \par
\codetitle{Meetings} Explicit mentions of meetings, including frequency, duration, time of day, quality, size, formal versus informal, and communication channels used. \bcreference{B13, C10} \par
\codetitle{Social Connections} Non-work communication with co-workers (e.g. to help facilitate work bonds).  This includes the difficulty of managing, forming, or maintaining informal and team relationships as well as feeling isolated and missing social connections. \bcreference{C1} \par
\codetitle{Team} Team characteristics such as team culture, team social activities, team productivity, and team mood. \par
\themetitle{Communication} \par \codetitle{Channels} Discussions of the use of various communication channels such as Teams chats and calls, Email, instant messaging, including comments about them such as using too many tools, difficulty of use, and benefits of different tools.  In addition, this includes comparisons of tools to working in office such as the lack of in-person communication or missing richness of communicating at a whiteboard. \bcreference{C14} \par
\codetitle{Communication Gaps} Challenges around communication such as it being difficult to connect with particular people (for example, because schedules are more flexible), hard to communicate, missing communication, and miscommunication.  This also includes lack of awareness of what others are working on. \bcreference{C4, C6} \par
\codetitle{Formal Communication} Formal communication such as scheduled chats that are work related. \par
\codetitle{Informal Communication} Unscheduled, informal, or ad-hoc communication that is work related.  This includes the inability to drop by someone's office or run into someone in the break room as well as the use of tools (e.g. Teams) for frequent, short interactions. \par
\themetitle{Well-being} \par \codetitle{Breaks} Taking or needing more or less breaks (including meal breaks or walks); Reasons for less breaks such as having meetings are back to back \bcreference{B10, C13} \par
\codetitle{Healthy Habits} e.g., diet, explicitly saying being "healthy", physical activity (working out), walking between meetings; too much time on the computer; too much time at home \bcreference{B15, C5} \par
\codetitle{Mental Health} work related stress; personal stress, anxiety; burnout; fatigue; loneliness \bcreference{B6} \par
\codetitle{No Commute} The impact of not having a commute.  This includes benefits such as less wasted time, but also negatives such as missing reading on the bus or calling relatives on the drive into work. \bcreference{B1} \par
\codetitle{Personal Comfort} The impact on working from home on personal comforts such as listening to music without headphones, wearing more comfortable clothes, or creating a more comfortable working space. \bcreference{B5} \par
\codetitle{Routine} Any mentions of routine.  May include "missing a routine" such as  "I miss having breakfast every morning" or the importance of maintaining a routine. \bcreference{C12} \par
\codetitle{Schedule Flexibility} The ability to and impact of working outside of the traditional "9-5" work day.  This includes working outside of non-work hours as well as doing non-work related activities (e.g. laundry) during traditional working hours. \bcreference{B3} \par
\codetitle{Work Hours} Whether the number of hours worked during the day stayed the same, went up or went down.  For instance, "I get the same amount done, but I'm working 12 hour days to do it." \bcreference{B9, C20} \par
\codetitle{Work-Life Balance} Changes in boundaries between work and non-work life and the ability (or lack) to not let work concerns or responsibilities interfere with non-work activities. \bcreference{B11, C2, C9} \par
\codetitle{Focus} The impact of working from home on the ability to focus, or the impact of various factors on focus time \bcreference{B7} \par
\codetitle{Interruptions and Distractions} Interruptions or distractions (or the lack of them) when working from home, whether work related or not. \bcreference{B8, C11} \par
\codetitle{Motivation} Various intrisic and extrinsic factors affecting motivation as well as differences or changes in motivation. \bcreference{C15} \par
\codetitle{Productivity} Discussion of perceived productivity and the impact of various factors on productivity \par
\themetitle{Work Environment} \par \codetitle{Connectivity} The challenges of connectivity such as internet speed and latency, "remoting in" to a machine at work to accomplish work, using secured machines, connections to remote machines going up and down, and the need to reboot machines remotely. \bcreference{C8} \par
\codetitle{Environment} Aspects of the physical work environment such as access to natural light, now having a window, dedicated (e.g. a study) vs non-dedicated (e.g., kitchen table) space, having privacy, temperature, more or less noise. \bcreference{B12, C17} \par
\codetitle{Ergonomics} The availability or absense of ergonomic furniture, often in comparison to the work office environment. \bcreference{C3} \par
\codetitle{Furniture} References to furniture that do not mention or allude to ergonomics.  Items such as whiteboards or bookcases are also included. \par
\codetitle{Hardware} Differences in displays (quantity and quality), and machines (also quantity and quality) as well as accessories such as mice, keyboards, headphones, and webcams. \bcreference{C7} \par

%% file: tables/codecount.tex
\begin{table}\centering

\caption{Counts and Ranks of the Codes in Survey 1. The columns under \emph{Counts} indicate the frequency of each code within questions ``Please share details about your answer to the previous question on how your productivity has changed.'' (Q14), ``What is \textbf{good} about working from home?'' (Q15), ``What is \textbf{bad} about working from home?'' (Q16), ``What \textbf{challenges} have you encountered working from home?'' (Q17), and all four questions combined (Total). The columns under \emph{Ranks} indicate the rank of each code with respect to the other codes for Q14, Q15, Q16, Q17, and all four questions combined (Total). The most frequent code is \#1.}
\label{tab:codecount}

\small
\resizebox{\textwidth}{!}{%
\begin{tabular}{lrrrrrlrrrrr}
\toprule
& \multicolumn{5}{c}{\textbf{Counts}} & &  \multicolumn{5}{c}{\textbf{Ranks (\#)}} \\
\cmidrule(r){2-6}\cmidrule(l){8-12}
Code & Q14 & Q15 & Q16 & Q17 & Total & &  Q14 & Q15 & Q16 & Q17 & Total \\
\midrule
\multicolumn{10}{l}{\textbf{Beyond Work}} \\ 
\hspace{2mm} Ecological Impact & 0 & 6 & 0 & 0 & 6 & & 30 & 18 & 30 & 31 & 32 \\
\hspace{2mm} Family, Children, and Pets & 65 & 61 & 38 & 61 & 225 & & 3 & 6 & 9 & 3 & 3 \\
\hspace{2mm} Food & 7 & 32 & 17 & 9 & 65 & & 22 & 10 & 16 & 23 & 22 \\
\hspace{2mm} House Work & 6 & 33 & 0 & 8 & 47 & & 26 & 9 & 30 & 25 & 25 \\
\hspace{2mm} Money & 0 & 11 & 1 & 1 & 13 & & 30 & 16 & 28 & 28 & 31 \\
\cmidrule{1-1} \multicolumn{10}{l}{\textbf{Collaboration}} \\ %
\hspace{2mm} Blocks & 3 & 0 & 14 & 1 & 18 & & 28 & 26 & 21 & 28 & 29 \\
\hspace{2mm} Collaboration & 12 & 3 & 38 & 16 & 69 & & 19 & 22 & 9 & 20 & 20 \\
\hspace{2mm} Meetings & 39 & 29 & 22 & 18 & 108 & & 7 & 11 & 14 & 18 & 12 \\
\hspace{2mm} Social Connections & 12 & 1 & 94 & 21 & 128 & & 19 & 24 & 1 & 13 & 10 \\
\hspace{2mm} Team & 7 & 1 & 8 & 1 & 17 & & 22 & 24 & 26 & 28 & 30 \\
\cmidrule{1-1} \multicolumn{10}{l}{\textbf{Communication}} \\ %
\hspace{2mm} Channels & 63 & 5 & 87 & 66 & 221 & & 4 & 19 & 2 & 2 & 4 \\
\hspace{2mm} Communication Gaps & 13 & 0 & 12 & 27 & 52 & & 17 & 26 & 22 & 9 & 24 \\
\hspace{2mm} Formal Communication & 0 & 4 & 16 & 14 & 34 & & 30 & 21 & 19 & 21 & 26 \\
\hspace{2mm} Informal Communication & 61 & 0 & 49 & 26 & 136 & & 5 & 26 & 4 & 10 & 9 \\
\cmidrule{1-1} \multicolumn{10}{l}{\textbf{Well-being}} \\ %
\hspace{2mm} Breaks & 19 & 21 & 17 & 20 & 77 & & 14 & 14 & 16 & 15 & 19 \\
\hspace{2mm} Commute & 46 & 229 & 5 & 0 & 280 & & 6 & 1 & 27 & 31 & 2 \\
\hspace{2mm} Healthy Habits & 7 & 27 & 48 & 21 & 103 & & 22 & 12 & 5 & 13 & 14 \\
\hspace{2mm} Mental Health & 17 & 45 & 22 & 9 & 93 & & 15 & 8 & 14 & 23 & 15 \\
\hspace{2mm} Personal Comfort & 3 & 48 & 0 & 7 & 58 & & 28 & 7 & 30 & 26 & 23 \\
\hspace{2mm} Routine & 7 & 0 & 12 & 4 & 23 & & 22 & 26 & 22 & 27 & 28 \\
\hspace{2mm} Schedule Flexibility & 21 & 71 & 1 & 25 & 118 & & 13 & 4 & 28 & 11 & 11 \\
\hspace{2mm} Work Hours & 25 & 16 & 45 & 20 & 106 & & 12 & 15 & 8 & 15 & 13 \\
\hspace{2mm} Work-Life Balance & 11 & 10 & 47 & 20 & 88 & & 21 & 17 & 6 & 15 & 16 \\
\cmidrule{1-1} \multicolumn{10}{l}{\textbf{Work}} \\ %
\hspace{2mm} Focus & 83 & 80 & 12 & 25 & 200 & & 2 & 3 & 22 & 11 & 5 \\
\hspace{2mm} Interruptions and Distractions & 193 & 91 & 87 & 35 & 406 & & 1 & 2 & 2 & 6 & 1 \\
\hspace{2mm} Motivation & 5 & 0 & 12 & 10 & 27 & & 27 & 26 & 22 & 22 & 27 \\
\hspace{2mm} Productivity & 31 & 24 & 16 & 17 & 88 & & 11 & 13 & 19 & 19 & 16 \\
\cmidrule{1-1} \multicolumn{10}{l}{\textbf{Work environment}} \\ %
\hspace{2mm} Connectivity & 34 & 0 & 34 & 99 & 167 & & 10 & 26 & 12 & 1 & 7 \\
\hspace{2mm} Environment & 39 & 63 & 26 & 57 & 185 & & 7 & 5 & 13 & 4 & 6 \\
\hspace{2mm} Ergonomics & 13 & 0 & 36 & 33 & 82 & & 17 & 26 & 11 & 7 & 18 \\
\hspace{2mm} Furniture & 16 & 2 & 17 & 33 & 68 & & 16 & 23 & 16 & 7 & 21 \\
\hspace{2mm} Hardware & 38 & 5 & 47 & 51 & 141 & & 9 & 19 & 6 & 5 & 8 \\
\bottomrule
\end{tabular}
} %
\end{table}